\documentclass[manuscript,iop, numberedappendix]{emulateapj}
\usepackage{graphicx}
\usepackage{threeparttable}
\usepackage{apjfonts}
\usepackage{layouts}
\usepackage{rotating}
\usepackage[usenames, dvipsnames]{color}

\usepackage{natbib}
\usepackage{color}
\definecolor{citeRGB}{rgb}{0,0.1,0.7}
\usepackage[hyperfootnotes=true,naturalnames=true,letterpaper,pdfstartview=FitH,pdfpagemode=UseNone,colorlinks=true,citecolor=citeRGB]{hyperref}
\usepackage[table]{xcolor}

\shorttitle{The 3D-HST Survey}
\shortauthors{Momcheva, Brammer, van Dokkum et al.}

\slugcomment{Submitted to The Astrophysical Journal Supplement Series}

\begin{document} 

\newcommand\XXX[1]{{\textcolor{red}{\textbf{x\ #1\ x}}}}
\newcommand\XX{{\textcolor{red}{\textbf{XX}}}}
\newcommand{\calwf}{{\tt calwf3}}
\newcommand{\kms}{${\rm km \; s^{-1}}$}
\newcommand\zeroth{0$^\mathrm{th}$}
\newcommand\aXe{{\tt aXe}}
\newcommand{\JH}{JH_{IR}}
\vspace{-1.5cm}
\title{The 3D-HST Survey:
Hubble Space Telescope WFC3/G141 grism spectra, redshifts, and emission line measurements for $\sim 100,000$
galaxies}

\author{Ivelina G. Momcheva,\altaffilmark{1} Gabriel B. Brammer,\altaffilmark{2} Pieter G. van Dokkum,\altaffilmark{1}\\
and\\
Rosalind E. Skelton,\altaffilmark{1,3} Katherine E. Whitaker,\altaffilmark{4,17}  
Erica J. Nelson,\altaffilmark{1}  
Mattia Fumagalli,\altaffilmark{5}  
Michael V. Maseda,\altaffilmark{6} 
Joel Leja,\altaffilmark{1} 
Marijn Franx,\altaffilmark{5}
Hans-Walter Rix,\altaffilmark{6} 
Rachel Bezanson,\altaffilmark{7,17} 
Elisabete Da Cunha,\altaffilmark{8}  
Claire Dickey,\altaffilmark{1} 
Natascha M. F\"orster Schreiber,\altaffilmark{9} 
Garth Illingworth,\altaffilmark{10} 
Mariska Kriek,\altaffilmark{11} 
Ivo Labb\'e,\altaffilmark{5} 
Johannes Ulf Lange,\altaffilmark{1} 
Britt F. Lundgren,\altaffilmark{12} 
Daniel Magee,\altaffilmark{10} 
Danilo Marchesini,\altaffilmark{13} 
Pascal Oesch,\altaffilmark{1} 
Camilla Pacifici,\altaffilmark{2} 
Shannon G. Patel,\altaffilmark{14} 
Sedona Price,\altaffilmark{11} 
Tomer Tal,\altaffilmark{10}  
David A. Wake,\altaffilmark{12,15} 
Arjen van der Wel,\altaffilmark{6} 
Stijn Wuyts\altaffilmark{16}}

\altaffiltext{1}{Department of Astronomy, Yale University, 260 Whitney Avenue, New Haven, CT 06511, USA}
\altaffiltext{2}{Space Telescope Science Institute, Baltimore, MD 21218, USA}
\altaffiltext{3}{South African Astronomical Observatory, Cape Town, 7935, South Africa}
\altaffiltext{4}{Department of Astronomy, University of Massachusetts, Amherst, MA 01003, USA}
\altaffiltext{5}{Leiden Observatory, Leiden University, Leiden, The Netherlands}
\altaffiltext{6}{Max Planck Institute for Astronomy, D-69117, Heidelberg, Germany}
\altaffiltext{7}{Steward Observatory, University of Arizona, Tucson, AZ 85721, USA}
\altaffiltext{8}{Centre for Astrophysics and Supercomputing, Swinburne University of Technology, Hawthorn, VIC 3122, Australia}
\altaffiltext{9}{Max-Planck-Institut f\"ur extraterrestrische Physik,  Garching, Germany}
\altaffiltext{10}{Department of Astronomy \& Astrophysics, University of California, Santa Cruz, CA, USA}
\altaffiltext{11}{Astronomy Department, University of California, Berkeley, CA 94720, USA}
\altaffiltext{12}{Department of Astronomy, University of Wisconsin-Madison, Madison, WI 53706, USA}
\altaffiltext{13}{Department of Physics and Astronomy, Tufts University, Medford, MA 02155, USA}
\altaffiltext{14}{Carnegie Observatories, Pasadena, CA 91101, USA}
\altaffiltext{15}{Department of Physical Sciences, The Open University, Milton Keynes, MK7 6AA, UK}
\altaffiltext{16}{Department of Physics, University of Bath, Claverton Down, Bath BA2 7AY, UK}
\altaffiltext{17}{Hubble Fellow}
\altaffiltext{18}{\url{http://3dhst.research.yale.edu}}
\email{ivelina.momcheva@yale.edu}

\begin{abstract}
We present reduced data and data products from the 3D-HST survey, a 248-orbit HST Treasury program. The survey obtained WFC3 G141 grism spectroscopy in four of the five CANDELS fields: AEGIS, COSMOS, GOODS-S, and UDS, along with WFC3 $H_{140}$ imaging, parallel ACS G800L spectroscopy, and parallel $I_{814}$ imaging. In a previous paper (Skelton et al.\ 2014) we presented photometric catalogs in these four fields and in GOODS-N, the fifth CANDELS field. Here we describe and present the WFC3 G141 spectroscopic data, again augmented with data from GO-1600 in GOODS-N (PI: B. Weiner).  The data analysis is complicated by the fact that no slits are used: all objects in the WFC3 field are dispersed, and many spectra overlap. We developed software to automatically and optimally extract interlaced 2D and 1D spectra for all objects in the Skelton et al. (2014)\ photometric catalogs. The 2D spectra and the multi-band photometry were fit simultaneously to determine redshifts and emission line strengths, taking the morphology of the galaxies explicitly into account. The resulting catalog has 98,663 measured redshifts and line strengths down to $\JH\leq 26$ and 22,548 with $\JH\leq 24$, where we comfortably detect continuum emission. Of this sample 5,459 galaxies are at $z>1.5$ and 9,621 are at $0.7<z<1.5$, where H$\alpha$ falls in the G141 wavelength coverage.
Based on comparisons with ground-based spectroscopic redshifts, and on analyses of paired galaxies and repeat observations, the typical redshift error for $\JH\leq 24$ galaxies in our catalog is $\sigma_z \approx 0.003 \times (1+z)$, i.e., one native WFC3 pixel. The $3\sigma$ limit for emission line fluxes of point sources is $1.5\times10^{-17}$\,ergs\,s$^{-1}$\,cm$^{-2}$. We show various representations of the full dataset, as well as individual examples that highlight the range of spectra that we find in the survey. All 2D and 1D spectra, as well as redshifts, line fluxes, and other derived parameters, are publicly available.$^{18}$
\end{abstract}

\section{Introduction}

Since its deployment in 1990 the Hubble Space Telescope ({\it HST}) has not only been used as an imager but also as a spectrograph. Space-based spectroscopy offers the same advantages as space-based imaging: spatial resolution that is difficult or impossible to achieve from Earth and access to wavelength regimes that are blocked by the atmosphere. Whereas dedicated HST spectrographs such as STIS and COS use a slit or a small aperture to isolate the light of an individual object, several of the imaging instruments on HST employ, or employed, a different technique. The NICMOS, ACS, and  WFC3 cameras were all equipped with dispersing grisms that can be placed in the light path in lieu of a filter. This technique is very efficient as it provides spectra of all objects in the imaging field simultaneously  \citep{Pirzkal04, Malhotra05, Straughn08, vanDokkum10}. Such slitless spectroscopy has limited appeal in ground-based astronomy, as the brightness of the sky greatly reduces the signal-to-noise ratio (S/N) compared to slit spectroscopy. However, the much fainter background from space makes slitless HST spectroscopy competitive with, and in several respects superior to, ground-based slit spectroscopy.

While the NICMOS grisms have left little mark on the field of galaxy formation, the ACS grisms were successfully used to obtain deep optical spectroscopy in several fields \citep[e.g., the PEARS and GRAPES surveys;][]{Pirzkal04, Malhotra05, Straughn08}. Among other successes, the ACS G800L data of GRAPES led to the spectroscopic identification of passively evolving galaxies at $z\sim 2$ in the Hubble Ultra Deep Field \citep{Daddi05}. This discovery was aided by a particular aspect of HST grism observations. As galaxies are spatially resolved at HST resolution, their spectra are spread over multiple pixels. Therefore, the S/N is strongly dependent on morphology, and is higher for galaxies that are more compact. As it turns out, the passively evolving galaxies in the Ultra Deep Field have extremely small sizes, yielding relatively high S/N spectra. The high spatial resolution of HST also enables the study of the spatial variation in spectral features; as discussed in detail in \citet{Nelson15} this opens up the possibility to study the spatial distribution of line emission at scales of $\sim 0\farcs 1$.

\begin{table*}[!Ht]
\centering
\caption{3D-HST Fields}\label{table:fields}
\begin{tabular}{lcccc}
\hline \hline
\noalign{\smallskip}
Field & RA & Dec & G141 Area & G800L Area \\
& (h m s) & (d m s) & (arcmin$^2$) & (arcmin$^2$) \\
\noalign{\smallskip}
\hline
\noalign{\smallskip}
AEGIS &14:18:36.00 &  +52:39:00.0 & 121.9  & 102.5 \\
COSMOS & 10:00:31.00 & +02:24:00.0 & 122.2 & 112.7  \\
GOODS-N & 12:35:54.98 & +62:11:51.3 & 116.0 &   84.1  \\
GOODS-S & 03:32:30.00 & -27:47:19.0  &  147.3 & 134.6 \\
UDS & 02:17:49.00 & -05:12:02.0 & 118.7 & 107.4  \\
\noalign{\smallskip}
\hline
\noalign{\smallskip}
Total & & & 626.1 & 541.3  \\
\noalign{\smallskip}
\hline
\end{tabular}
\end{table*}

The grism mode of the WFC3 camera's near-IR channel is realizing the full potential of space-based slitless spectroscopy. Although the sky background from space is lower than that from Earth at all wavelengths, the differences are more pronounced in the near-IR than in the optical: HST's near-IR background is similar to that of a 30m class telescope on Earth. As a result, the per-object sensitivity of WFC3 grism spectroscopy {\em without} slits is similar to that of ground-based spectrographs on 10m telescopes {\em with} slits (as will be quantified later in this paper). With the added benefits of superb spatial resolution and highly efficient multi-plexing, the WFC3 camera is an excellent spectroscopic survey instrument at near-IR wavelengths. It is complementary to ground-based multi-object spectrographs such as MOSFIRE \citep{McLean12} and KMOS \citep{Sharples13, Sharples14}: these ground-based spectrographs have much higher spectral resolution ($R\sim 3500$ for MOSFIRE versus $R\sim 100$ for the WFC3/G141 grism) but cannot match the continuum sensitivity or observing efficiency of WFC3.

The 3D-HST Treasury program \citep{vanDokkum11,Brammer12,Skelton14} has obtained 2-orbit depth WFC3/G141 grism observations over four large sky areas comprising a total of 124 pointings. The G141 grism has a wavelength coverage of $1.1-1.7\,\mu$m, approximately corresponding to ground-based $J$ and $H$ (including the region in between these bands, which is inaccessible from the ground due to H$_2$O absorption). The main aim of the survey is to obtain a large, representative spectroscopic sample of galaxies at $0.7<z<3$, the epoch when most of the stars in the present-day Universe were formed. As we show below, a typical single 2-orbit WFC3/G141 grism observation provides redshifts of $\sim 130$ galaxies at $0.7<z<3$ down to $H_{160}\leq24$ in a 4.6\,arcmin$^2$ area. The survey also obtained ACS/G800L grism observations in parallel, covering $0.5-0.9\,\mu$m, as well as short direct imaging exposures in the WFC3 $JH_{140}$ and ACS $I_{814}$ bands. 

The survey fields were chosen to coincide with those of the CANDELS Multi-Cycle Treasury project \citep{Grogin11,Koekemoer11}. CANDELS has obtained WFC3 and ACS imaging of five fields, comprising a total area of $\sim 0.25$ degree$^2$ (see Table 1). These fields have a wealth of complementary imaging data at other wavelengths from ground- and space-based observatories \citep[see][and references therein]{Grogin11,Brammer12,Skelton14}, and have quickly become the "standard" deep, moderately wide areas for studies of the distant Universe. The four fields observed by the 3D-HST Treasury program are AEGIS, COSMOS, GOODS-S, and UDS. The GOODS-N field had already been observed in a Cycle 17 program (GO-11600; PI Weiner), using a very similar observing strategy.  We have included the GOODS-N data in our analysis and data release, and throughout this paper we discuss the combined grism data set for all five fields. The footprint of 3D-HST\footnote{Except in cases where we specifically refer to ``data from the 3D-HST Treasury program'', the terms ``3D-HST'' and ``the 3D-HST project'' almost always refer to the combination of the CANDELS imaging and all the other ancillary space- and groud-based imaging datasets as presented in \citet{Skelton14}, plus the grism spectroscopy of the 3D-HST Treasury program, GO-11600, and other datasets (Table 2), in all five CANDELS fields.} is slightly smaller than that of CANDELS; approximately 70\,\% of the CANDELS WFC3 area is covered by grism spectroscopy from our program or the Weiner program.

In many cases the grism spectra can stand on their own, particularly for galaxies that have bright (redshifted) emission lines that falls between $1.1\,\mu$m and $1.7\,\mu$m \citep[see, e.g.,][]{Atek10, Straughn11}. However, the value of the grism spectra can be enhanced by combining them with broad- and medium-band photometry at other wavelengths, which is possible in the CANDELS fields \citep[see][]{Skelton14}. We have developed an integrated approach, where the ground- and space-based imaging data are optimally combined with the G141 grism spectroscopy. The combined grism and photometric dataset was used to derive redshifts, measure emission lines, and determine other parameters of {\em all} galaxies in a photometric catalog (not just those with bright lines), down to well defined magnitude limits. These steps can be summarized as follows:  
\begin{enumerate}
\item{We obtained and reduced the available HST/WFC3 imaging in the fields, using the same pixel scale and tangent point as those used by the CANDELS team. The WFC3 imaging includes the CANDELS data and also the Early Release Science data, the HUDF09 Ultra Deep Field campaign, and various other programs.}
\item{Source catalogs were created with SExtractor \citep{Bertin96}, detecting objects in deep combined $J_{125}$ + $JH_{140}$ + $H_{160}$ images.}
\item{These source catalogs, along with the detection images, associated segmentation maps and PSFs,  were used as the basis to measure photometric fluxes at wavelengths $0.3\,\mu$m -- $8\,\mu$m from a large array of publicly available imaging datasets. The resulting SEDs are of high quality, particularly in fields with extensive optical and near-IR medium band photometry.}
\item{The catalogs and segmentation maps were blotted to the original (interlaced) coordinate system of the WFC3 and ACS grism data, and spectra are extracted for each object that is covered by the grism. No source matching is required, and the photometric SEDs can be combined directly with the grism spectroscopy.}
\item{The interlaced 2D spectra and SEDs were fitted simultaneously to measure redshifts, allowing a limited range of emission line ratios.}
\item{With the redshift determined, emission line fluxes were measured from the 2D spectra with no prior on line ratios.}
\item{Stellar population parameters were determined by fitting stellar population synthesis models to the SEDs, using the redshifts as input.}
\item{Mid-IR photometry was obtained from Spitzer/MIPS imaging. These data, combined with rest-frame UV emission measurements from the SEDs, were used to determine star formation rates of the galaxies.}
\item{The set of images, PSFs, and catalogs was used to measure structural parameters of the objects in the WFC3 and ACS bands, following the methodology of \citet{vdWel12}.}
\end{enumerate}
Steps 1--3 have been described in \citet{Skelton14}. This paper also presents the photometric catalogs and images, and provides photometric redshifts and stellar population parameters that are based on the photometry alone (i.e., not on the grism information). Step 8 is discussed in \citet{Whitaker12} and, specifically for the 3D-HST survey, in \citet{Whitaker14}. Step 9 is described in \citet{vdWel14}, who measured the structural parameters of objects in the $J_{125}$ and $H_{160}$ WFC3 bands. All these datasets have been made available for download from the 3D-HST website.
In this paper we discuss steps  4, 5, 6, and 7, and describe the full 3D-HST spectroscopic data release. These steps are inextricably linked to the previously published datasets. The detection and segmentation maps of \citet{Skelton14} are used as inputs for the grism reduction and their WFC3 images are used to create the grism image model.

\begin{figure*}[th]
\centering
\includegraphics[width = \textwidth]{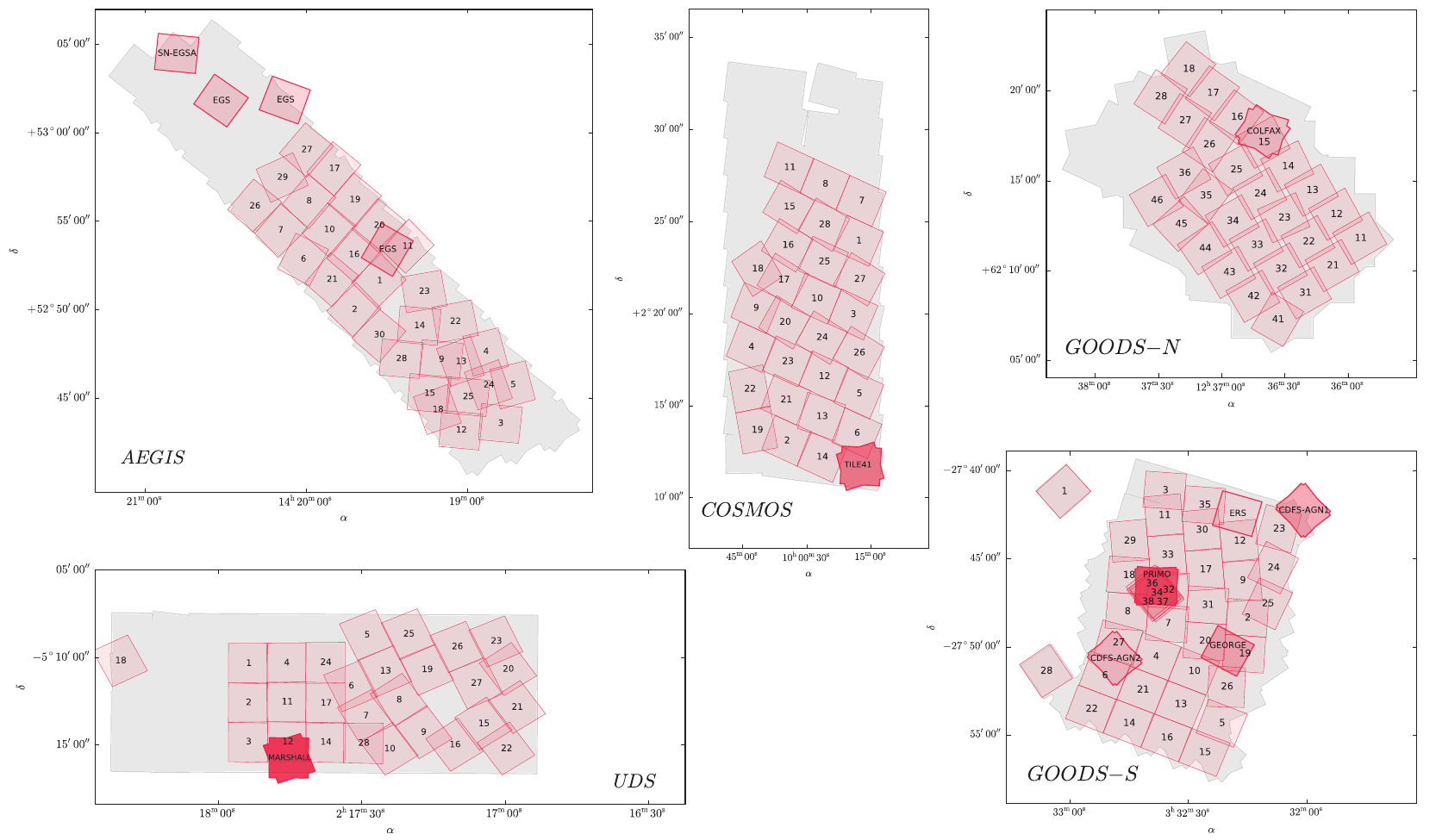}
\caption{\footnotesize Layout of the WFC3 G141 observations. The primary WFC3 G141 pointings are shown with red outlines with the pointing ID numbers as defined in the Phase II file. Observations in the GOODS-N field are from the AGHAST survey (PI: Weiner). Additional pointings are marked with the program or pointing names. The color is proportional to the grism depth, ranging from $\sim$5\,ks for 3D-HST to 60\,ks in the GOODS-S HUDF/PRIMO area. See Table \ref{table:hstdata} for details. }
\label{fig:g141_layout}
\end{figure*}

\begin{figure*}[th]
\centering
\includegraphics[width = \textwidth]{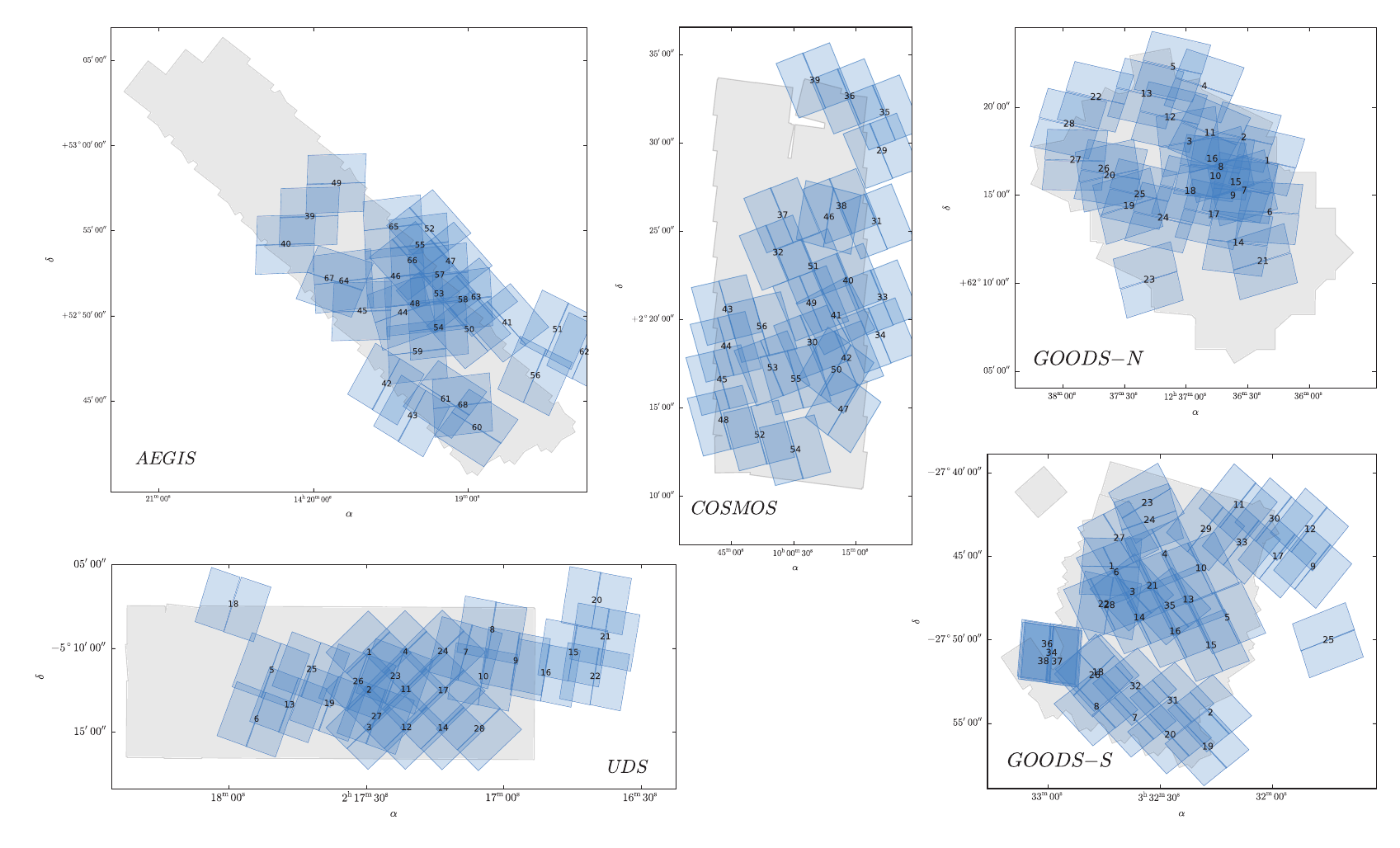}
\caption{\footnotesize Layout of the ACS G800L observations. The observations in GOODS-N are from GO-13420 (PI: Barro). The AEGIS, COSMOS and GOODS-N pointings are numbered differently from their WFC3/IR parallels. The pointing numbers shown in this figure are also used in the data release.}
\label{fig:g800l_layout}
\end{figure*}

\begin{table*}[]
\centering
\caption{WFC3 and ACS Grism Observations in the 3D-HST/CANDELS Fields in Cycles 17 to 21}\label{table:hstdata}
\begin{tabular}{lllccccll}
\hline
\hline
\noalign{\smallskip}
 Field & Instrument & \multicolumn{3}{ c }{Number of Orbits} & Proposal ID & \textit{HST} Cycle & Survey/Pointing & PI  \\
 &  & G800L & G102 & G141  & & & & \\
\noalign{\smallskip}
\hline
\noalign{\smallskip}
AEGIS & WFC3        &  \nodata &  \nodata & 2  & 13063     & 20    & SN EGSA       & Riess     \\
     & WFC3        &   \nodata &  \nodata & 6\textsuperscript{a}  & 12547     & 19    &       & Cooper       \\
     & ACS, WFC3    & 62    & \nodata & 62  & 12177      & 18   & 3D-HST & van Dokkum  \\
\noalign{\smallskip}
\hline
\noalign{\smallskip}
COSMOS & WFC3		& \nodata & 12	& 12	    & 12461      & 19	& SN TILE 41 & Riess \\ 
	& ACS, WFC3  & 56    &  \nodata & 56  & 12328      & 18    & 3D-HST & van Dokkum  \\       
                   
\noalign{\smallskip}
\hline
\noalign{\smallskip}
GOODS-N  & WFC3    &  56    & 56 & \nodata  & 13420      & 21   &  & Barro  \\
	           & WFC3    &  \nodata    & \nodata & 4  & 12461      & 19   & SN COLFAX & Riess  \\
	           & WFC3    & \nodata    & \nodata & 56  & 11600      & 17   & AGHAST & Weiner  \\
\noalign{\smallskip}
\hline
\noalign{\smallskip}
GOODS-S & ACS, WFC3    & 76    & \nodata & 76  & 12177      & 18   & 3D-HST & van Dokkum  \\
		 & WFC3	  & \nodata   & 12 & 12 & 12190 & 18 & CDFS-AGN 1 \& 2 & Koekemoer \\
		 & ACS, WFC3     & 6 & \nodata & 6+15 & 12099 & 18  & GEORGE, PRIMO & Riess \\
		 & WFC3     & \nodata & 2  & 2  & 11359   & 17 & ERS & O'Connell \\
\noalign{\smallskip}
\hline
\noalign{\smallskip}
UDS & WFC3    & \nodata    & 10 & \nodata  & 12590      & 19   & IRC0218A & Papovich  \\
	& ACS,WFC3    & 56    & \nodata & 56  & 12328      & 18   & 3D-HST & van Dokkum  \\
	& WFC3    & \nodata    & \nodata & 18  & 12099      & 18   & MARSHALL & Riess  \\
\noalign{\smallskip}
\hline
\noalign{\smallskip}
Total & 	& 312 & 92 & 383 & & & & \\
\noalign{\smallskip}
\hline
\noalign{\smallskip}
\end{tabular}
\begin{tablenotes}
\item[](a) The full program is 24 orbits in 12 pointings, however only 6 orbits overlap with the 3D-HST/CANDELS footprint.
\end{tablenotes}
\end{table*}

The utility of the data products that are described in this paper has already been demonstrated in a large number of studies, and we highlight several examples that illustrate particular aspects of the grism data. \citet{Nelson12}, \citet{Schmidt13}, and \citet{Wuyts13} analyze H$\alpha$ emission line maps of galaxies at $z\sim 1$, which are very difficult to obtain by any other means. Furthermore, \citet{Nelson13, Nelson15} show that emission lines in 3D-HST can be traced to large radii and extremely faint levels by stacking thousands of galaxies. The H$\alpha$ emission is more extended than the stellar continuum, suggesting that galaxies grow inside-out. \citet{Fumagalli12} study the evolution of the H$\alpha$ equivalent width, and find that it increases rapidly with redshift. \citet{Price14} show that the Balmer decrement of galaxies at $z\sim 1.5$ increases with stellar mass, and derive expressions for the relation between continuum extinction and the extinction towards H{\sc ii} regions. \citet{Brammer13} use the deepest G141 that were in existence at the time to constrain the spectrum of a $z\sim 12$ galaxy candidate in the Ultra Deep Field.\footnote{These deep spectra of galaxies in the Hubble Ultra Deep Field were released in 2013 \citep{vanDokkum13deep} and are available from the 3D-HST website.} Despite the relatively shallow depth of our survey we also obtain information on absorption lines of galaxies out to fairly high redshift. This is demonstrated in \citet{vanDokkum10} and particularly in \citet{Whitaker13}, who spectroscopically confirm the existence of a large population of galaxies with old stellar populations at $z\sim 2$.

Lastly, we list several 3D-HST results that do not use particular spectral features but utilize the large, homogeneous dataset of galaxies with reliable redshifts that the survey provides. \citet{vanDokkum13} and \citet{Patel13} describe the evolution of Milky Way like galaxies from $z\sim 2.5$ to the present, using number density-matched samples. \citet{vdWel14} combine the 3D-HST catalogs with CANDELS photometry to study the evolution of the mass-size relation with redshift. \citet{Whitaker14} provide a new measurement of the relation between star formation and stellar mass (the ``star formation main sequence''), and find that there is a turnover in the relation at low masses. \citet{Nelson14} and \citet{vanDokkum14, vanDokkum15} study the formation and evolution of the cores of massive galaxies. A full list of 3D-HST papers can be found on the 3D-HST website.\footnote{\url{http://3dhst.research.yale.edu/Publications.html}}

The structure of the paper is as follows. In \S\ref{sec:data} we describe the data that are now part of the 3D-HST project. \S\ref{sec:data_red} describes the data reduction, including the interlacing procedure that we use instead of drizzling. The extraction of the 2D and 1D spectra is discussed in \S\ref{sec:extract}. The redshift fits are described in \S\ref{sec:redshift_fits}, along with a discussion of their accuracy. We fit the spectra twice, once in conjunction with the photometry to determine redshifts, and then a second time to measure emission line fluxes and equivalent widths. These line flux fits are described in \S\ref{sec:line_fits}. The catalog entries are described in \S\ref{sec:catalogs}. In \S\ref{sec:properties} we highlight the properties of the spectroscopic sample. The paper is summarized in \S\ref{sec:summary}.
Magnitudes throughout are on the AB system.

\section{Data} 
\label{sec:data}

Most of the data described in this paper were obtained by the 3D-HST Treasury Survey. 3D-HST was allocated 248 orbits of HST time during Cycles 18 and 19. We obtained 2-orbit depth observations using the ACS/G800L and WFC3/G141 grisms in parallel. These observations cover 124 pointings in four of the five deep fields observed by CANDELS (AEGIS, COSMOS, GOODS-S, and UDS) and constitute the largest effort to acquire space-based near-infrared spectra in these fields.  A detailed description of the 3D-HST observations is presented in \citet{Brammer12}. The fifth CANDELS field, GOODS-N, had already been observed with WFC3/G141 prior to 3D-HST, in a Cycle 17 program as part of AGHAST (A Grism H$\alpha$ SpecTroscopic Survey; GO-11600, PI: Weiner), using a very similar observing strategy. ACS/G800L observations in GOODS-N were taken as part of GO-13420 in Cycle 21 (PI: Barro). For completeness, we have included the ACS/G800L and WFC3/G141 GOODS-N data in our analysis and data release, and throughout this paper we discuss the combined grism data set for all five fields. Table \ref{table:fields} lists the coordinates of each field and total areas covered with each instrument. 

A comprehensive list of all ACS and WFC3 grism observations in the five 3D-HST/CANDELS fields taken in Cycles 17 through 21 is presented in Table \ref{table:hstdata}. For reference, in addition to the ACS/G800L and WFC3/G141 data, we also summarize the available archival WFC3/G102 observations in the same fields. While the 3D-HST WFC3/G141 observations cover a relatively wide area to a shallow, 2-orbit, depth, other programs have obtained deep observations - up to a depth of 15 orbits (PRIMO, GO-12099, PI Riess, \citealp{Rodney12}) - over a single pointing.  The major source of these deep grism observations are the CANDELS supernovae follow-up programs (GO: 12099 \& 12461; PI: Riess) in addition to data  from  GO-12190 (CDFS-AGN; PI: Koekemoer),  GO-11367, (PI: O'Connell) and GO-12547 (EGS, PI: Cooper). We made a subset of these deep data (in the Hubble Ultra Deep Field) publicly available in a previous data release\footnote{See \citet{vanDokkum13deep}.} but they are not part of the dataset described in this paper: here we concentrate on the shallow, 2-orbit depth wide-field data.

\subsection{WFC3 Observations}

The WFC3 G141 grism  has spectral coverage from 1.1 to 1.65 $\mu m$ (at $>30\%$ throughput) and a peak transmission of 48\% at 1.45$\mu$m. The G141 dispersion is 46.5 \AA\,pixel$^{-1}$ ($R\;\sim\;130$) in the primary ($+1^\mathrm{st}$) spectral order. However, in practice, the spectral resolution for each (resolved) object is different as it is largely determined by its morphology. The uncertainties of the wavelength zeropoint and the dispersion of the G141 grism are $8$ \AA~ and $0.06$ \AA~pixel$^{-1}$ respectively \citep{Kuntschner10}.  The field of view of the WFC3 IR channel is $136\arcsec\times123\arcsec$.

The layout of the WFC3/G141 observations in the CANDELS fields is shown in Figure~\ref{fig:g141_layout}, overlaid on the $H_160$ imaging footprint. Across the five fields, 70\% of the CANDELS area is covered with at least two orbits of WFC3/G141 data. In AEGIS, COSMOS and UDS, 60\% of the CANDELS imaging area has complimentary G141 grism data, while 70\% of GOODS-N and 86\% of GOODS-S have G141 coverage. The total area of the G141 observations is 626 arcmin$^2$ (Table \ref{table:fields}).

The observations for the 3D-HST survey started October 30, 2010 and ended March 22, 2012. Two pointings in AEGIS, 1 and 22, were re-observed on April 21, 2013 and November 30, 2012, respectively. Each pointing of the 124 3D-HST pointings was observed for two orbits, with 4 paired $JH_{140}$ direct and G141 grism exposures. Typical total exposure times in each pointing are 800\,s in $JH_{140}$ and 5000\,s in G141. The four pairs of direct+grism exposures are separated by small telescope offsets to improve the sampling of the PSF, to enable the identification of hot pixels and other defects not flagged by the default pipeline processing, and to dither over some WFC3 cosmetic defects such as the ``IR blobs'' (Pirzkal et al. 2010). The sub-pixel dither pattern used throughout the survey is shown in Figure~3 of \citet{Brammer12}. 

The 56 orbits from the AGHAST program in GOODS-N are divided into 28 pointings, each with two-orbits depth. The observations were carried out between September 16, 2009 and September 26, 2010. Due to high background and scattered light artifacts, nine of the AGHAST pointings were partially re-observed between April 19 and 24, 2011.  Analogous to 3D-HST, each two orbit observation was split in four sets of G141 grism images and $JH_{140}$ direct exposures. The dither patterns of AGHAST and 3D-HST are slightly different, but they both sample the WFC3 PSF on a grid that is $0.5\times 0.5$ the native pixel size. The typical exposure time per pointing is 800\,s in $JH_{140}$ and $\sim$5200\,s in G141. Further information about AGHAST can be found on the survey website.\footnote{\url{http://mingus.as.arizona.edu/~bjw/aghast/}}

\begin{table}[h]
\centering
\caption{ACS Pointings Outside the CANDELS/3D-HST Footprint}\label{table:lost_acs}
\begin{tabular}{lccccc}
\hline \hline
\noalign{\smallskip}
Field & RA & Dec & ACS         & WFC3   \\
      &    &     & Pointing ID & Primary \\
\noalign{\smallskip}
\hline
\noalign{\smallskip}
AEGIS  &  	14:18:46.129 &	+52:49:27.29 & 41 & 3 \\  
	     & 14:18:26.632 &	+52:49:17.39 & 51 & 13 \\
 	     & 14:18:16.253 &	+52:47:58.13 & 62 & 24 \\
GOODS-S  & 03:31:49.881 &	-27:45:28.50 & 9  & 9 \\
 		 & 03:31:50.754 &	-27:43:14.19 & 12 & 12 \\
 		 & 03:31:45.353 &	-27:49:48.84 & 25 & 25 \\
UDS      & 02:16:45.756 &	-05:10:12.60 & 15 & 15 \\
 	     & 02:16:40.606 &	-05:07:05.53 & 20 & 20 \\
 	     & 02:16:38.757 &	-05:09:16.03 & 21 & 21 \\
 	     & 02:16:40.999 &	-05:11:38.08 & 22 & 22 \\
\noalign{\smallskip}
\hline
\noalign{\smallskip}
\end{tabular}
\end{table}

\subsection{ACS Observations}

Exposures with the ACS G800L grism, accompanied by $I_{814}$ direct imaging, were taken in parallel with the primary WFC3 exposures.  ACS coverage of the GOODS-N fields was done in program GO-13420 (PI: Barro) as parallels to their WFC3/G102 primary observations. The G800L grism has a wavelength coverage from 0.55\,$\mu$m -- 1.0\,$\mu$m with a  dispersion of 40\,\AA\,pixel$^{-1}$ in the primary first order.   The total exposure times in each pointing/visit are 480\,s in $I_{814}$ (1299\,s in GO-13420) and between $\sim$2800\,s (GOODS-N) and $\sim$3500\,s (AEGIS) in G800L.  Figure~\ref{fig:g800l_layout} shows the layout of the pointings in all five fields. Unlike the WFC3 pointings, the ACS pointings do not have a regular pattern but an effort was made to maximize the overlap between the two grisms. Fully 86.5\% of the WFC3 grism observations also have ACS grism coverage. Within each pointing, four pairs of $I_{814}$ direct images and G800L grism images were taken in a sequence. As a result of the larger ACS field of view ($202\arcsec\times202\arcsec$) there is larger overlap between the pointings, with some areas covered up to a depth of 8 orbits. 

Several of the 3D-HST ACS pointings fall completely outside of the footprint of the CANDELS/3D-HST WFC3 imaging. As described below, the WFC3 mosaics are used as WCS reference images for aligning the direct $I_{814}$ images. Pointings that fall outside these mosaics cannot be aligned to the same WCS and cannot be processed in the same manner as the rest of the pointings. The 10 pointings which are outside the footprints are listed in Table \ref{table:lost_acs}. These pointings are processed throughout the preliminary reduction steps only.

\section{Data Reduction}
\label{sec:data_red}

\begin{figure*}[th]
\centering
\includegraphics[width = \textwidth]{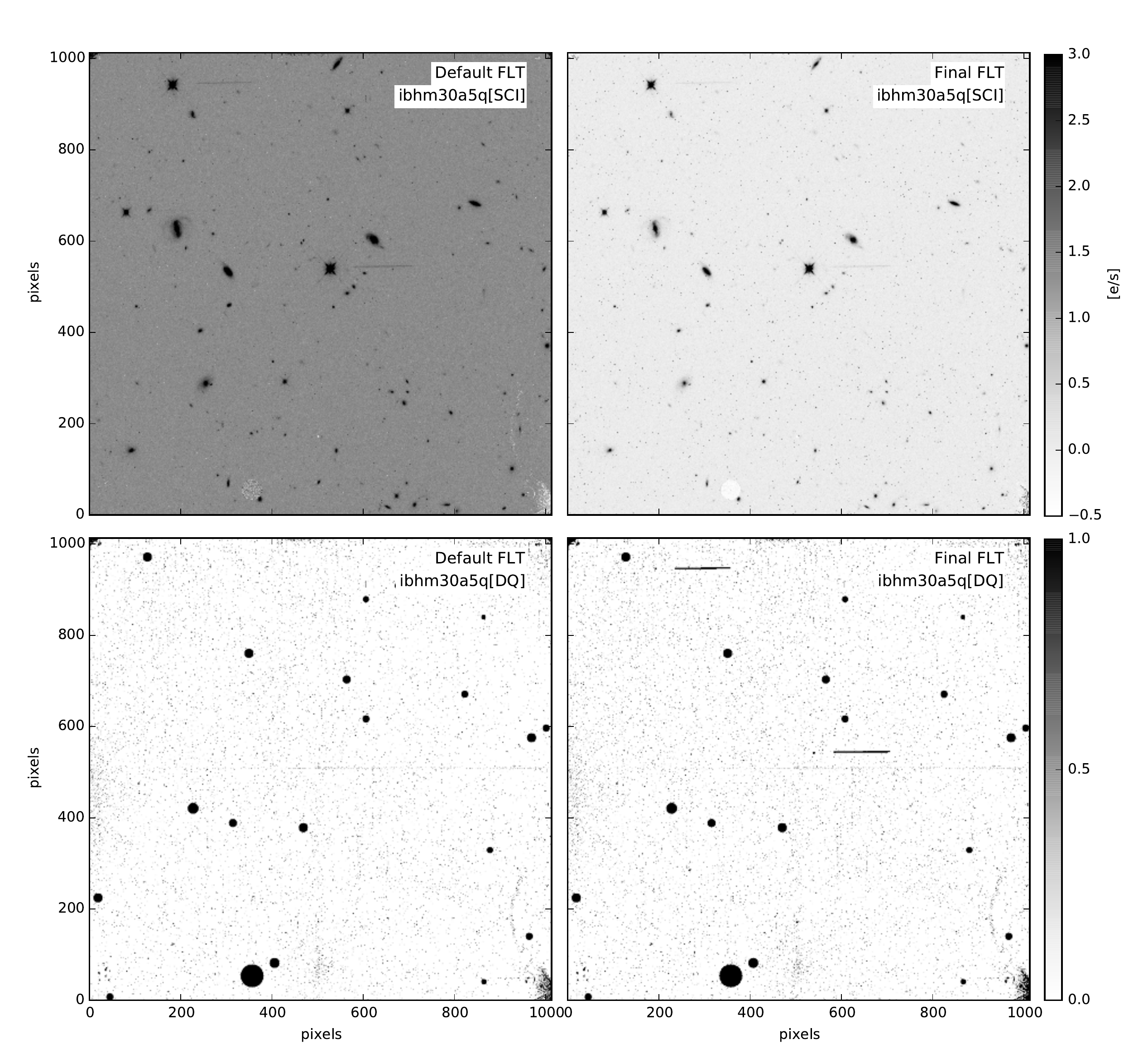}
\caption{\footnotesize Original default \textsc{flt} ({\it left}) and final processed \textsc{flt} ({\it right}). For each \textsc{flt} we show both the science image ({\it top}) and the data quality array ({\it bottom}). The main difference between the two science images is the background subtraction. The main difference between the data quality arrays is the persistence masking.}
\label{fig:steps_direct}
\end{figure*}

\begin{figure*}[th]
\centering
\includegraphics[width = \textwidth]{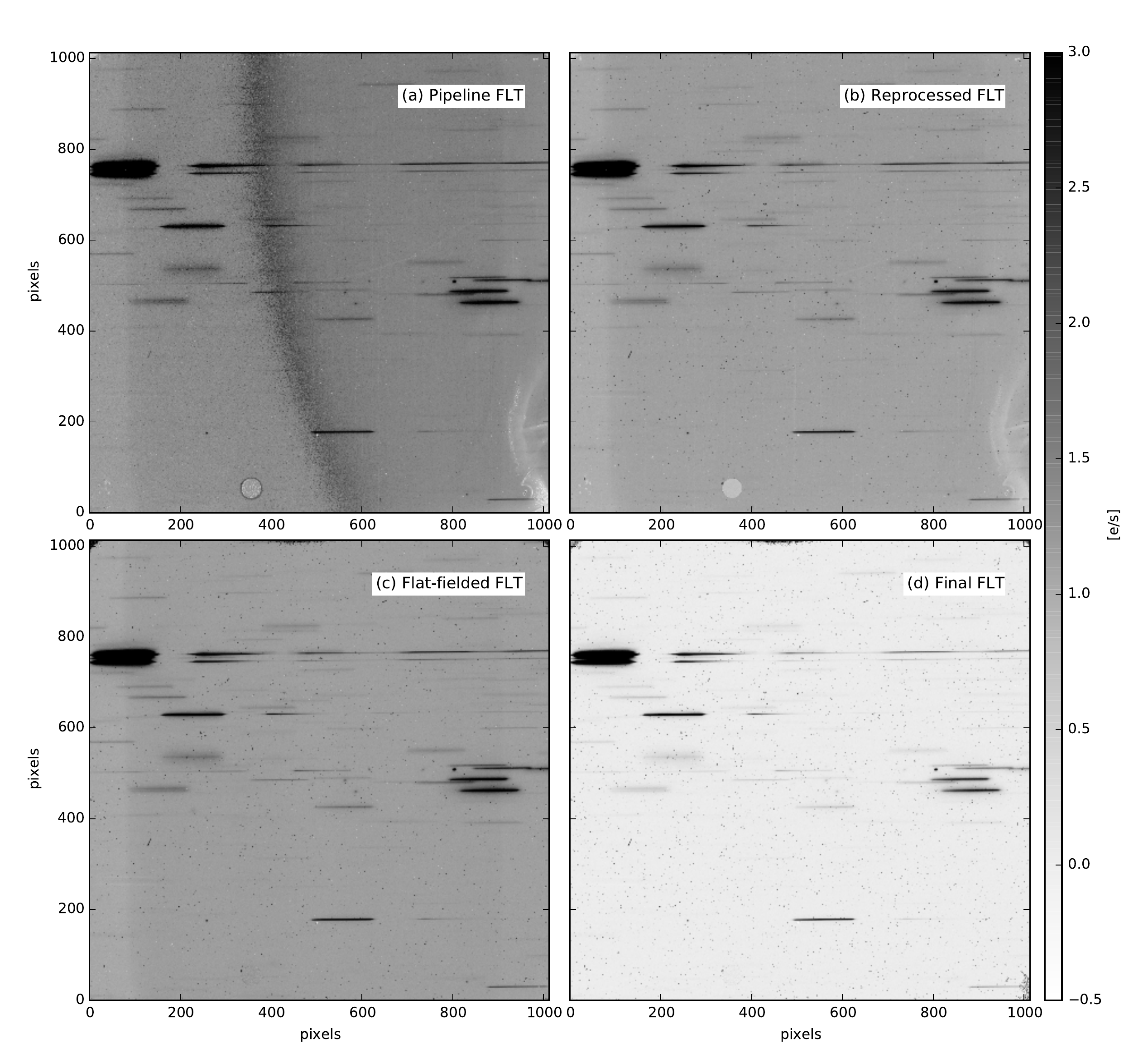}
\caption{\footnotesize Steps in the reduction of a G141 image. We use as an example one of the \textsc{flt} images of pointing AEGIS-01: ibhj39uuq\_flt.fits (also shown in Figure~\ref{fig:earthshine}). Shown are: (a) the default \calwf\ pipeline-processed image which, in this case, has a high Earthshine background component; (b) the re-processed \calwf\ image with the last four reads removed; (c) the flat-fielded re-processed frame; and (d) the background-subtracted final image. The \calwf\ reprocessing is done only for a small subset of all \textsc{flt} images; most of the pipeline-processed \textsc{flt} files resemble the panel b, not panel a.}
\label{fig:steps_grism}
\end{figure*}

An early version of our reduction pipeline was described in \citet{Brammer12}. At that time, the pipeline used custom pre-processing steps such as alignment, flat-fielding and sky-subtraction followed by extraction of the grism spectra using the \aXe\ software package \citep{Kummel09}. We have made major changes to the reduction procedures since then and here we describe our final approach to the data processing. 

Some changes have been made to the pre-processing steps, as discussed below, but the main difference between our current pipeline and that described by \citet{Brammer12} takes place after the pre-processing. In particular, we no longer use the \aXe\ package. \aXe\ drizzles the data onto a grid with linear sampling in the wavelength and spatial direction prior to extraction of 2D and 1D spectra. Drizzling introduces correlations between pixels and smooths the data, and in an effort to optimally use the information in the grism spectra  we have developed an approach which uses the original WFC3/IR pixels without resampling.  We do not drizzle the data, but instead place the original pixels of four dithered exposures on a (distorted) output grid whose pixels are exactly half the native pixel size. This interlacing approach, discussed in detail in \S\ref{sec:interlace}, retains the independence of adjacent pixels and the full resolution of the data. The distortions are encoded in the software that is used to analyze the spectra, and in a pixel-to-wavelength conversion table that is unique to each object and supplied with our data release.

In the following subsections we describe the pre-processing steps for the WFC3 and ACS data, the interlacing procedure (\S\ref{sec:interlace}), and the creation of the masks and direct images that are used in the extraction and analysis of the spectra (\S\ref{maskred.sec}). The extraction itself is described in \S\ref{sec:extract}.

\subsection{WFC3 $JH_{140}$ Images}
\label{wfc3red.sec}

We download the raw (\textsc{raw}) images, the calibrated (\textsc{flt}) images and the association tables (\textsc{asn}) for all observations from the Mikulski Archive for Space Telescopes (MAST\footnote{\url{http://archive.stsci.edu}}). The calibrated images are processed with the \calwf\ pipeline; a detailed description of the \calwf\ processing steps is given by \citet{Koekemoer11}.  We also obtain the persistence images\footnote{\cite{Long13}; \url{http://archive.stsci.edu/prepds/persist/}} made available by STScI (\textsc{persist}). These images provide estimates for the total IR persistence that affects a given exposure, both from sources internal to the 3D-HST visit and also external sources from prior observations.

The reduction of the direct images is described in detail in \citet{Skelton14}. Here we give a brief summary of the relevant steps. The main difference in the image preparation steps relative to \citet{Skelton14} is the full integration of {\tt TweakReg} and {\tt AstroDrizzle} \citep{Gonzaga12} in the reduction, while previously it was used only for the final alignment and drizzling steps. Note that we do not apply the advanced processing of individual reads that is described in Sect.\ \ref{sec:read_removal}; the direct images are comprised of only four samples and eliminating one or more of these would lead to a significant loss in the integration time.

All \textsc{flt} images are first inspected for satellite trails and artifacts, as well as for regions of elevated background due to Earthshine. Affected regions identified in the inspection are masked and given a data quality flag of 2048 (the cosmic ray data quality flag) so that they are treated as pixels without information in the subsequent processing steps.  For persistence masking, we apply a conservative threshold, requiring that the predicted persistence is less than 0.6 times the values in the \textsc{flt} error extension.  We grow the persistence-masked area slightly and then set the 4196 bit in the data quality extension for the masked pixels. These are later treated as cosmic rays and are not used in the final mosaics. Finally, we add a component to the \textsc{flt} uncertainties to account for crosstalk from pixels where the total number of deposited elections is greater than $2\times10^4\,{\rm e}^-$ .  Time-dependent sky flats were created from the science exposures, which account for the appearance of new IR ``blobs'' with time since the installation of WFC3. 

We run {\tt AstroDrizzle} first to identify hot pixels and cosmic rays not flagged by the \calwf\ calibration pipeline. This step produces a preliminary combined $JH_{140}$ science image of each pointing. We subtract the background from this image in the following way. A preliminary source detection is done with SExtractor \citep{Bertin96}. The detected sources are used to create a mask, and we fit a second order polynomial background and subtract it from the \textsc{flt} exposure.  Using {\tt TweakReg} we align each \textsc{flt} image to the reference frame of the \citet{Skelton14} mosaics by providing a reference list of object positions derived from the \citet{Skelton14} photometric catalogs. These alignment corrections beyond the commanded dither positions are typically small, of order 0.1 pixels.

Figure \ref{fig:steps_direct} illustrates the differences between the default \textsc{FLT} image and the final processed  \textsc{FLT}. The most notable difference between the science images (top) is the background subtraction which removes the pedestal of $\sim 2 {\rm e}^-$\,s$^{-1}$ in the default image. In the data quality arrays (bottom), the persistence, caused by the spectra of the two bright stars in the frame, has been masked.  

\subsection{WFC3 G141 Images}

Following the \calwf\ pre-processing, we apply several steps to improve the grism data quality. These steps are removing satellite trails, persistence masking,  flat-fielding, sky-subtraction, astrometric alignment, and final cosmic-ray and bad pixel rejection. Some of these steps were also described in \citet{Brammer12}; these are briefly summarized with an emphasis on any differences that we implemented since that paper.  Figure~\ref{fig:steps_grism} demonstrates various stages of our processing of one of the \textsc{flt} images (of pointing AEGIS-01); details are provided below.

\subsubsection{Removing Satellite Trails and Earthshine}
\label{sec:read_removal}

The grism images occasionally contain satellite trails and other cosmetic blemishes, which we identify by visually inspecting all grism exposures in a manner similar to the direct images. When processing the direct images we simply masked these blemishes, but we use a more sophisticated approach for the grism data. We make use of the fact that a single WFC3 exposure is comprised of multiple independent samples, which are generated by multiple non-destructive MULTIACCUM reads during the exposure. Therefore, a single WFC3 image is really a sum of independent images, which can be recovered by analyzing the individual reads.\footnote{We use the term ``sample'' and ``read'' interchangably, although it is more correct to use the term ``read'' for the process that, through differencing, produces a sample.} While the short direct images typically only have four non-destructive reads, the longer grism images have 12--15 100\,s reads. 

Satellites move across the WFC3 field-of-view quickly and typically only affect a single read.  Rather than masking areas of the detector, we remove the read (or, sometimes, multiple reads) that is affected by extraneous light. In order to remove the affected reads, we use the \textsc{ima} files (intermediate MULTIACCUM files) produced by \calwf, which contain the individual calibrated reads from the exposure. We average the count rates in all of the clean reads in the calibrated \textsc{ima} files and use this averaged image in place of the \textsc{FLT}.  As this process bypasses the \calwf\ up-the-ramp cosmic ray rejection step, the cosmic rays in these (few) reprocessed exposures must be identified separately based on comparison with the other dithered exposures (with {\tt AstroDrizzle}). We note that the final exposure time of the reprocessed exposures are reduced by the duration of the rejected reads (typically by $1\times 100$\,s). 

A number of pointings, specifically in GOODS-N, are affected by scattered Earth light or ``Earthshine". This light is observed when the telescope points near the bright Earth limb and its light reaches the detector through an unintended path in the optics. As a result of this, the background level in the leftmost $\sim200$ columns of the detector can be increased to levels up to twice that of the rest of the detector (see Figure\ 6.17 of the WFC3 Data Handbook). As with the satellite trails, we examine the individual reads and remove those affected by bright Earthshine from the sequence (reads with low level Earthshine are not removed, as they can be corrected in the background subtraction step, discussed below).  Removing reads decreases the effective exposure time of the image by $\sim 100$ seconds per read removed; however, the exposure is not completely lost. 

Further information about the removed reads is provided in the Appendix (Table \ref{table:popped_reads} lists all removed reads; Figures~\ref{fig:earthshine} and \ref{fig:satellite} demonstrate the removal of reads due to Earthshine and a satellite in two different pointings). In total, 30 pointings have each had one read removed  and one pointing has had two reads removed due to satellites crossing the WFC3 field of view during the exposure. The effect of these readout removals is minimal because satellites only affect one of the four \textsc{flt}s in a pointing and the loss of a single read only constitutes a loss of $\sim12\%$ of its exposure. Earthshine, on the other hand, can have a significant effect on the depth of a pointing. Earthshine typically appears at the beginning or at the end of the exposure and lasts for  multiple reads. Twenty-three \textsc{flt} images in 12 pointings in GOODS-N as well as two of the AEGIS pointings are affected by Earthshine.  Panels a and b of Fig.\ \ref{fig:steps_grism} show the removal of Earthshine in pointing AEGIS-01, where four of the 12 readouts are removed. Across all \textsc{flt}s, between one and eight reads have been removed in each affected image which results in significant loss of depth in some of these pointings. Some were partially re-observed, but, within our framework, the additional data cannot easily be combined with the original observations (see \S~\ref{sec:interlace} below). Even though the final depth in these 25 affected pointings is lower than the rest of the survey, they only constitute 8\% of the data.

\subsubsection{Correcting for the Effects of a Time-variable Background}

A significant background component in the G141 grism is the emission of metastable He at 1.083~$\mu\mathrm{m}$ \citep{Brammer14}, which is negligible in the Earth's shadow but increases sharply when the spacecraft is outside the shadow. Unlike the Earthshine, which only appears on the edge of the image close to the bright Earth limb, the He emission elevates the background flux across the whole detector. The strength of the He line background depends on the position of the telescope relative to the bright Earth limb and can therefore vary significantly within a single exposure. This time-variable background results in a non-linear increase in the background counts during the exposure, and therefore a non-linear increase in the background counts in subsequent reads. The \calwf\ pipeline uses such non-linearity to identify and filter out cosmic rays during the exposure: a cosmic ray hit in between two reads leads to an increase in the flux of a pixel that is inconsistent with the expectation from the gradual accumulation of charge during the exposure. As the non-linear background variation mimics the behavior of cosmic rays,  \calwf\ flags the majority of pixels in these images as cosmic rays and corrupts the \textsc{flt} products. 

To avoid this inadvertent cosmic ray removal and mitigate the effects of the time-variable He background, we redistribute the total counts in the \textsc{ima} files so they are distributed evenly over the individual samples. We first subtract the median per-second count rate of each read and then add back the average per-second count-rate of the entire exposure.  We then run \calwf\ on the corrected \textsc{ima} files using only the final cosmic-ray identification step to produce the final \textsc{flt}.   Exposures that were otherwise rendered unusable due to the variable backgrounds are recovered, albeit with somewhat lower signal-to-noise than unaffected exposures because the overall background count rates are higher. By redistributing the charge we also retain the ability of \calwf\ to identify cosmic rays using the up-the-ramp sampling. 

\subsubsection{Grism Flat-fielding and Background Subtraction}

Following \citet{Brammer12} we first divide the G141 grism exposures by the $JH_{140}$ imaging flat-field calibration image.  This neglects the wavelength dependence of the flat-field (which is at most a few percent across the field) in favor of greatly reduced computational complexity.
Panel c in Fig.\ \ref{fig:steps_grism} shows the flat-fielded \textsc{flt} image in our example pointing.  The main effect is the removal of the ``wagon wheel'' in the lower right corner of the frame.

At each pixel in the grism exposures the background is the sum of different spectral orders sampled at different wavelengths.  There is significant structure in the background across the detector resulting from vignetting of the spectral orders, and this structure must be removed to enable extraction of clean spectra of objects.  Using on-orbit science observations, \cite{Kummel11} created a single master sky image that can be used with the \aXe\ software to remove the grism sky background. However, in \cite{Brammer12} we noted significant variation in the spatial structure of the grism backgrounds and created four separate master sky images that helped to account for the variation. 

As described in \cite{Brammer14}, we now understand that the observed variation in the background structure is mainly due to three distinct sources: the zodiacal continuum, scattered light, and the He emission line. \citet{Brammer14_ISR} create master sky images\footnote{\url{http://www.stsci.edu/~brammer/grism_sky/}} for each of these three physically-motivated background components individually.  We fit a linear combination of these component images to each exposure, requiring the zodiacal component to be constant throughout a given visit and allowing for a variable contribution from the emission line component (see above).  This technique removes much of the background structure in the grism images.  Following \cite{Brammer12} we subtract a final masked column average to create the final background-subtracted images to remove low-level residuals not accounted for by the three-component fits. 

This final step in the pre-processing sequence is shown in panel d of Fig.\ \ref{fig:steps_grism}. These final images have uniform and low background. The final quality of this example \textsc{flt} file is representative of all the data in the survey.

\subsection{ACS $I_{814}$ Images}
\label{acsred.sec}

We download the CTE-corrected (\textsc{flc}) images and association tables (\textsc{asn}) for all observations from MAST. The calibrated images were processed on the fly by the {\tt calacs} pipeline. A detailed description of the  {\tt calacs} steps is given in the ACS Data Handbook (Chapter 3). In brief, the  {\tt calacs} pipeline does all the calibration steps including bias-subtraction, cross-talk correction, dark-subtraction, flat-fielding, cosmic ray rejection, charge transfer efficiency (CTE) correction, shutter shading correction, and masking of bad and saturated pixels. The final images are in units of electrons. 

\subsection{ACS G800L Images}

As for the $I_{814}$ images, the CTE-corrected \textsc{flc} images and association tables are obtained from MAST. The images are then processed with AstroDrizzle to identify cosmic rays; cosmic-ray affected pixels are flagged in the \textsc{flc} images. A model for the grism background is obtained by carefully masking all detected objects and taking the median of all G800L exposures, after scaling each detector of each exposure so that the average sky values are matched. This background model, multiplied by the appropriate detector- and exposure-dependent scaling factor, is then subtracted from the individual exposures. The ACS grism images are not flat-fielded. \citet{Pirzkal02} show that applying a direct-imaging flat to the grism observations introduces $\pm10\%$ large scale differences. Without the flat-fielding, these differences are much smaller, $\sim5\%$ across the detector. 

The individual exposures for each pointing are combined by rounding the offsets between exposures to the nearest integer. As no interpolation is used this step retains the noise properties of the data, at the expense of also retaining the geometric distortions in the frame. In the following Section we discuss the rationale of this approach in the context of the WFC3 G141 grism data. 

The reduced ACS data are part of the 3D-HST data products, and publicly available. However, in this paper we limit the spectral extractions and redshift fitting to the WFC3 G141 data. The procedures described below can be applied in the same way to the ACS data: although the ACS data are shifted and summed and the WFC3 data are interlaced, the final product is similar: distorted frames whose noise properties are preserved, with similar pixel size. We note here that a key advantage of the G141 data over ground-based near-IR spectroscopy, namely the low near-IR background from space compared to the ground, does not apply in the same way to the optical ACS spectra. 

\subsection{Interlacing}
\label{sec:interlace}

The traditional method of combining dithered images from {\it HST} is through ``drizzle'' image processing which allows for the recovery of resolution in under-sampled images as well as the correction of geometric distortions \citep{Fruchter02}.  The drizzle algorithm works particularly well when a large number of images are being combined. However, in the limit of few images, it is prone to producing correlated noise. The reason for these noise correlations is that pixels from the individual input images contribute to multiple pixels in the resampled output grid. The amount of this resampling ``diffusion'' can be controlled in the drizzle algorithm; however, some diffusion is usually necessary to avoid uneven coverage of the output grid. The net effect is a slight smoothing, resulting in a loss of resolution and correlations between adjacent output pixels.

Drizzling is particularly problematic for spectra, as correlated noise can mimic emission or absorption features. Furthermore, the correlated noise is difficult to properly take into account when fitting the spectra, again leading to confusion between noise and real spectral features. 
Lastly, for spectra, correcting the geometric distortions is not strictly necessary, as long as the mapping between pixels and wavelength is known.

\begin{figure}[ht]
\centering
\includegraphics[width=0.45\textwidth]{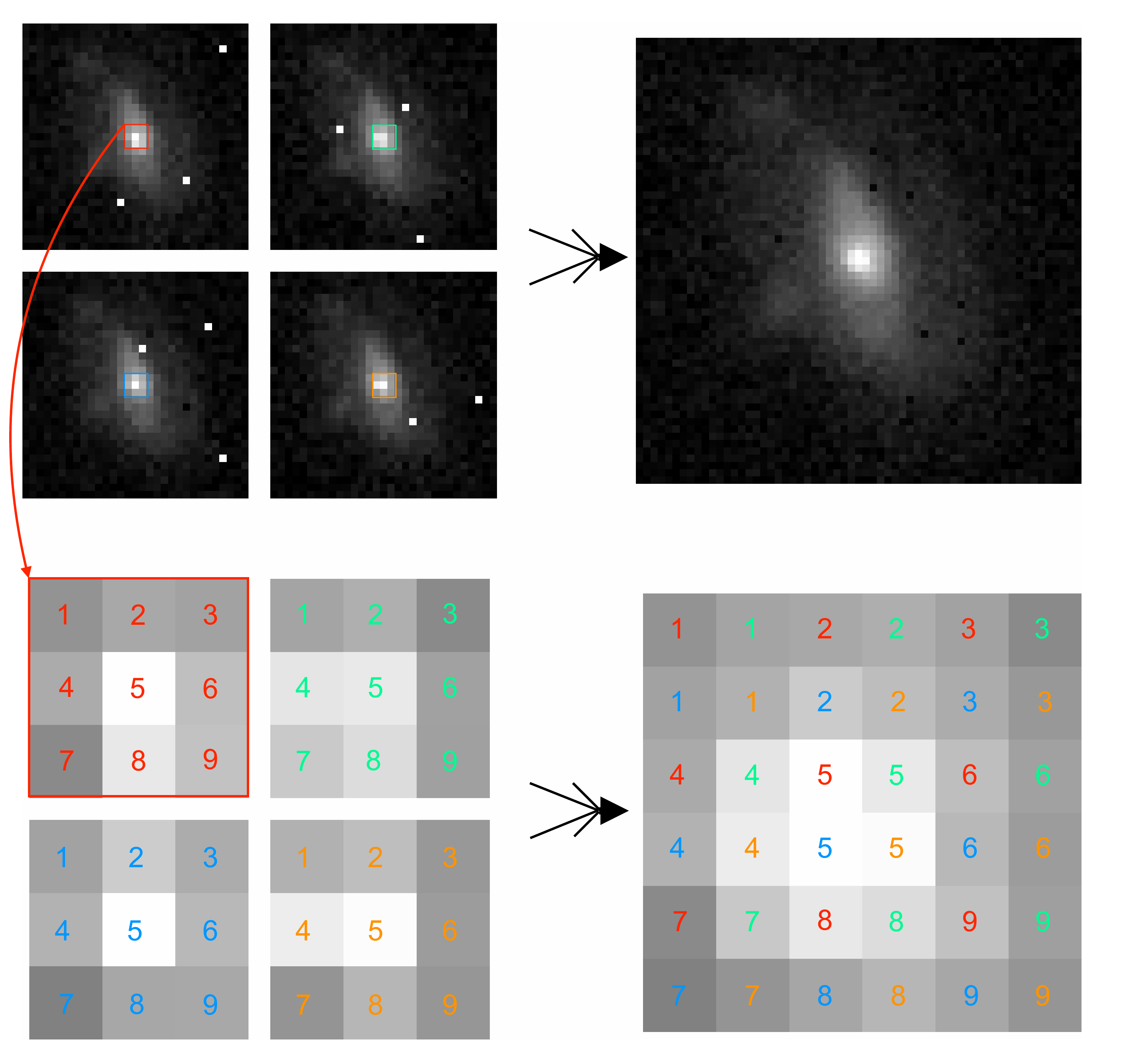}
\caption{\footnotesize Illustration of interlacing with a small section of a $JH_{140}$ direct image. The same process is also used to interlace the grism exposures. The top left shows the four individual exposures. These are combined to produce the interlaced image in the top right.  In the bottom row we show the same procedure for a $3\times3$ pix part of the core of the galaxy to demonstrate how the pixels from the individual images are arranged in the final interlaced grid.
}
\label{fig:interlacing1}
\end{figure}

\begin{figure}[ht]
\centering
\includegraphics[width=0.45\textwidth]{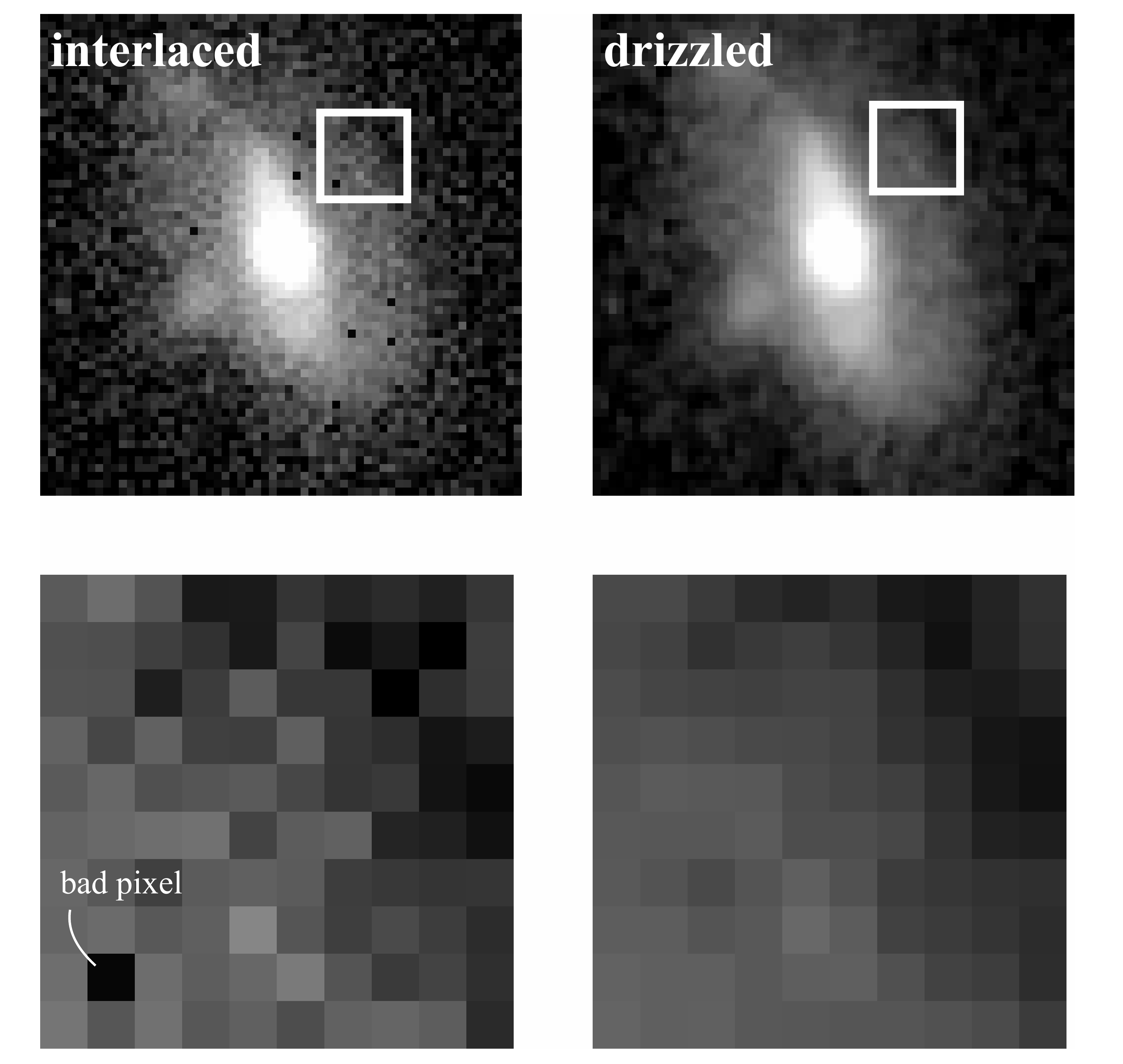}
\caption{\footnotesize Comparison between interlacing and drizzling. Compared to the drizzled image, the interlaced image has higher resolution as the pixels were not interpolated. Flagged pixels (due to cosmic rays and chip defects) are retained as single pixels in the interlaced image, whereas they are interpolated over in the drizzled image. 
}
\label{fig:interlacing2}
\end{figure}

Most 3D-HST pointings (exceptions described below) are comprised of eight images---four direct and four dispersed, observed with a 4-point dither pattern \citep[see Figure 3 of ][]{Brammer12}. The dithers between the images sample the WFC3 pixels at half-pixel intervals.  With this optimal sub-pixel sampling, we {\em interlace}, rather than drizzle, the original exposures into an output mosaic used for the spectral extractions.

We combine the exposures of each visit into a single output frame by placing the original images onto a subgrid of pixels that are exactly half their original size. The procedure is illustrated in detail in Fig.\ \ref{fig:interlacing1} using a portion of a $JH_{140}$ image, and compared to a standard drizzling approach in Fig.\ \ref{fig:interlacing2}. As the input images have exactly half-pixel offsets by design, this results in a one-to-one correspondence between input and output pixels \citep[e.g.,][]{vanDokkum00} and offers the key benefit of preserving the individual pixel errors. Adjacent pixels come from sections of the original images that are $\sim$10 pixels apart and are entirely uncorrelated. Furthermore, interlacing improves the sampling of the PSF by a factor of two without having to interpolate; it therefore produces the highest resolution images that are attainable with the WFC3 camera. Both the G141 and the direct $JH_{140}$ images are interlaced in the same manner.The output G141 images have a pixel size of $\sim$23\,\AA\ $\times$ $0\farcs 06$. 
Interlacing is possible because (a) the relative pointing errors between the images of a given set are small ($\sim0.1$ pixels) and (b) the dithers between images are small ($\leq10$ pixels) and the relative distortion on these scales in WFC3 and ACS is small.

\begin{figure*}[th]
\centering
\includegraphics[width = \textwidth]{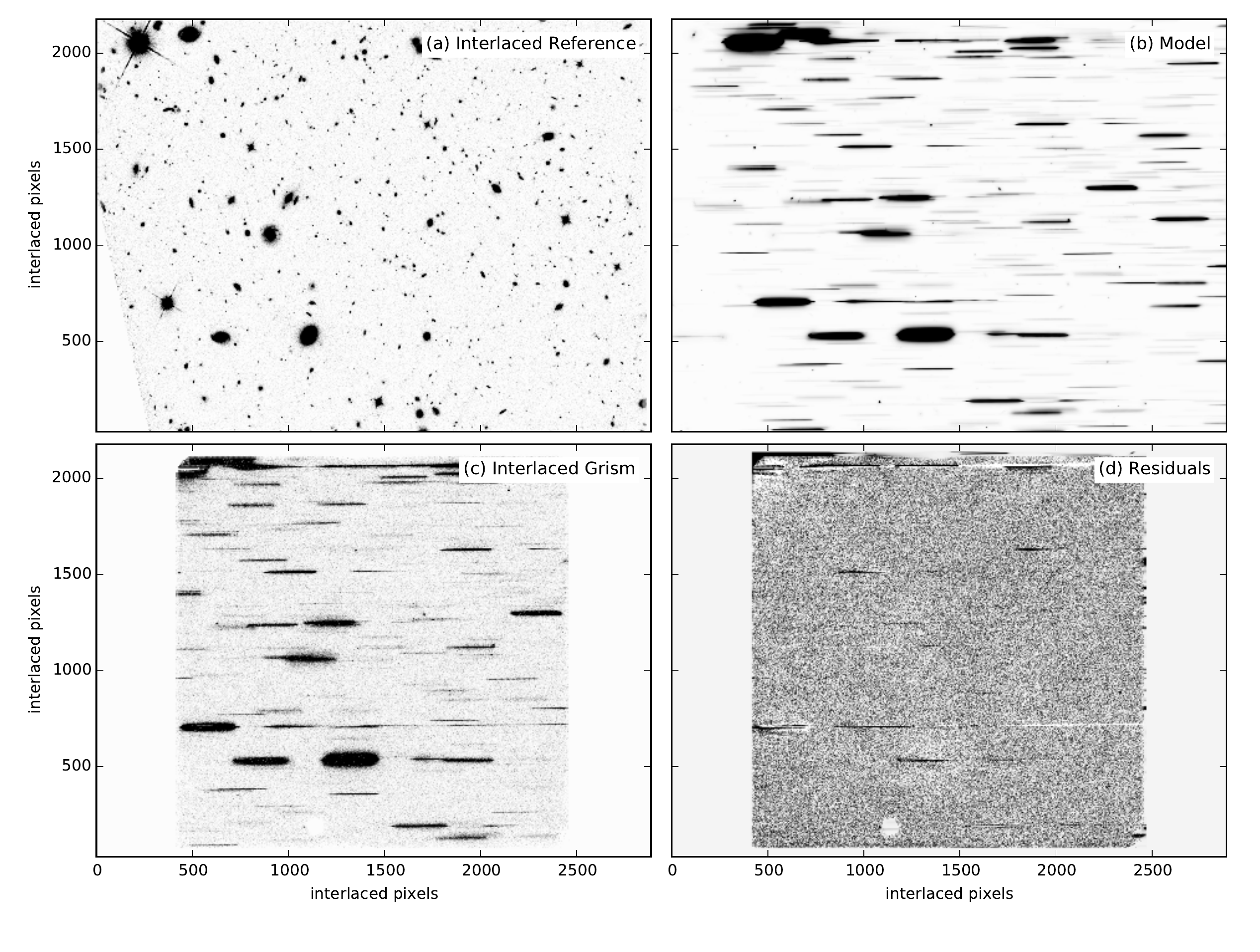}
\caption{Full contamination model of the COSMOS-04 pointing. The panels show: (a) the interlaced direct reference image, created from the CANDELS+3DHST $J_{IR}$ = $J_{125}$ + $JH_{140}$ + $H_{160}$ mosaic of the COSMOS field; (b) the contamination model created using the direct image and model spectra for all the objects; (c) the observed interlaced grism image; and (d) the residuals after subtracting the contamination model from the interlaced grism image.}
\label{fig:contam_model}
\end{figure*}

The primary shortcoming of this approach is that if one or more images in a dither sequence are missing, the combined image will have empty pixels.\footnote{Note that this is a shortcoming in the data, not the method, and drizzled images are similarly affected. However, it is less obvious in drizzled images as the missing pixels are effectively interpolated over.} This only affects one of our pointings, AEGIS-20, which only has three direct and two grism exposures. Areas that were masked in all four exposures will also have no information. This only affects four 3D-HST pointings: GOODSS-15 where one dither position was repeated, UDS-15 which has many masked pixels affected by the Mars-crossing asteroid 1036 Ganymed, and UDS-25 and UDS-26, which are both affected by long term persistence. Small portions of these images were masked entirely. Finally, all pixels flagged due to other reasons will also be empty in the final image.  These are typically less than 1\% of all pixels in the image.  We note that empty pixels in the grism images 
are handled trivially in the fits to the 2D spectra described below: empty pixels do not contribute to the $\chi^2$ or likelihoods of the fits. We stress that drizzled versions of the same data have the same missing information. The only difference is that the missing information is interpolated over in the drizzling process.

Another short-coming of the interlacing approach (also true for \aXe) is that only observations taken at the same rotation angle can be combined. Observations within a single visit are always taken at the same rotation angle, but re-observations of failed visits are frequently done at a different rotation angle. Within our dataset, the re-observations of GOODSN-11, 14 and 23 are done at a different angle from the original visits and cannot be combined with the original observations. Furthermore, the re-observed data in GOODSN-11 contains only two dither positions which means that only half of the pixels in the image are filled. These data can still be valuable for certain applications and are included in the data release accompanying this paper. However, the lower information content of the spectra from these three pointings  decreases the accuracy of the redshift fits. The majority of the spectra from these pointings were flagged in the visual inspections and they are excluded from the analysis in \S~\ref{sec:redshift_fits} and \ref{sec:line_fits}.

\begin{figure*}[th]
\centering
\includegraphics[width = \textwidth]{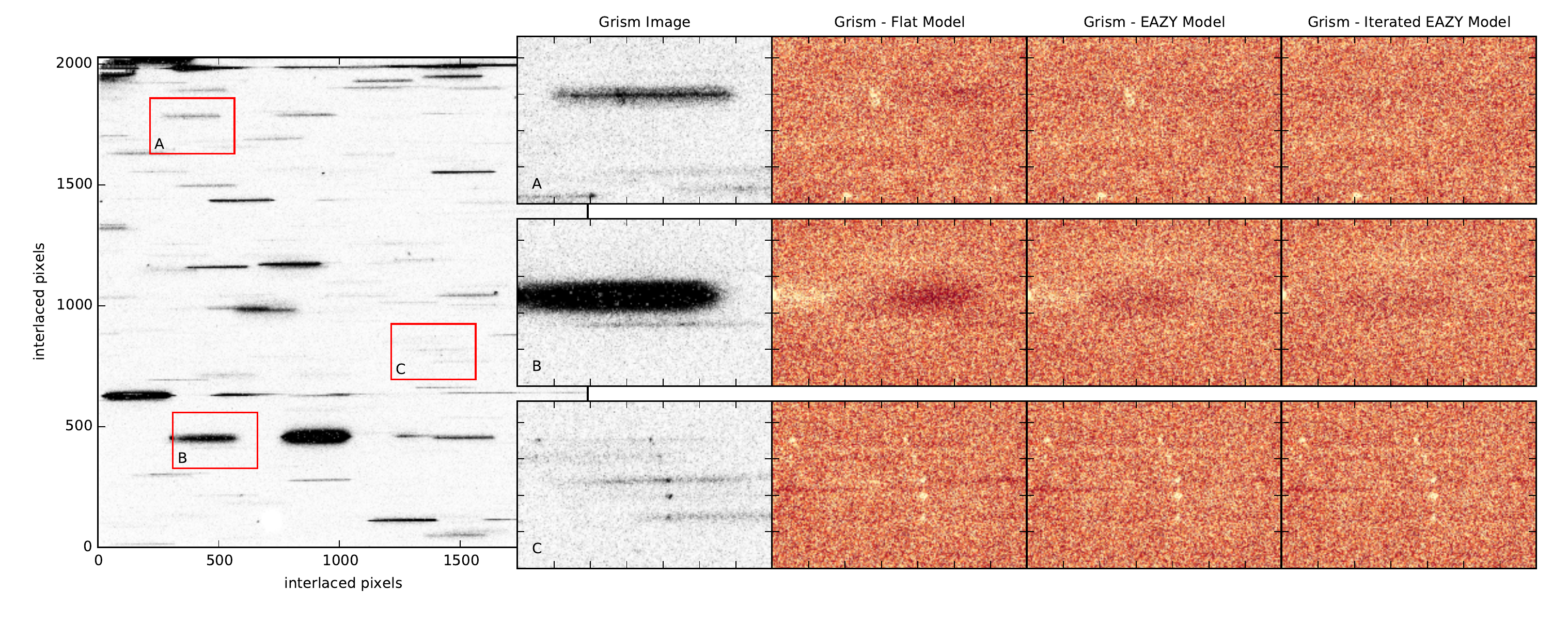}
\caption{\footnotesize Illustration of the quality of our contamination modeling. For demonstration purposes we model and subtract all spectra, including that of the object of interest. The pointing is COSMOS-04. The full interlaced grism image is shown for context, with the panels showing zoomed-in portions of the image. The orange panels show the residuals after subtracting three different versions of the contamination model: a flat spectrum (left); the best-fitting EAZY model template (middle); and the best-fitting EAZY template for faint objects plus empirical spectra for bright objects (right). The final model is excellent, with the only significant residuals emission lines of faint sources.}
\label{fig:model_demo}
\end{figure*}

\subsection{Reference Image, Catalog and Segmentation Map}
\label{maskred.sec}

Before we can extract spectra from the interlaced frames we require a reference image, which provides the positions and morphologies of all objects in a given grism pointing. This image sets the wavelength reference for all sources and is used to create a model that accounts for the contamination from overlapping objects. The reference image must be accompanied by a catalog which defines positions and magnitudes for all objects within the pointing as well as a segmentation map which defines the pixels which belong to each object in the catalog. In \citet{Brammer12} we used the direct $JH_{140}$ images as reference and we ran {\tt SExtractor} to create a catalog and a segmentation map for each grism pointing. This approach posed two main challenges: (1) repeat objects, which appeared in multiple pointings, could not be directly coadded as they would have different segmentation polygons; and (2) when we later matched the catalog objects to external photometric catalogs the matches were not always unique.

Here we use the data products of \citet{Skelton14} to create the reference images, catalogs and segmentation maps. Note that we will now make a distinction between the {\it direct image} ($JH_{140}$ for 3D-HST and AGHAST), which was used to align the exposure WCS, and the {\it reference image}, which may be a deeper astrometrically aligned image in a different filter.  The reference image is used to determine the distribution of light in each source, which defines the spatial morphology of the two-dimensional object spectra. The reference image is a sum of the $J_{125}$, $JH_{140}$ and $H_{160}$ WFC3 images. The reference mosaic is created for the whole field by coadding the \citet{Skelton14} mosaics (before PSF matching) in the three bands: the individual images are scaled to the $JH_{140}$ AB zero-point and then coadded with the inverse variance maps used as weights. Although we use the \citet{Skelton14} catalogs we do not use their flux measurements as our ``standard'' magnitudes. The reason is that the catalogs  include areas where the 3D-HST $JH_{140}$ and the CANDELS $H_{160}$ images do not overlap. As a result, there is not one consistent flux/magnitude measurement that exists for all objects in the catalogs. To remedy this, we run {\tt SExtractor} on the coadded image, in dual image mode with the \citet{Skelton14} detection images and the same settings, in order to determine uniform fluxes and magnitudes that are defined for all objects. These magnitudes, which we refer to as $\JH$, are used throughout the paper to determine the depth of the fits and to apply magnitude cuts to the grism catalogs. 

For each grism pointing, we create a reference image by ``blotting'' the full mosaic to the frame of each of the grism \textsc{flt} images, where the WCS alignment has been performed in the preparation steps and the {\tt AstroDrizzle.ablot} utility transforms the rectified mosaics into the distorted \textsc{flt} frame.   The individual blotted images are then interlaced in the same manner as the grism exposures. The same procedure is also used to blot the master segmentation map into the \textsc{flt} frames.  Finally, a reference catalog is created for all objects which fall within the blotted segmentation map, with object pixel positions in the distorted frame computed with the {\tt skytopix.rd2xy} task. 

One significant benefit of using an external reference image is that it no longer needs to be limited to the size of the WFC3 field of view.  The blue edge of the first order grism spectrum is offset by $\sim65$ (interlaced) pixels to the right from the object position in the direct image and the zeroth order is offset by $\sim380$ (interlaced) pixels to the left of this position (see Figure~8.6 in the WFC3 Instrument Handbook). Objects that fall off the left (blue) side of the direct image will still be dispersed onto the detector and objects to the right (red) of the image can create 0$^\mathrm{th}$ order spectra within the grism exposures. Such objects need to be taken into account in the contamination model and can also yield scientifically useful spectra.  To account for these objects, we make the blotted reference images larger than the original \textsc{flt} frames by 215 original pixels (430 interlace pixels) on each side along the x-axis. We also add 45 pixels on each side along the y-axis to account for objects along the top and bottom edge of the image. Figure~\ref{fig:contam_model}a,c shows how the interlaced reference image produced from the mosaic compares to the interlaced grism image. The final interlaced images are $2888\times2208$ pixels.

\begin{figure*}[th]
\centering
\includegraphics[width = \textwidth]{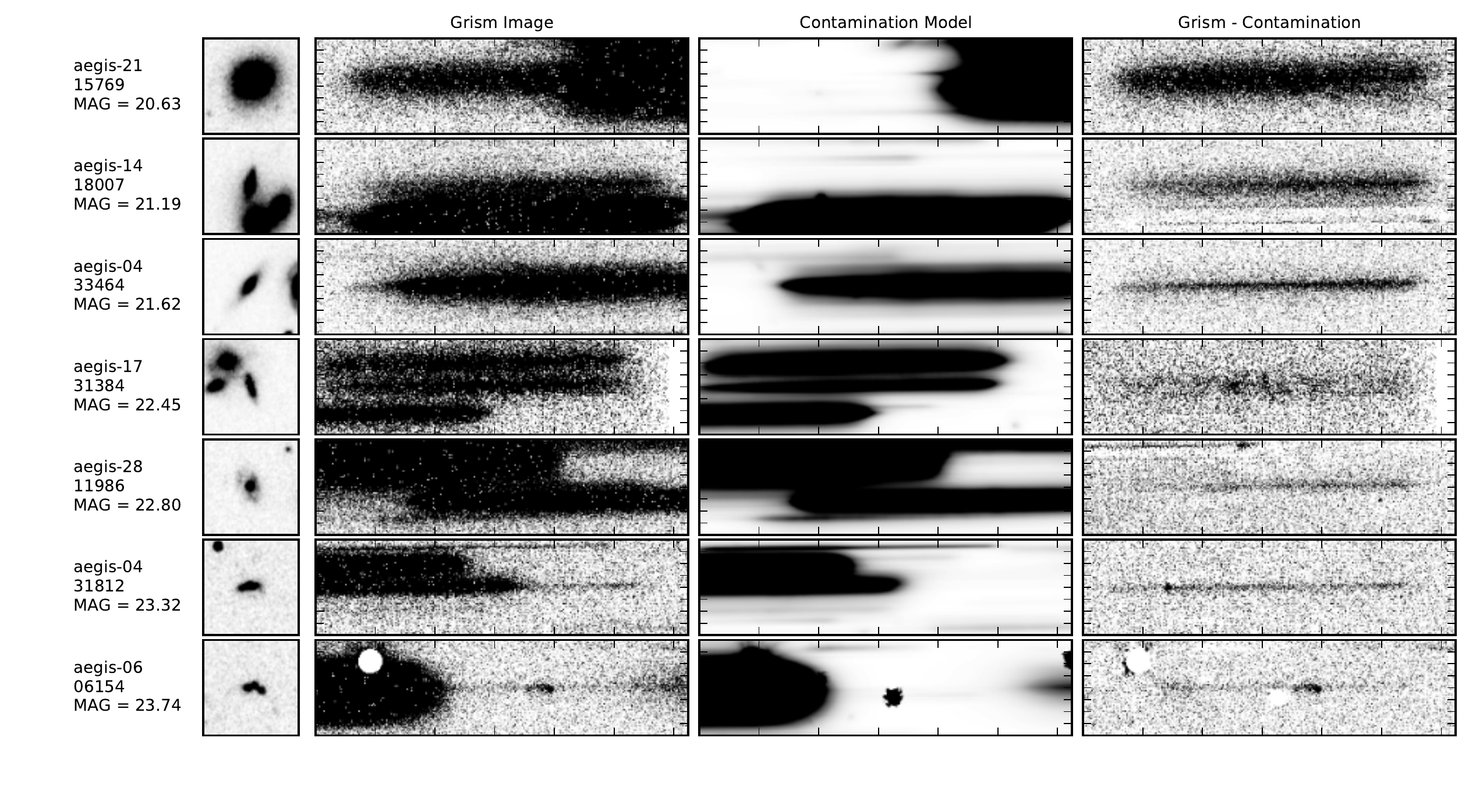}
\caption{\footnotesize Illustration of contamination subtraction for a range of objects of different magnitudes with substantial contamination. Objects shown here were chosen to have high contamination where the modeled contaminating flux is 70\% or more of the expected object continuum flux somewhere along its spectral trace.  Creating a quantitatively accurate contamination model is critical because $\sim$50\% of objects at $\JH\leq24$ have contamination at this level.}
\label{fig:various_contam}
\end{figure*}

\section{Contamination Model and Spectral Extractions}
\label{sec:extract}

Following the preparation steps, we extract the two-dimensional spectra of individual objects from the interlaced mosaic images. A key element of the extraction of slitless spectra is creating a model that identifies which pixels constitute the spectrum of a given object, which pixels belong to neighboring sources, and areas where spectra overlap. Our goal is not only to simply identify areas of the image with overlapping spectra, but to create a quantitative model that accurately accounts for overlapping spectra from sources dispersed onto the same or neighboring pixels. The basis of this contamination model is an estimate of the contribution of every source in the direct image to the grism image. The contributions of the individual objects are independent and can be co-added to create a complete model of the grism image. For each object in the grism image then, the contamination model consists of the co-added contributions of all other objects. We refer to this as the {\it contamination model} to distinguish it from spectral models described in \S\ref{sec:redshift_fits} and \S\ref{sec:line_fits}. The  accuracy of the contamination model determines the quality of the extracted spectra. Since our goal is to extract high-quality spectra for all objects in the footprint of the survey, the fidelity of the contamination model is of paramount importance. In this section we describe the approach to creating the quantitative contamination model, the steps of the extraction, and show examples of the final two-dimensional (2D) and one-dimensional (1D) reduced spectra.

\subsection{General Considerations}
\label{sec:contam_overall}

The grism dispersion varies across the instrument field. The dispersion is described in configuration (\textsc{conf}) files provided by STScI, such that for a given $x$ and $y$ pixel position in the observed direct image frame, one can determine the position of the dispersed spectrum, the ``trace'', of each spectral order in the observed grism exposure, as well as the wavelength along that trace. The position of the trace and the wavelength solution along the trace are described by low
order polynomials, where the polynomial coefficients are themselves order-$n$ polynomials that encode the position-dependence of the trace calibration ($n\leq6$). The dispersion varies smoothly across the field of view and the edge-to-edge variations are small. The WFC3  dispersion of the main, $+1^{st}$ order varies between 44.7 \AA\ per pixel and 47.8 \AA\ per pixel across the field ($\pm3\%$ over 1014 pixels).   The resulting traces and dispersion are smooth functions of the $x$ and $y$ position in the image and the spectra are slightly tilted (by $\sim 0.5^{\circ}$) with respect to the detector rows. It is important to note that both the position of the trace and the wavelength along the trace are defined within the coordinate system of the distorted image; adapting them to the interlaced images, which are also distorted (but padded and re-sampled) is therefore straightforward. 

As described in the previous section, we pad the reference images in order to account for objects dispersed within the grism frame. In order to model the spectra for these objects we assume that the trace and dispersion polynomials continue their smooth behaviour outside the field of view of the instrument. We find that this extrapolation  is sufficiently stable to enable modeling the spectra of these outlying objects. 

The HST grisms are not equipped with order-blocking filters and, therefore, multiple spectral orders are dispersed onto the detector for each object (depending on its position in the frame). The primary, +$1^\mathrm{st}$, order contains most of the power of the dispersed spectra, followed by the \zeroth\ order, and then by the $-1^\mathrm{st}$ and the higher orders. As a result, for most objects, only the \zeroth\ and +$1^\mathrm{st}$ orders are visible in the grism images, with higher orders only visible for bright objects such as stars.
The \zeroth\ order is only slightly dispersed and appears similar to the undispersed images of objects in the direct images, though offset in position.  These spectra can appear much like compact emission line sources and are an important component of the contamination model (the \zeroth-order position is fairly well calibrated to a precision of $\sim$1 pixel in the current configuration files).  For the contamination model we include all of the orders with entries in the configuration files ($-1^\mathrm{st}$ through +$3^\mathrm{rd}$), although we note that the positions and intensities of the higher orders are generally less well calibrated than the \zeroth\ and +$1^\mathrm{st}$ orders.\footnote{For reference, the dispersion of the ACS G800L grism is 38.8\,\AA\ per pixel in the center of the frame; the G800L spectra are tilted by $\sim 2^{\circ}$; and the contamination model contains the $-3^\mathrm{rd}$ to +$3^\mathrm{rd}$ orders.}

\subsection{The Contamination Model}
\label{sec:contam_model}

The inputs for creating the contamination model are the reference direct image (panel a in Fig.\ \ref{fig:contam_model}, see \S~\ref{maskred.sec}) and the segmentation map, both projected into the distorted interlaced image frame, the {\tt SExtractor} $\JH$ catalog, also projected into the distorted frame, and the interlaced grism image (panel c in Figure~\ref{fig:contam_model}). 

To model the two-dimensional spectrum of a given object, we first compute the trace and dispersion parameters for each spectral order at the center coordinates of that object.  These parameters define the dispersion of a single pixel in the reference image into a long skinny, one pixel wide spectrum in the cross-dispersion direction. The full two-dimensional model is then built by shifting and adding this elemental spectrum, after scaling it by the observed flux in the reference image, for each pixel within the segmentation region.  The entire process is analogous to a convolution of the two-dimensional thumbnail in the reference image with an assumed one-dimensional object spectrum.

The two main considerations in creating the models for the individual objects is the treatment of their spatial and spectral light distributions.  For modeling the {\em spatial} distribution
we use a single reference image to define the morphology of an object (constructed from the available $J_{125}$ + $JH_{140}$ + $H_{160}$ mosaics as described in \S\ref{maskred.sec}).  In favor of computational simplicity, this neglects any wavelength dependence of the source morphology, which may be complex for well-resolved objects (e.g., extended line emission and compact continuum emission).  The measurement of the relative sizes and morphologies of continuum and line components of distant galaxies is in itself an important scientific diagnostic largely unique to \textit{HST} slitless spectroscopy \citep[e.g.,][]{Nelson13}. Furthermore, compact objects are susceptible to the change in PSF size as a function of wavelength; the WFC3/IR PSF at $1.7~\mu\mathrm{m}$ is $\sim20\%$ larger than at $1.0~\mu\mathrm{m}$. In general, we find that cross-dispersion residuals are small for all but the most compact objects. 

We model the {\em spectral} distribution in the following way.
To first order, the full contamination model can be computed by assuming flat source spectra normalized to the observed flux in the reference image.  While this would typically be sufficient for contamination \textit{masking} (see Figure~\ref{fig:model_demo}), our goal is to generate a high-fidelity, \textit{quantitative} contamination model that can be subtracted from the observed spectra.  For every object in the 3D-HST photometric catalogs from \cite{Skelton14}, we obtain the best-fit galaxy EAZY SED template determined from the photometric redshift fit (with emission lines removed, as these would not be at the correct observed wavelengths based on the imprecise photometric redshift estimates).  In some sense this is similar to the \aXe\ fluxcube model that measures fluxes directly from reference images in multiple bands to model the broadband spectrum shape.  Here, however, the galaxy template is obtained from the fit to all of the available photometric bands, \textit{HST} and ground-based; the EAZY fits accounts to some extent for line contributions to the broad-band fluxes; and, most importantly, the EAZY spectrum is a full stellar population synthesis model, not a polynomial.  Once we create a complete contamination model of the pointing from the EAZY templates, we then refine it for objects brighter than $\JH=22$ based on the object spectrum itself, extracted along the central pixel of the trace.  To the extent that contaminating objects are not aligned exactly along the trace, this refinement step will then include emission lines and will correct any mismatch between the EAZY template and the observed spectrum.

The overall quality of the contamination model is illustrated in Fig.\ \ref{fig:contam_model}. The full contamination model of the pointing (panel b) is subtracted from the interlaced grism image (panel c) to highlight the differences between the modeled and observed data (panel d). Significant residuals are only seen at the position of the spectra of the two bright stars in the pointing in rows $\sim700$ and $\sim2100$, as well as several fainter ones across the image. Overall, the quantitative agreement is excellent. Figure~\ref{fig:contam_model} also demonstrates that that, as a result of the padding of the reference image, the model extends across the full width of the grism image. Without the padding, the spectra in the left-most 300 pixels of the grism image would not have been modeled. We note that the model image (panel b) is only created for quality control and visualization purposes. The actual contamination model is a set of arrays of wavelength, flux and position for each object within the pointing, which, combined with the reference image thumbnails and segmentation maps, can be used to reproduce the positions and fluxes of all spectra. 

A detailed demonstration of the iterations on the contamination model is shown in Fig.\ \ref{fig:model_demo}. To illustrate the differences between the models we show the residuals for three objects, modeled in three different ways. The first set of panels shows the residuals from a model where flat ($f_\lambda$) spectra are assumed for all objects.  The second set of panels shows the residuals for models computed with spectra defined by the EAZY templates. The rightmost set of panels show the final model where the spectra of bright objects are refined based on the observed spectrum itself. Three representative galaxies are shown.
Galaxy A is a bright galaxy with an emission line.  The flat model does not include the slope of the spectrum or the emission line. The EAZY model decreases the residuals in the continuum, but only the iterative approach models both the continuum and the emission line well. Galaxy B is a bright continuum source: the EAZY model decreases the residuals, but the iterative model is slightly better. Panel C shows faint emission-line sources. In this case all models do equally well. These sources are below the magnitude limit for the iterative model and their lines are not included in any of the models.
The combination of the EAZY first-guess approach with the subsequent iteration for bright objects correctly captures the spectral energy distribution for the majority of objects in the survey and accounts for most of the contaminating flux. The fact that emission lines in faint objects are not included in the contamination model could affect redshift fits of overlapping objects. However, since the redshift and emission line fits are done in 2D (\S\ref{sec:redshift_fits} and \S\ref{sec:line_fits} below), this is only a serious issue if the traces of two objects with the same morphology are exactly aligned, which is exceedingly rare.

A final illustration of the contamination model is provided in Fig.\ \ref{fig:various_contam}, where we specifically show objects which have modeled contaminating flux that is at least 70\% of the expected continuum flux at a point along the trace. Approximately 50\% of objects with $\JH\leq24$ have contamination at this level, and 95\% of objects with $24< \JH<26$. If we were unable to subtract the contaminating flux, the vast majority of spectra in the survey, especially at faint magnitudes, would not be usable. However, as show, the contamination-subtracted spectra are clean, with little or no residual flux from other objects.

We note here that the contamination modeling procedure as described here is a \textit{static} model.  That is, for the redshift and emission line fits for individual objects described in \S\ref{sec:redshift_fits} and \S\ref{sec:line_fits}, the static model of all contaminating sources is subtracted before performing the fit.  Future generations of the fitting software could perform the fitting in an iterative process, updating the static model of the full field after performing the detailed fits to each spectrum individually.

\subsection{Extracting 2D and 1D Spectra}
\label{sec:extract_2d}

The 2D spectra are the main ``basic'' data product of the survey. With the contamination model in hand the spectral extraction is relatively straightforward. For each object we use the configuration file to determine the pixels which contain the first order dispersed spectrum. For uniformity, we extract the same number of pixels, 312, along the dispersion axis for each object, which results in non-uniform wavelength grids for the extracted spectra as the dispersion $\Delta\lambda/\mathrm{pix}$ varies across the field. The extent of the spectrum in the spatial direction is chosen to be three times the {\tt SExtractor} FLUX\_RADIUS (with a minimum value of 26 interlaced pixels or $1.56\arcsec$ enforced for small objects),\footnote{We also provide cutouts with a fixed size in the spatial direction of 80 interlaced pixels. These cutouts are particularly useful for stacking analyses, and the analysis of spectra that are neighboring the object of interest.} and a two-dimensional cutout is extracted from the interlaced grism image. Cutouts are then extracted from the same pixels in the interlaced error image and the static contamination model generated for all objects other than the object of interest, and a separate extension is created for the object model itself. Square cutouts of the interlaced reference image and segmentation map are also extracted with the same cross-dispersion dimensions.  

With additional extensions defining the wavelength grid and sensitivity curve, the 2D FITS files therefore contain all of the necessary information for detailed modeling of the spectra with the \texttt{threedhst}\footnote{\url{https://github.com/gbrammer/threedhst}} analysis software, such as the redshift and emission line fits described in \S\ref{sec:redshift_fits} and \S\ref{sec:line_fits}. We emphasize that modeling the interlaced spectra directly in 2D is critical to the success of the fits and ultimately the quality of the catalogs derived from the spectra.

3D-HST is tiled in such a way that adjacent pointings occasionally overlap. For objects which fall within the overlap region we can obtain two (or more) independent 2D spectra, typically observed at different rotation angles. Additionally, the HUDF area of the GOODS-S field was observed in five separate visits. We refer to these as {\it duplicate spectra}. Such repeat spectra can be co-added \citep[see ][]{Brammer13} and in \citet{vanDokkum13deep} we released the full depth grism spectra in the HUDF. In the current release all spectra have consistent 2-orbit depth and repeat observations are not co-added. In \S~\ref{sec:z_duplicates} and \ref{sec:line_fits} below we use the duplicate spectra to quantify the accuracy of redshift and emission line flux measurements. 

While the analysis of ground-based slit spectra is typically performed on 1D extractions, collapsing the data to 1D leads to important information losses in the case of the slitless grism spectra. Galaxies often have complex morphologies in the 2D spectra \citep[see, e.g.,][]{Nelson15}, and several distinct clumps of line emission would be degraded to a single broad feature in a 1D spectrum.  Although we do not use 1D spectra in the fits we provide them for convenience and to facilitate plotting and visualization. The 1D spectra are optimally-weighted \citep{Horne86}.  We limit the extraction window to the region where the average flux in the dispersion direction is greater than 10\% of its maximum to stay within the high S/N area of the spectrum.  The \texttt{threedhst} analysis software that operates on the 2D spectra also provides the capability of extracting 1D spectra in apertures specified in integer pixels relative to the center of the spectral trace.

\begin{figure*}[ht]
\centering
\begin{minipage}[b]{0.49\linewidth}{
\includegraphics[width=\textwidth]{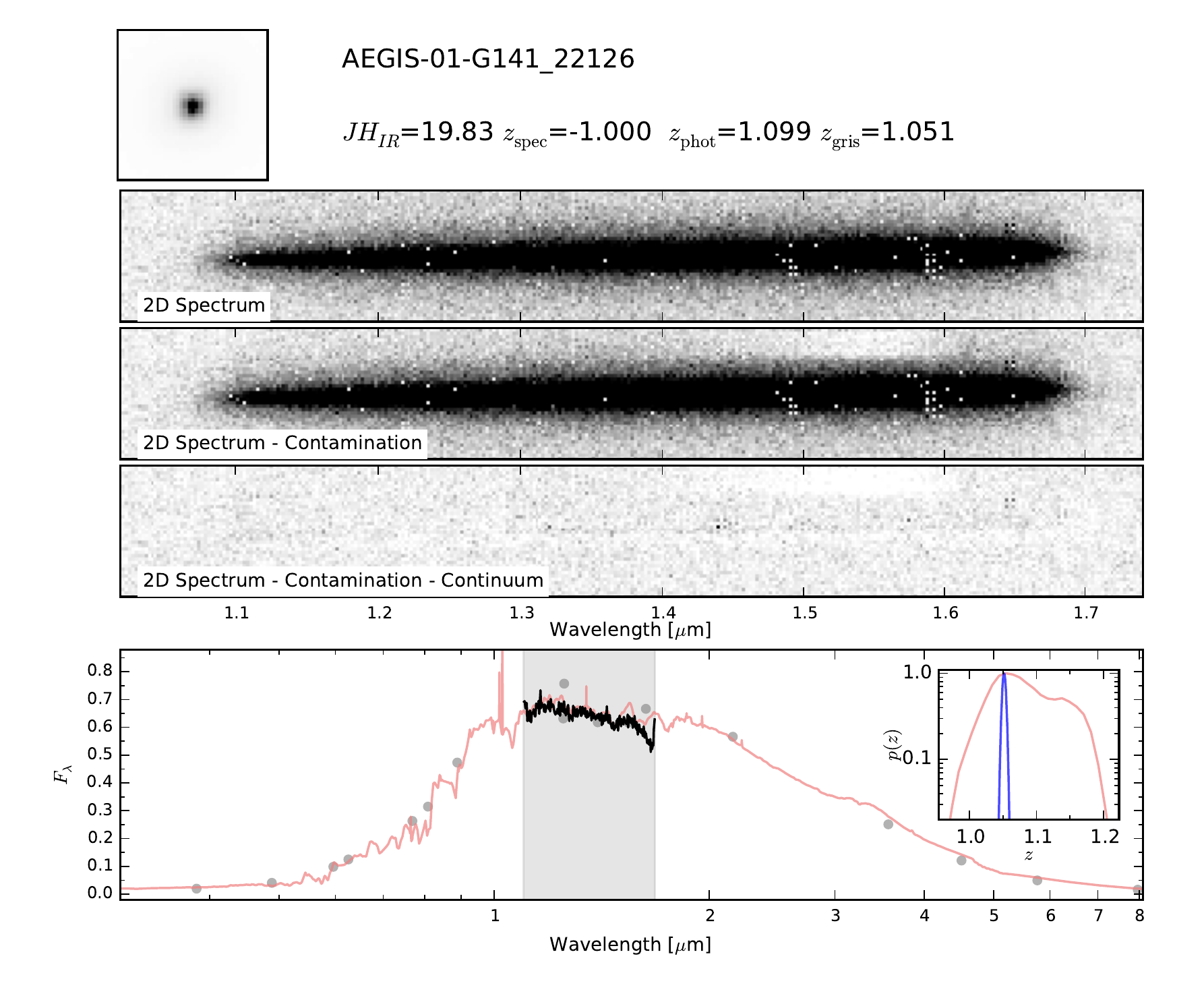}}
\end{minipage}
\begin{minipage}[b]{0.49\linewidth}{
\includegraphics[width=\textwidth]{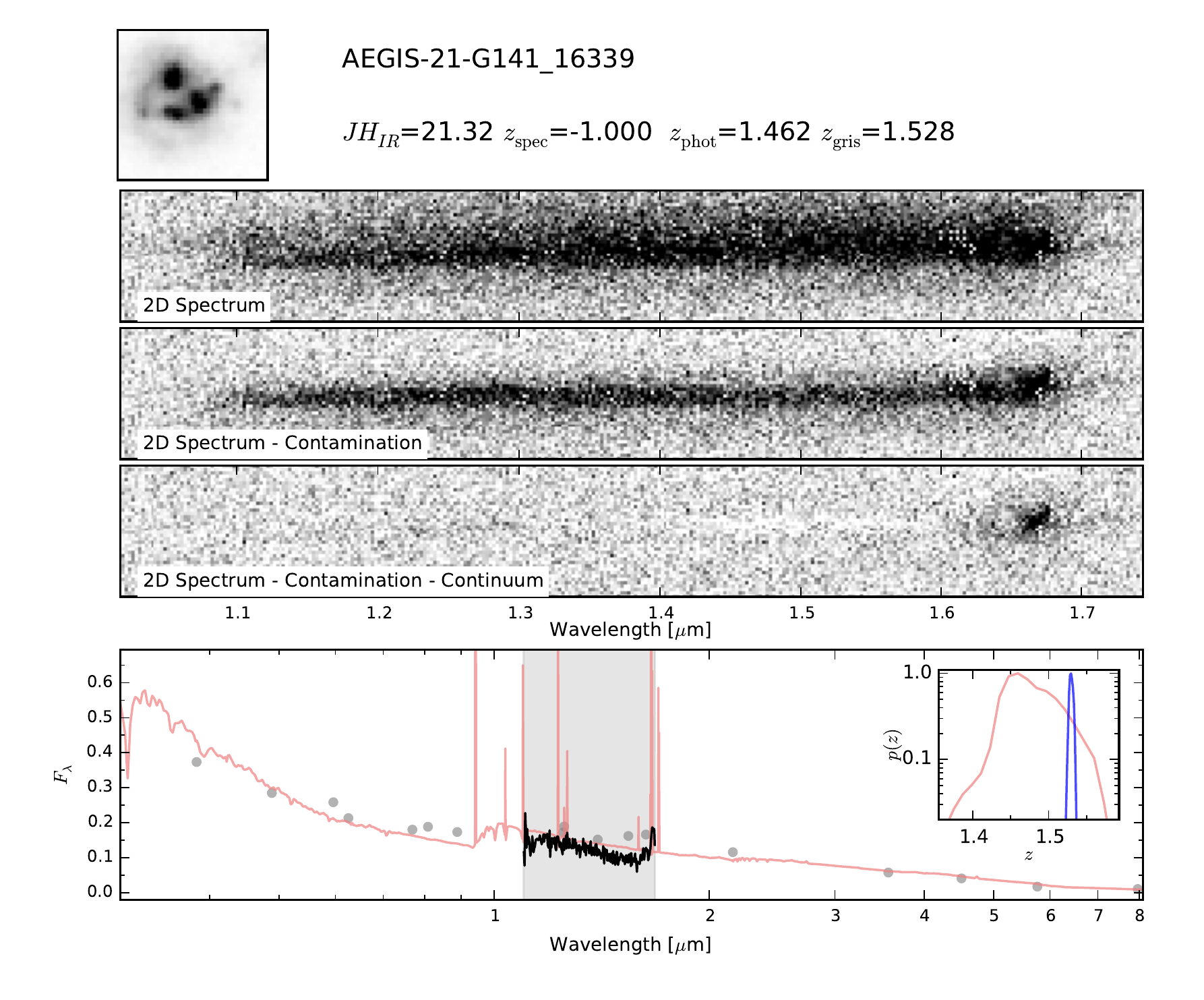}}
\end{minipage}
\begin{minipage}[b]{0.49\linewidth}{
\includegraphics[width=\textwidth]{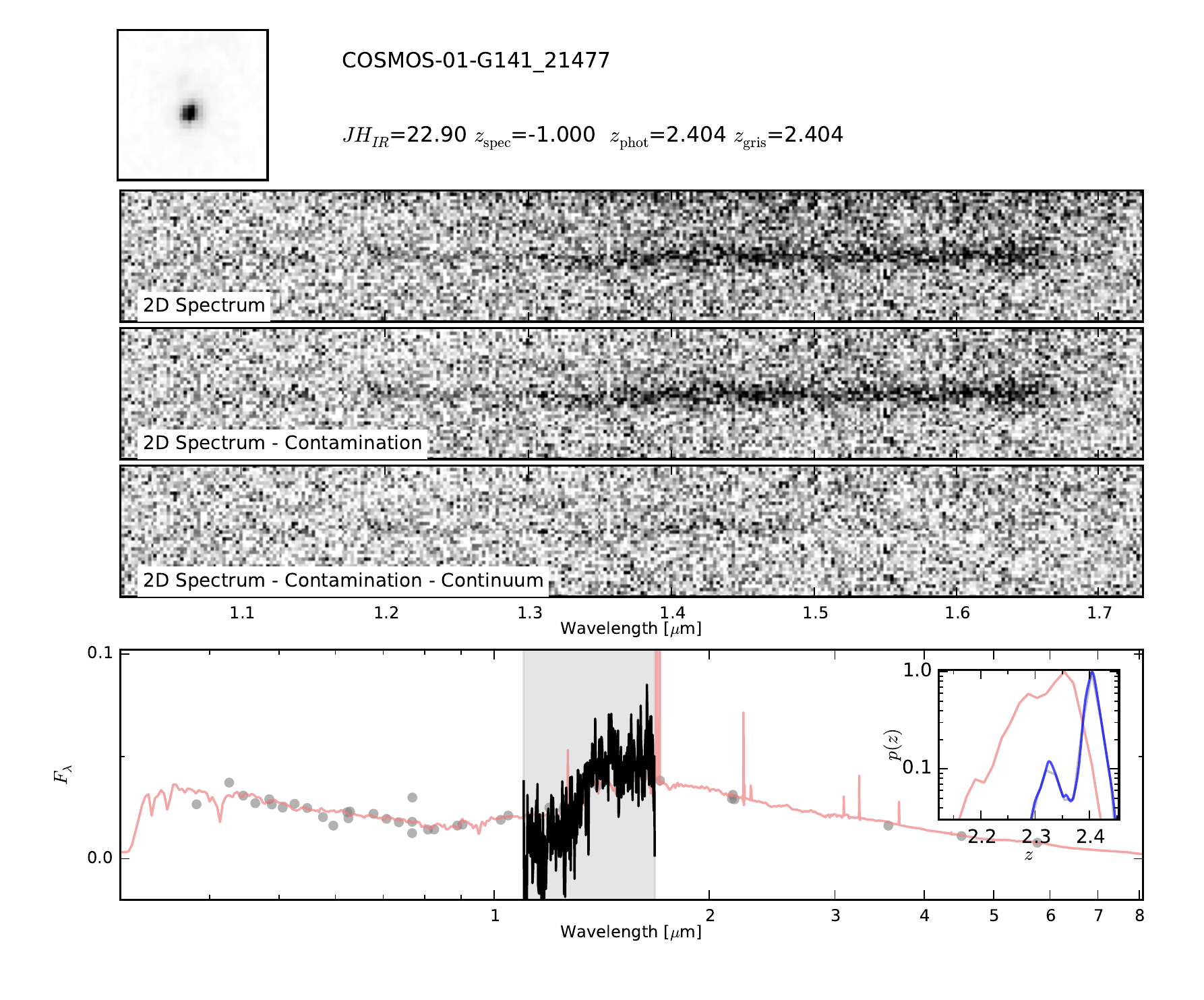}}
\end{minipage}
\begin{minipage}[b]{0.49\linewidth}{
\includegraphics[width=\textwidth]{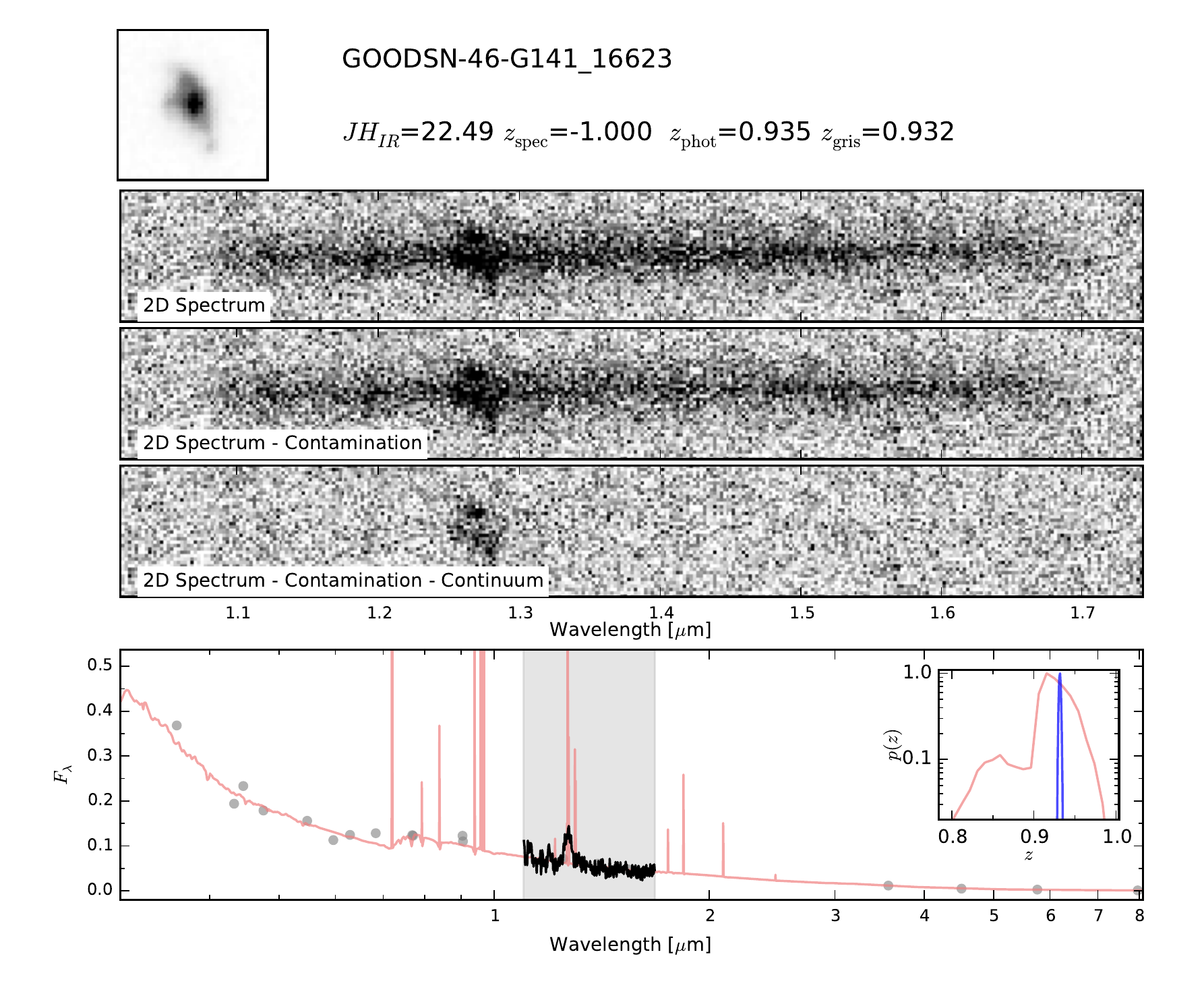}}
\end{minipage}
\begin{minipage}[b]{0.49\linewidth}{
\includegraphics[width=\textwidth]{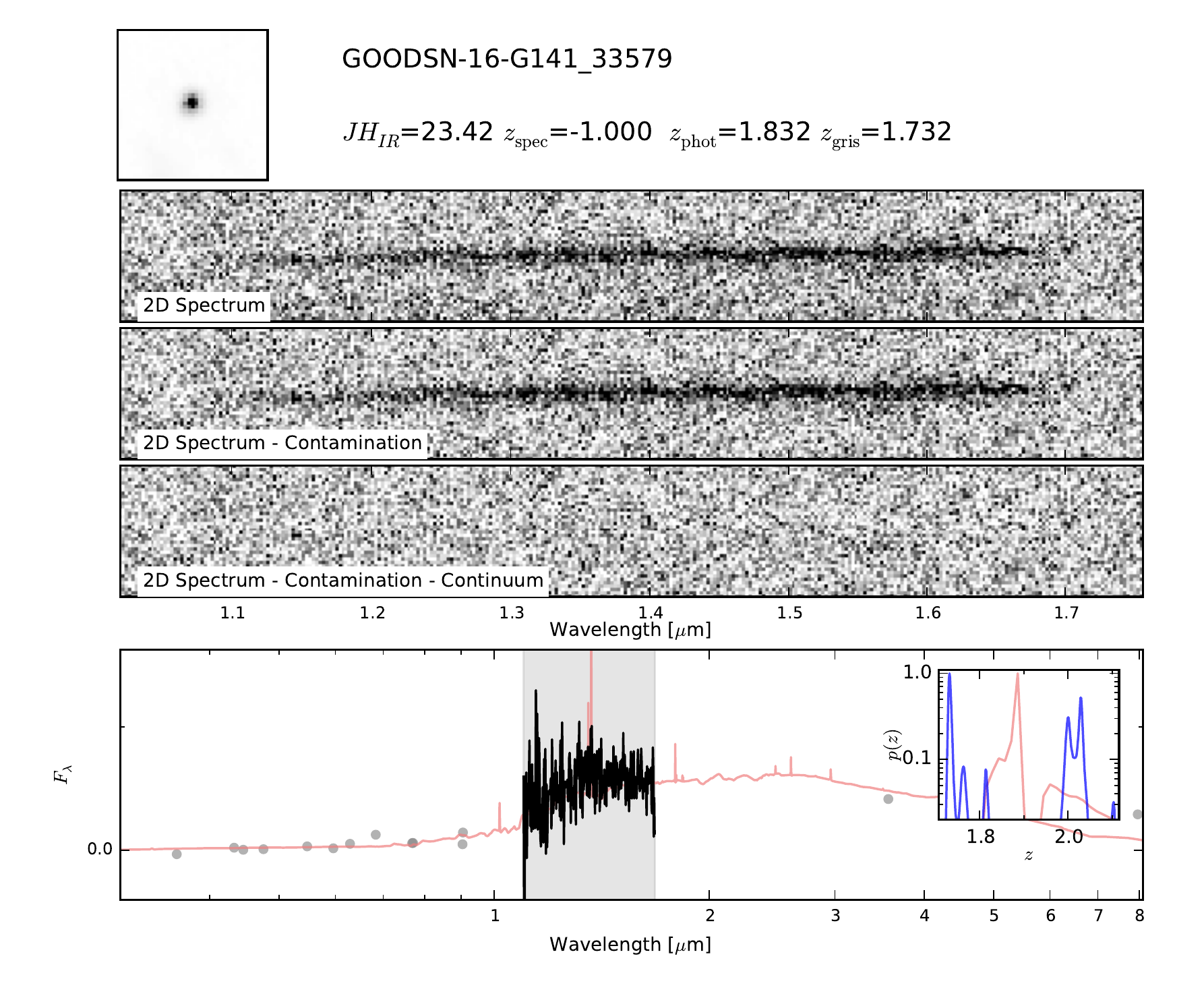}}
\end{minipage}
\begin{minipage}[b]{0.49\linewidth}{
\includegraphics[width=\textwidth]{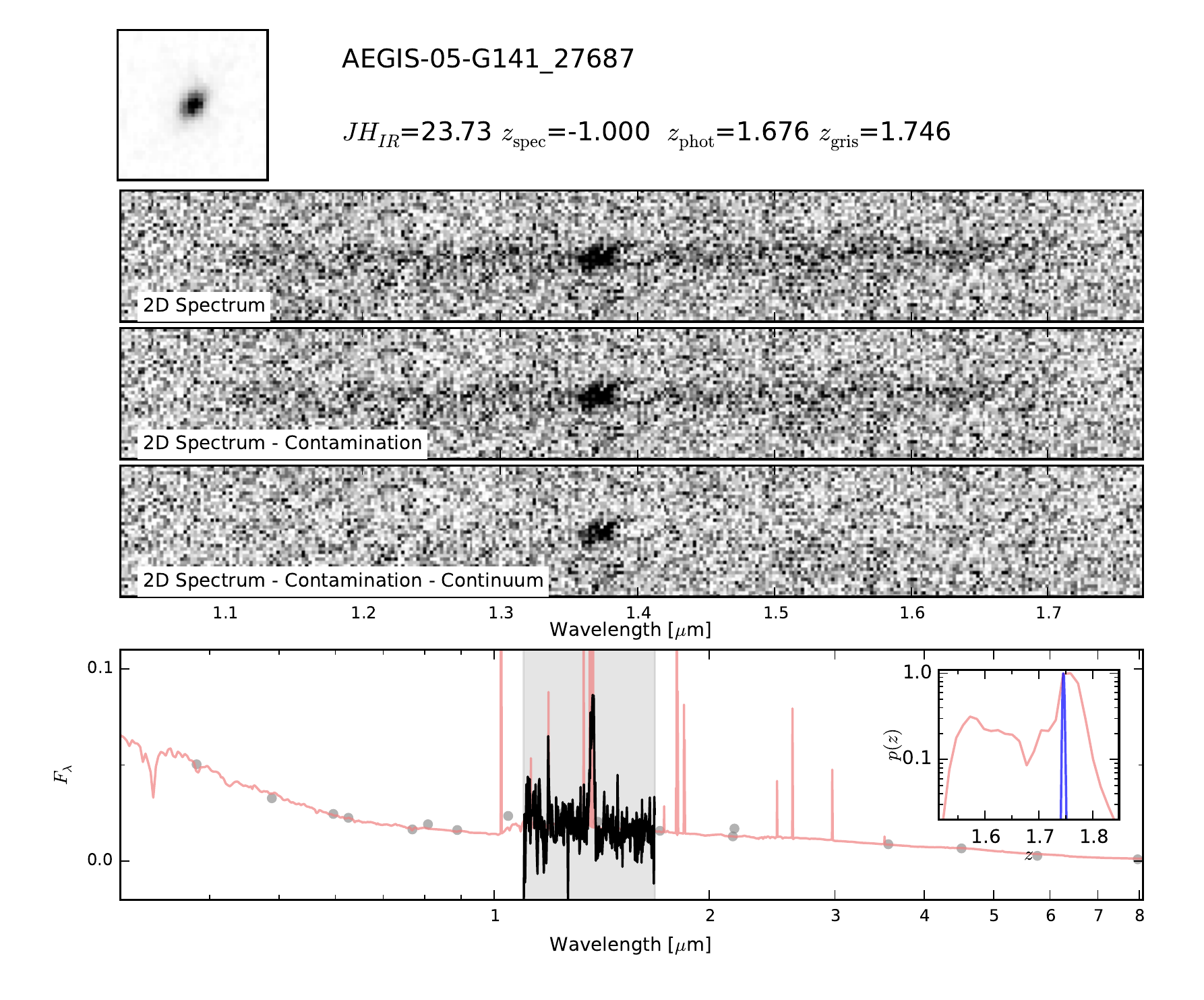}}
\end{minipage}
\caption{\footnotesize Examples of grism spectra. In the left column we show three quiescent galaxies with decreasing $\JH$ magnitudes (top to bottom) and in the right column we show three emission-line objects. For each object we show the reference direct image (top), the 2D grism spectrum extracted from the G141 image, the contamination-subtracted 2D spectrum and the continuum-subtracted 2D spectrum (labeled). In the bottom panel in each plot we show the optimally-extracted 1D spectrum (black line), overlaid on the photometric points (gray circles) and the best-fit redshift template (red line). The inset shows the redshift probability distributions for the photometry fit alone (red), for the grism fit alone (gray) and for the joint photometry+grism fit (blue).
}
\label{fig:examples}
\end{figure*}

With the data release described in this paper, we deliver spectra for all objects in the \citet{Skelton14} catalogs that fall within the 3D-HST grism pointings. A total of 246,052 2D and 1D spectra have been extracted. In Figure~\ref{fig:examples} we show several examples of 2D and 1D spectra. For each object we show the reference image as well as the 2D interlaced spectrum, the 2D contamination-subtracted spectrum, and the 2D spectrum with the continuum model subtracted (see below). The latter spectra are essentially 2D maps of line emission \citep[see][]{Nelson15}. Also shown are the 1D spectrum and the broad band SED.\footnote{All data products in Fig.\ \ref{fig:examples} are provided in the data release.} The objects were chosen to span a range in $\JH$ magnitude, redshift and color. In all examples, the extractions are clean and contaminating spectra from nearby objects are well subtracted. Note that the emission lines have a similar morphology as the reference images of the objects.

\section{Redshift Fits}
\label{sec:redshift_fits}

Most large scale extragalactic surveys using low-resolution slitless spectroscopy have focused on selecting samples of emission line objects \citep[e.g.,][]{Zamorano94, Gallego96, Colbert13, Morris15}. However, 3D-HST has broad science goals which require measurements for galaxies whether or not emission lines are present. In this section we describe our approach to fitting redshifts in a uniform manner for the complete sample of galaxy spectra.

\subsection{Methodology}

\begin{figure*}[ht]
\centering
\begin{minipage}[b]{0.49\linewidth}{
\includegraphics[width=\textwidth]{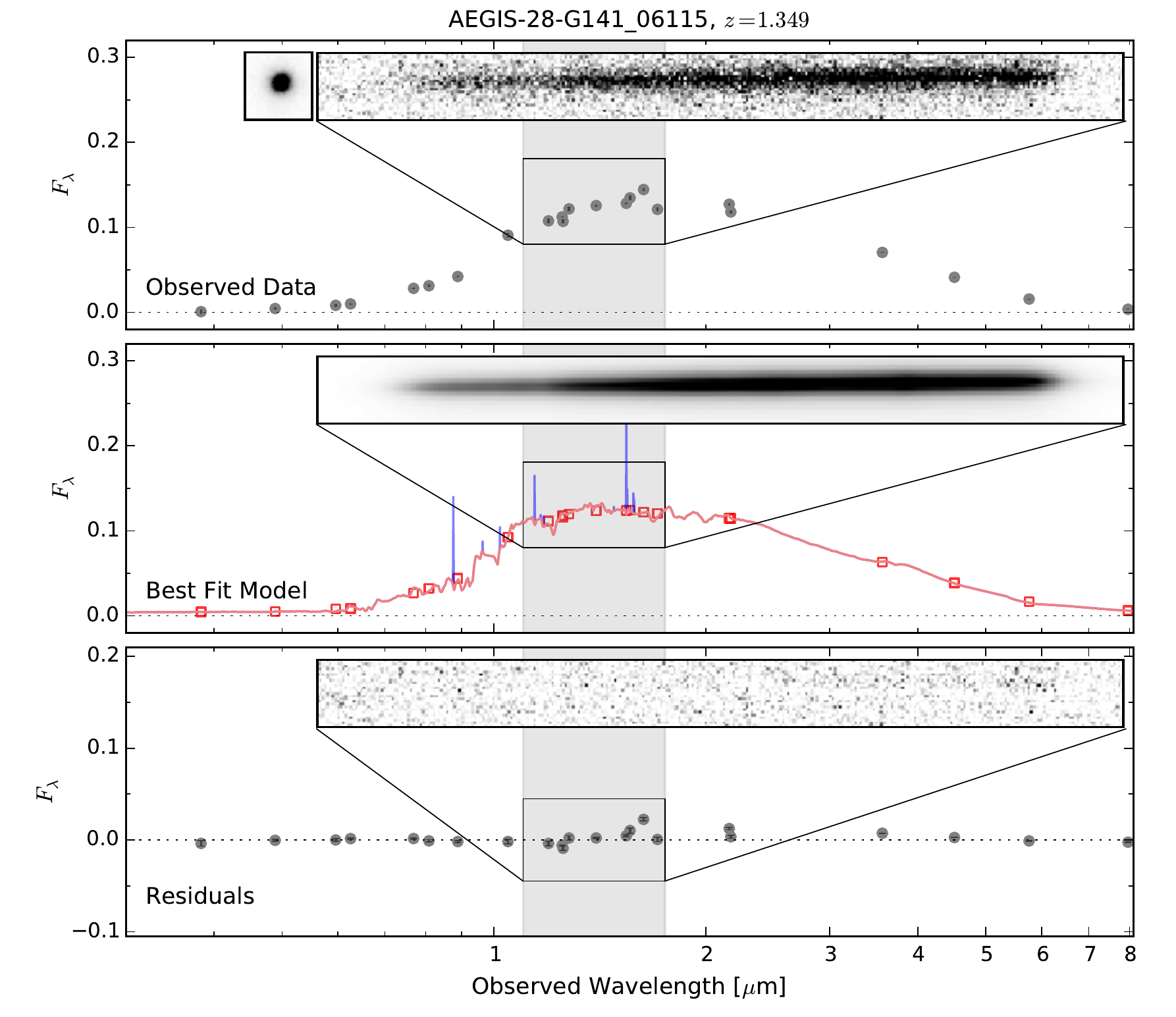}}
\end{minipage}
\begin{minipage}[b]{0.49\linewidth}{
\includegraphics[width=\textwidth]{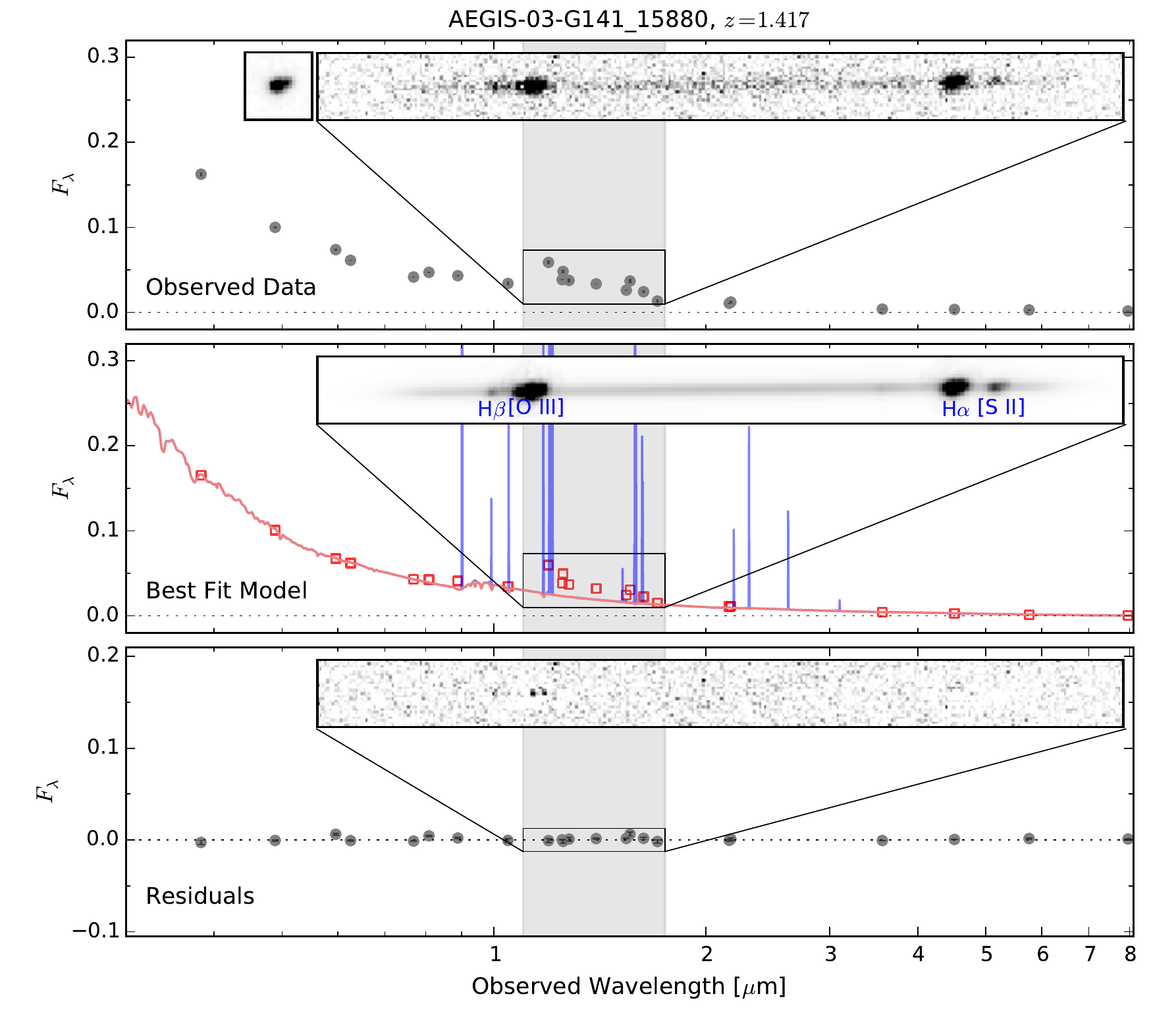}}
\end{minipage}
\caption{\footnotesize Examples of the redshift fitting procedure for two objects: AEGIS-03-G141\_06115, a quiescent  galaxy at $z=1.349$ and AEGIS-28-G141\_15880, a star forming galaxy at $z=1.417$. In the top panels we show the observed data: the photometric fluxes (gray circles, errors are smaller than points), the 2D image of the object (square inset) and the 2D grism spectrum (inset). In the middle panels we show the best-fit linear combination of templates (red lines, with the emission lines added in blue). We also show the best-fit 2D model of the spectrum, which is a convolution of the template and the 2D direct image.  In the bottom panels we show the residuals (gray points, inset), which are minimized in the fitting process. Note that the 2D spectrum and the photometry are modeled simultaneously, in a single fit.
}
\label{fig:redshift_fits}
\end{figure*}

Due to the low resolution and limited wavelength coverage of the slitless \textit{HST} spectra, the information content of the spectra is often insufficient for robust redshift determinations from the spectrum alone. For example, absorption features are rarely detected with sufficient S/N in individual spectra and common emission line redshift indicators are unresolved (e.g., [\ion{O}{2}]$\lambda\lambda$3727,3729 and H$\alpha$+[\ion{N}{2}]). Furthermore, because of the fixed bandpass of the grism, the rest-frame wavelength coverage is smaller for higher redshift objects: at $z=0.7$ the grism captures $\sim4100$~\AA\ of the optical and NIR spectrum, while at $z=2.0$, only $\sim2300$~\AA\ of the UV/optical spectrum falls within the grism wavelength range. With such a limited wavelength coverage, it is common that only one prominent spectral feature appears within the G141 spectral coverage.

For the 3D-HST redshift measurements, we combine the additional information from the (multi-band) photometric catalogs with the information in the (2D) spectrum and fit both components simultaneously.  We note that the rich photometric datasets alone often provide photometric redshift precision of $\sim$1--3\%, thereby leading to little ambiguity in identifying lines detected in the spectra.  For the combined photometry + redshift fits we use a modified version of the EAZY templates. Emission lines are removed from the standard templates, and we supplement these continuum templates with an additional pair of emission line templates. The emission templates contain \textit{just} emission lines, with line ratios taken from the average SDSS star-forming galaxy spectra of \cite{Dobos12}.  The pair of templates brackets extreme values of the H$\beta$/[\ion{O}{3}] ratios to allow for flexibility in the fits, as this ratio has been observed to evolve significantly with redshift \citep[e.g.,][]{SteidelKBSS}.

After computing a scalar normalization factor $s$ between the photometry and spectrum (which are generally offset from one another, due to aperture effects), the data provided to the fit are the individual photometric measurements $f_i$ (with uncertainties $\sigma_i$) along with the scaled flux in \textit{each pixel} in the contamination-subtracted two-dimensional grism science and error spectra ($G_{xy}$ and $\epsilon_{xy}$).  A 2D spectrum model ($T_{xy,j}$) is computed for each redshifted template, $j$, as described in \S\ref{sec:extract_2d}, which is also convolved with the photometric bandpasses ($T_{i,j}$).  The final likelihood of the fit at redshift, $z$, is then 
\begin{eqnarray}
T_j & = &\sum_j^{N_\mathrm{templ}} \alpha_j \cdot T_{i,j}, \label{eq:tj} \\
T_{xy} & = &\sum_j^{N_\mathrm{templ}} \alpha_j \cdot T_{xy,j}, \label{eq:txy} \\
\epsilon_{xy}^2 & = & g_{xy}^2 + \gamma \cdot c_{xy}^2, \label{eq:eps} \\ 
\chi^2 & = & \sum_i^{N_\mathrm{filt}} \frac{\left(f_i - T_i\right)^2}{\sigma_i^2} + \sum_{xy}^{2\mathrm{D}} \frac{\left(s\cdot G_{xy} - T_{xy}\right)^2}{\epsilon_{xy}^2}, \label{eq:redshift_chisq} \\
\mathcal{L} & = & e^{-\chi^2/2} \nonumber,
\end{eqnarray}
\noindent where the individual  non-negative template normalizations $\alpha$ in Eqs.\ref{eq:tj}--\ref{eq:txy} are computed with the ``NMF'' algorithm of \cite{ShaNMF} \citep[see also][]{Blanton07, Brammer08}.  The per-pixel spectrum uncertainties $\epsilon_{xy}$ in Eq.~\ref{eq:eps} include an additional term to reduce the contribution of pixels with estimated contamination count rates $c_{xy}$ (with $\gamma=1$; higher values would result in more aggressive contamination masking).   Eq.~\ref{eq:redshift_chisq} demonstrates the power of fitting the interlaced 3D-HST spectra:  at no point in the processing steps outlined above are the pixels $G_{xy}$ in the 2D spectrum resampled and the per-pixel spectrum uncertainties $\epsilon_{xy}$ are preserved from the instrumental noise model.  Missing pixels (hot pixels, etc) in the 2D interlaced spectrum are simply excluded from the fit.

As the generation of the convolved G141 template spectra is the computational bottleneck for the redshift fit, it is done in three stages to reduce the computational time. In the first two steps we fit the photometry and spectrum separately, over a complete but coarse redshift grid that is just sufficient to identify strong emission lines in the spectrum.  We multiply the likelihoods from the separate fits. We finally carry out a third fit to the spectrum and photometry simultaneously (Eq.~\ref{eq:redshift_chisq}) on a fine wavelength grid, informed by the redshift range where the joint likelihood is the highest. To allow more freedom in this final step we assume that some small fraction ($1\times10^{-4}$) of the overall likelihood is evenly distributed between $z=0$ and $z=4$.  In the limit of low S/N in the spectrum ($G_{xy}, T_{xy} \rightarrow 0$), the contribution of the spectrum to the likelihood in Eq.~\ref{eq:redshift_chisq} will be effectively constant with redshift, and the fit will be dominated by the first term from the photometry.  That is, it will be similar to the EAZY photometric redshift fit, although not identical due to the adoption of different continuum and emission line templates as described above. 

Examples of the redshift fitting procedure are shown in Figure \ref{fig:redshift_fits} for two objects, a quiescent and a star forming galaxy, both at similar redshifts, $z\sim1.4$. In the top two panels we show the observed data: the photometric fluxes (gray points), the 2D image of the object and the 2D grism spectrum (insets). The model is evaluated on how well it fits both the photometric points and the 2D grism spectrum. In the middle panels we show the best-fit combination of templates (red lines, with the emission lines added in blue). The template is then projected into the observed space as we derive the observed fluxes in the photometric bandpasses (red squares) and convolve it with the 2D direct image to create the model 2D spectrum (inset). In the bottom panels we show the residuals, which are minimized in the fitting process.

The best estimate for the galaxy redshift determined from the photometry+spectrum fit, $z_\mathrm{grism}$, is taken to be the redshift where $\mathcal{L}$ is maximized.  We save the full likelihood distribution $\mathcal{L}(z)$, from which we calculate the central redshift 68\% and 95\% confidence interval for each object.

The redshift fits described in this section have been done for all objects with $\JH < 26$. We require that a 2D spectrum has at least one non-zero column in order to be fit. These fits result in a total of 98,668 spectral fits for 75,386 unique objects in the \citet{Skelton14} catalog. The limiting magnitude of these fits is well below where we can expect to detect continuum emission at sufficient S/N \citep{Brammer12}. At $\JH = 26$, a pure emission line object (EW=$\infty$) will have a line flux $f\sim9\times10^{-17}~\mathrm{erg}~\mathrm{s}^{-1}~\mathrm{cm}^{-2}$; similarly, lines detected at S/N~$>$~5 will have EW$_\mathrm{obs} > 5200$~\AA.  In the remainder of this paper we focus our attention on the brighter $\JH < 24$ sample of 30,621 2D spectra, which we have visually inspected in its entirety (\S\ref{sec:inspections}) and assigned quality flags to. The catalogs described in \S~\ref{sec:linematched} contain the full sample at $\JH < 26$. While fainter galaxies with $24 < \JH < 26$ can be useful for scientific analyses, we caution that the quality of the redshift fits for those fainter sources has not been fully validated from visual inspections.

\subsection{Visual Inspections}
\label{sec:inspections}

We visually inspected the 30,621 spectra and redshift fits of all 23,564 objects with $\JH<24$ and WFC3/G141 coverage in the 3D-HST survey. The first goal of the visual inspections was to identify any systematic problems in the reductions and redshift fits, which led to the finding that the fits for bright, red, low-redshift galaxies are unreliable (see end of this section). No other systematic problems were identified. The secondary goal of the inspections was to assign a quality flag for spectra affected by known failure modes such as the incomplete masking of 0$^{\rm th}$-order spectra which can mimic emission lines, residuals from the spectra of very bright stars which may not be subtracted properly, and instances where corrupted photometric measurements lead to errors in the spectral fit. The inspections were done by 12 individuals such that each spectrum was seen twice and given two separate flags. The primary criterion of the inspections was ``is the redshift clearly affected by a systematic error in the spectrum?''.  These inspections are therefore somewhat permissive in favor of completeness.  For example, a spectrum with substantial contamination residuals from an overlapping object would {\em not} be marked as ``bad'' if the fit identifies an unambiguous emission line.  Furthermore, reliable redshift measurements can be obtained for objects at the edge of the field with less than a complete spectrum if the available coverage includes an emission line.  Redshift fits where the grism spectrum contributes little to the fit and the final redshift probability is identical to that from the photometry alone are also marked as ``good". These inspections do not attempt to flag ``bad'' spectra because the criteria for what constitutes a bad spectrum will heavily depend on the particular application. For these reasons we caution against blindly using the spectra for objects identified to have reliable redshift measurements  for other purposes.

Individual classifications are either ``good'', ``bad'', or ``unclear''. Of the 30,621 spectra classified, 2,824 are flagged as ``bad'' by at least one user and 900 are flagged as ``unclear'' by at least one user. Overall 3,540 (11.5\%) spectra have at least one flag set to ``bad'' or ``unclear''. This selection is a bit more permissive than other cuts used for grism spectra. For example, \citet{Fumagalli12} require that more than 75\% of the spectrum falls on the detector ($\sim12.5\%$ of all spectra are excluded) and that the average contamination is lower than 10\%, which results in removing 60\% of the initial sample. In general, we find that a successful redshift fit does not depend on the fraction of the spectrum which falls on the detector or the amount of contamination (to a point). However, these requirements can be useful in selecting spectra for stacking or other purposes where the successful redshift fit is not the
only requirement. 

At this stage we reconcile redshift measurements for duplicate objects, so that for all these objects we have a primary measurement, which appears in catalogs with one line per object, and a secondary measurement. Out of the 30,621 spectra fit down to $\JH=24$, 7,057 are repeat observations. If there are two (or more) spectral fits for a given object, we first remove those that have at least one flag set to ``bad'' or both flags set to ``unclear''. Amongst the remaining, we chose the fit with the narrowest $p(z)$ (as measured by the 68\% confidence interval) to include in the catalog (if only one spectrum remains, it is the default choice). If all spectra for a given object have at least one flag set to ``bad" or both flags set to ``unclear", none of them are included in the catalogs (there are only 149 such objects in the full five fields).

The subjective classifications are combined into a single flag in the redshift catalogs (\texttt{use\_subj}). Objects have \texttt{use\_subj=0} if either one of the classifications is ``bad'', or if both are ``unclear''. Spectra with no ``bad'' classifications and at least one ``good'' classification are assigned \texttt{use\_subj=1}. Of all classified objects, only 1686 (7.1\,\%) have \texttt{use\_subj=0}. Limiting the sample to the 21,876 non-stellar objects according to the \cite{Skelton14} criteria (\texttt{star\_flag=0}), 1314 objects (6.0\,\%) have \texttt{use\_subj}=0.

As shown in the next subsection, redshifts derived from the combined fit to the photometry plus the G141 spectrum ($z_{\rm grism}$) are generally more precise than redshifts derived from the photometry alone ($z_{\rm phot}$). However, there are exceptions: for some bright, red galaxies at low redshift we find that the photometric redshift provides somewhat more accurate redshifts than the grism redshifts. This is likely due to subtle issues with our reddest, oldest templates; as discussed in \citet{Fumagalli15b} templates based on popular stellar population synthesis libraries do not always accurately reproduce the broad absorption features at $\lambda_{\rm rest}>0.8~\mu\mathrm{m}$ in old galaxies. We find that the photometric redshifts are preferable to the grism redshifts for the small subset of galaxies whose grism redshifts do not fall in the 95\,\% confidence interval of the photometric redshifts and either have $z_{\rm grism}<0.55$ or have $z_{\rm phot}<0.65$ and $(U-B)_{\rm rest}>0.9$.

The final \texttt{use\_zgrism} flag is set to 1 if the subjective flag is 1, the object has \texttt{star\_flag=0}, and the criteria to remove bright, red, low redshift galaxies with incorrectly identified spectral features are met. This results in 18,927 objects with \texttt{use\_zgrism}=1.

\subsection{Redshift Accuracy and Precision}
\label{sec:redshift_accuracy}

Understanding the accuracy of the grism redshifts is of paramount importance, especially since this is, to our knowledge, the first time joint photometric and spectral fits have been done for a complete, magnitude-limited, sample of objects (irrespective of their spectral characteristics). In this section we test the grism redshifts in three different ways: by comparing them to spectroscopic redshifts, by comparing the redshifts of projected pairs, and by comparing repeat redshift measurements. The redshift accuracy may depend on the characteristics of the objects (e.g., their magnitudes, colors, and the presence of strong emission lines), and we investigate this as well. The accuracy of the photometry and photometric redshift fits, which we use jointly with the grism spectra, to a large extent predetermines the accuracy of the redshifts in this work. A detailed analysis of the photometric redshift errors is outside the scope of this worn and is presented in a companion paper, \citet{Bezanson15}.

\subsubsection{Ground-based Spectroscopic Redshifts}
\label{sec:ground_z}

\begin{figure*}
\centering
\includegraphics[width = \textwidth]{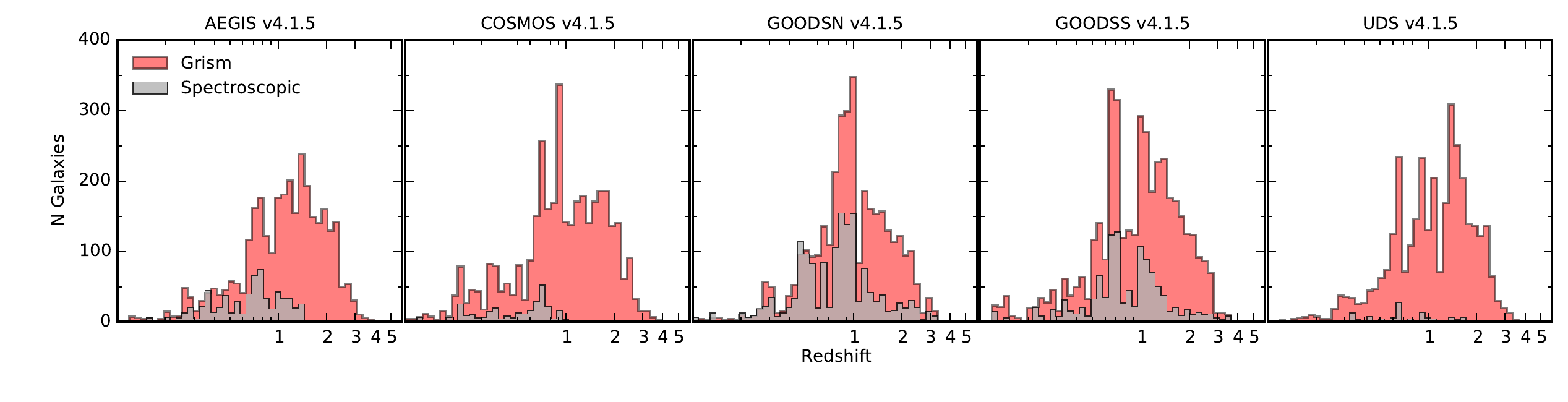}
\caption{\footnotesize Distributions of grism redshifts for objects with $\JH\leq24$ (red histogram) and ground-based spectroscopic redshifts (gray histogram) for the five deep extragalactic fields.
}
\label{fig:redshift_hist}
\end{figure*}

Comparison to ground-based spectroscopic redshifts is the most common method for determining the accuracy and precision of a redshift sample. In Fig.\ \ref{fig:redshift_hist} the spectroscopic redshift distribution is compared to the grism redshift distribution (for objects with $JH_{IR}<24$). The spectroscopic redshifts come from the entire CANDELS area \citep[see][]{Skelton14}, and are not limited to the area with grism coverage.  There are a total of 5,361 ground-based redshifts in these fields  \citep[see \S 5.1 of ][for references to the sources of the spectroscopic redshifts]{Skelton14}.
It is clear that the 3D-HST survey provides a major step forward in the spectroscopic coverage of the CANDELS fields. In all five fields the 3D-HST survey now provides the vast majority of redshifts $z>1$, and in COSMOS and UDS it is even the most important source of redshifts at low redshift. The field with the largest number of ground-based spectroscopic redshifts is GOODS-N, followed by GOODS-S. It is reassuring that the 3D-HST grism redshifts show the same redshift peaks as the ground-based spectroscopic redshifts. We illustrate the ability of our redshifts to identify large scale structures in section 8.

\begin{figure*}[th]
\centering
\includegraphics[width = \textwidth]{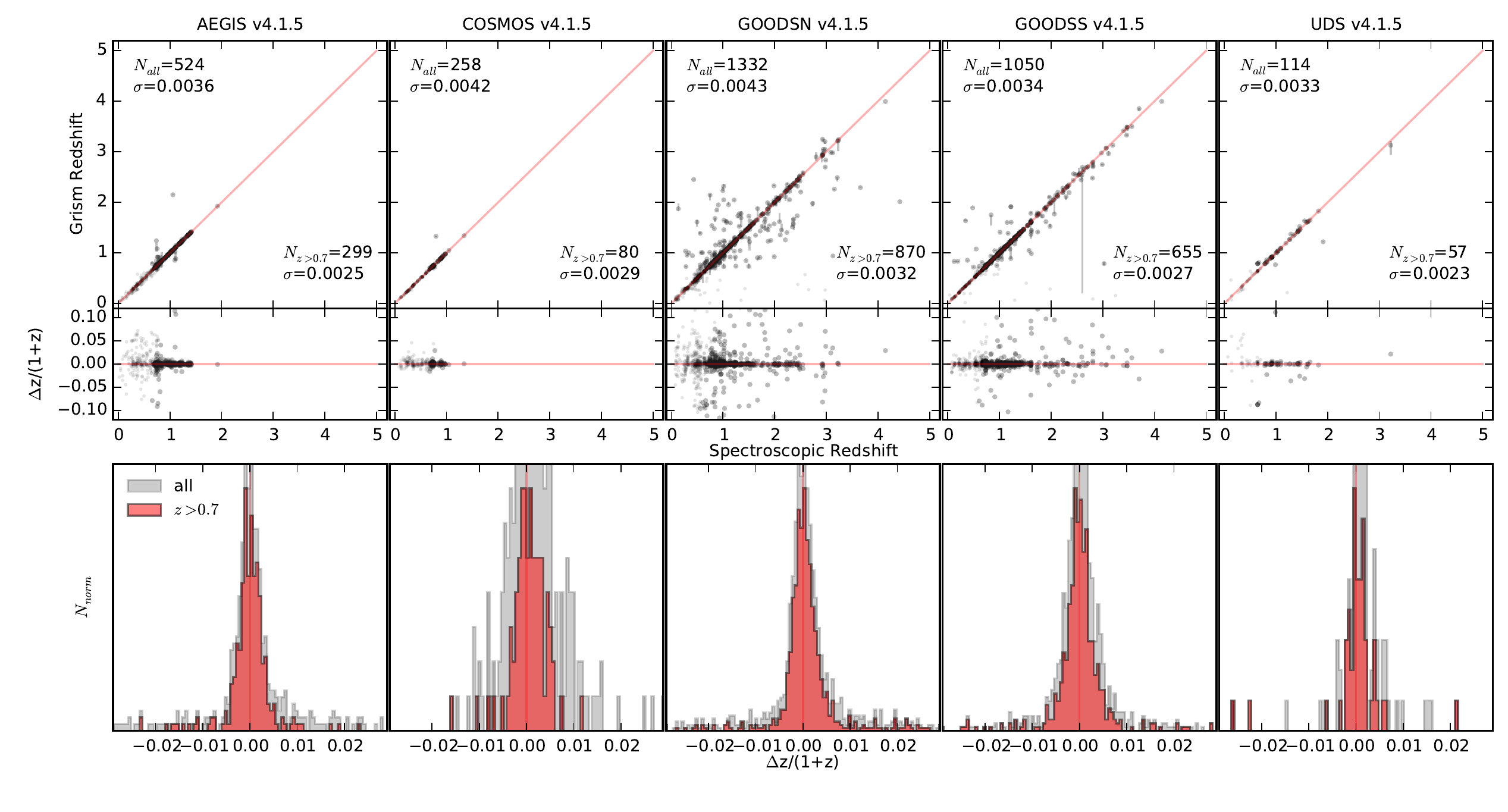}
\caption{\footnotesize Comparison between grism and spectroscopic redshifts. {\it Top row: } spectroscopic vs.\ grism redshifts for 3,278 objects. Objects with $z_{spec}, z_{grism}<0.7$ are plotted with smaller symbols. In the top left corner the total number of objects in each field is listed, along with the NMAD scatter. In the lower right corned of each panel the same statistics are listed for the $z>0.7$ sample. 
{\it Middle row: } $\Delta z/(z+1)$ as a function of redshift. {\it Bottom row: } Distributions of $\Delta z/(z+1)$ for the full sample (gray) and for the $z>0.7$ sample (red). The distributions are centered on zero and symmetric in all fields.}
\label{fig:redshift_comp}
\end{figure*}

In Figure~\ref{fig:redshift_comp} we compare the ground-based spectroscopic redshifts to the grism redshifts for the 3,278 objects where both exist. There is an excellent overall agreement between the two. We specifically focus our attention on the $z>0.7$ sample where H$\alpha$ enters the G141 coverage and the redshift accuracy may be expected to be the highest. We use the normalized median absolute deviation (NMAD), $\sigma_{NMAD}$, to quantitatively characterize the scatter in $\Delta z/(z+1)$ \citep[for a definition see][]{Brammer08}. For a Gaussian distribution, NMAD is equal to the standard deviation, but it is less sensitive to outliers. In all fields the $\sigma_{NMAD}$ for the $z>0.7$ sample is between 0.0023 and 0.0032. At $1.4\mu m$, this scatter corresponds to 42~\AA, i.e., approximately one original grism pixel (46~\AA). Uncertainties at this level are expected from morphological effects alone: if the luminosity-weighted center of the gas distribution in galaxies differs from that of the stars by $\sim 1$\,kpc it will introduce an error in the emission line redshift of $\sim 1000$ km/s \citep[see, e.g.,][]{Nelson15}.  The scatter in $\Delta z/(z+1)$ is centered at zero and symmetric about zero, indicating that there are no systematic offsets between our redshift measurements and the ground-based samples.

In each field there are clear outliers that are many $\sigma$ removed from the one-to-one line. This appears particularly pronounced in GOODS-N, although that is largely simply due to the large number of ground-based spectroscopic measurements in that field. These outliers are a mix of objects with large formal uncertainties in their grism redshifts, possible object confusion, errors in the ground-based spectroscopic redshifts, AGN, and genuine outliers with no straightforward explanation. Object confusion may occur due to the matching of our HST catalogs to ground-based spectroscopic measurements. Errors in spectroscopic redshifts may come from mis-identified lines or from low S/N spectra. It is difficult to disentangle these effects, particularly since the original ground-based catalogs and spectra are usually not available.

\begin{figure*}
\centering
\includegraphics[width = \textwidth]{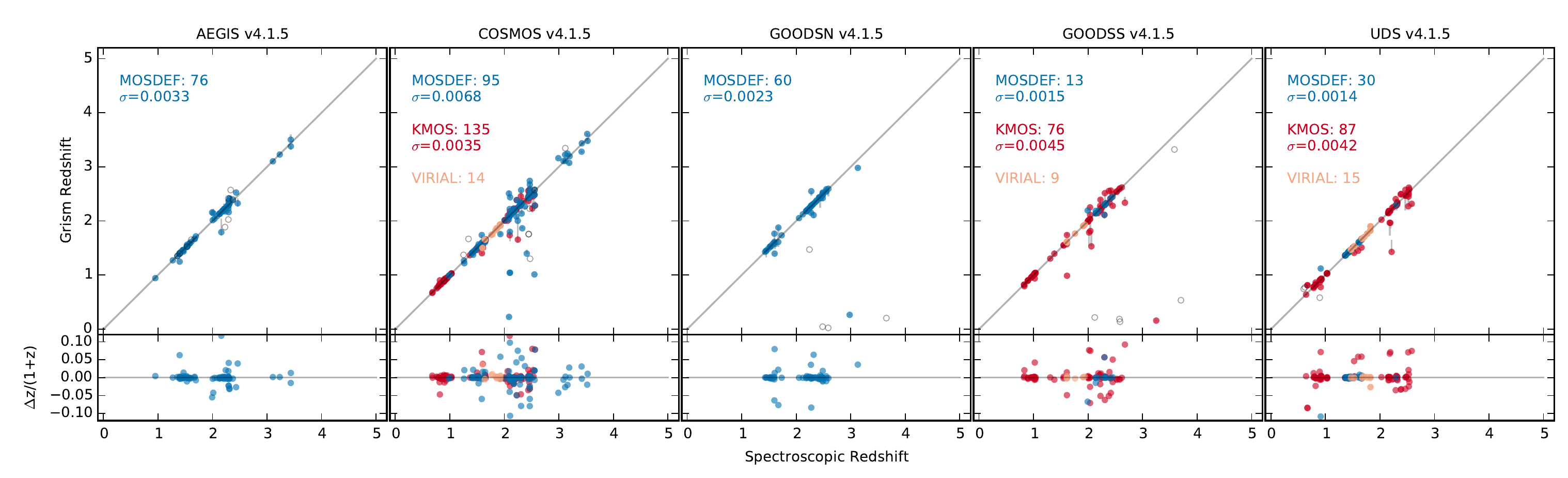}
\caption{\footnotesize Comparison between grism redshifts and spectroscopic redshifts from surveys that use the 3D-HST catalogs for target selection. Shown are 274 measurements from the MOSDEF survey \citep{Kriek15}, 298 from KMOS$^{3D}$ \citep{Wisnioski15} and 38 from VIRIAL \citep{Mendel15}. Only objects with $\JH<24$ are shown. Measurements are color-coded to indicate the survey. Open points have {\tt use\_grism} = 0. {\it Top row:} Spectroscopic vs.\ grism redshift. In each panel, we show the number of objects from each survey and the NMAD scatter in $\Delta z/(1+z)$. The outlier clipped NMAD scatter is in brackets. {\it Bottom row:} $\Delta z/(z+1)$ as a function of redshift. 
}
\label{fig:redshift_ground}
\end{figure*}

Three surveys have used preliminary catalogs from 3D-HST to select galaxies for ground-based near-infrared spectroscopic  follow-up: MOSDEF \citep{Kriek15}, KMOS$^{3D}$ \citep{Wisnioski15}, and VIRIAL \citep{Mendel15}. MOSDEF uses the multi-object slit spectrograph MOSFIRE on the Keck telescope; KMOS$^{3D}$ and VIRIAL use the multi-object integral field spectrometer KMOS on the Very Large Telescope. Since the selection of these objects is not independent of the grism redshifts, they were not included in Figure \ref{fig:redshift_comp}. Nevertheless, a comparison between their redshifts and those from the grism is informative as they rely strictly on rest-optical features observed in the near-infrared, and probe a regime of galaxy parameter space that is underrepresented in optically-selected ground-based spectroscopic surveys. In Figure \ref{fig:redshift_ground} we show a comparison between the grism redshifts and those from the first year campaigns of these ground-based collaborating surveys. The sample comprises 581 objects with $\JH\leq24$. The scatter, $\sigma_{NMAD}$, is $\sim0.0015$ to 0.0045, comparable to the the scatter from the larger ground-based sample. All three surveys only use the 3D-HST preliminary grism redshift and do not specifically target objects with detected emission lines in the grism, hence it is expected that the scatter is slightly larger relative to the ground-based sample. The scatter in the COSMOS MOSDEF sample is particularly large, likely because this sample has a larger fraction of faint $\JH>23.5$ targets (40\% vs. 15--30\% for the other fields).  If we limit the COSMOS sample to $\JH<23.5$, the scatter is consistent with the other fields. While both KMOS$^{3D}$ and MOSDEF target primarily star-forming galaxies, VIRIAL is specifically focused on quiescent objects without emission lines. Strikingly, even in that comparison our redshift accuracy is excellent, $\sigma_{NMAD} = 0.0034$ for the full sample.

\subsubsection{Pair Analysis}

While comparison to spectroscopic redshifts is the most straightforward method of determining redshift accuracy, such a comparison can be limited because the spectroscopic sample is not fully representative of the parent grism sample. The sample used in \S~\ref{sec:ground_z} is dominated by low redshift bright objects: 70\% of objects are at $z<1$ and 56\% are brighter than $\JH=22$. Furthermore, they are also likely dominated by bluer galaxies (emission lines make for easier redshift identification) and biased towards specific sub-populations (Lyman break galaxies, for example). Therefore, the redshift accuracy we quote in the previous section may not apply to the full grism sample. In this section we use an empirical method to determine the redshift accuracy, first proposed by \citet{pairs}. The method uses pairwise redshift differences to estimate the width and shape of the error distribution for the full sample. In brief, the method takes advantage of the fact that galaxies are strongly clustered in real space and projected pairs have a high likelihood of being at the same redshift. The errors in grism redshifts, limited by the WFC3/IR grism resolution, are larger than the true virial motions of galaxies within structures; this aspect makes this method \citep[originally used by ][on photometric catalogs]{pairs} applicable to our sample. 

Following the methodology described in \citet{pairs}, we choose primary (tracer) galaxies with given characteristics (magnitude range, redshift range, color) as well as a secondary (companion) population. For each tracer galaxy, we identify all galaxies within the companion population that are within $r=25\arcsec$ and determine $(z_t-z_s)/(1+z_{mean})$, where $z_t$ and $z_s$ are the corresponding redshifts of the tracer and companion. In the analysis presented here, both $z_t$ and $z_s$ are grism redshifts. To subtract the background contribution of uncorrelated pairs, we assign the companions random positions from our master catalog. We then identify anew all pairs for each tracer galaxy. The distribution of $(z_t-z_s)/(1+z_{mean})$ for the randomized companions gives us the background level of projected pairs which are not physically associated. This background is then subtracted from the observed distribution. The physically bound pairs comprise the residual above the background. The characteristic error of the grism redshifts, $\sigma_{girsm}$, is approximated by $\sigma_{Gauss}/\sqrt{2}$, where $\sigma_{Gauss}$ is the width of the Gaussian that is the best fit to the residual distribution. 

\begin{figure*}
\centering
\includegraphics[width = \textwidth]{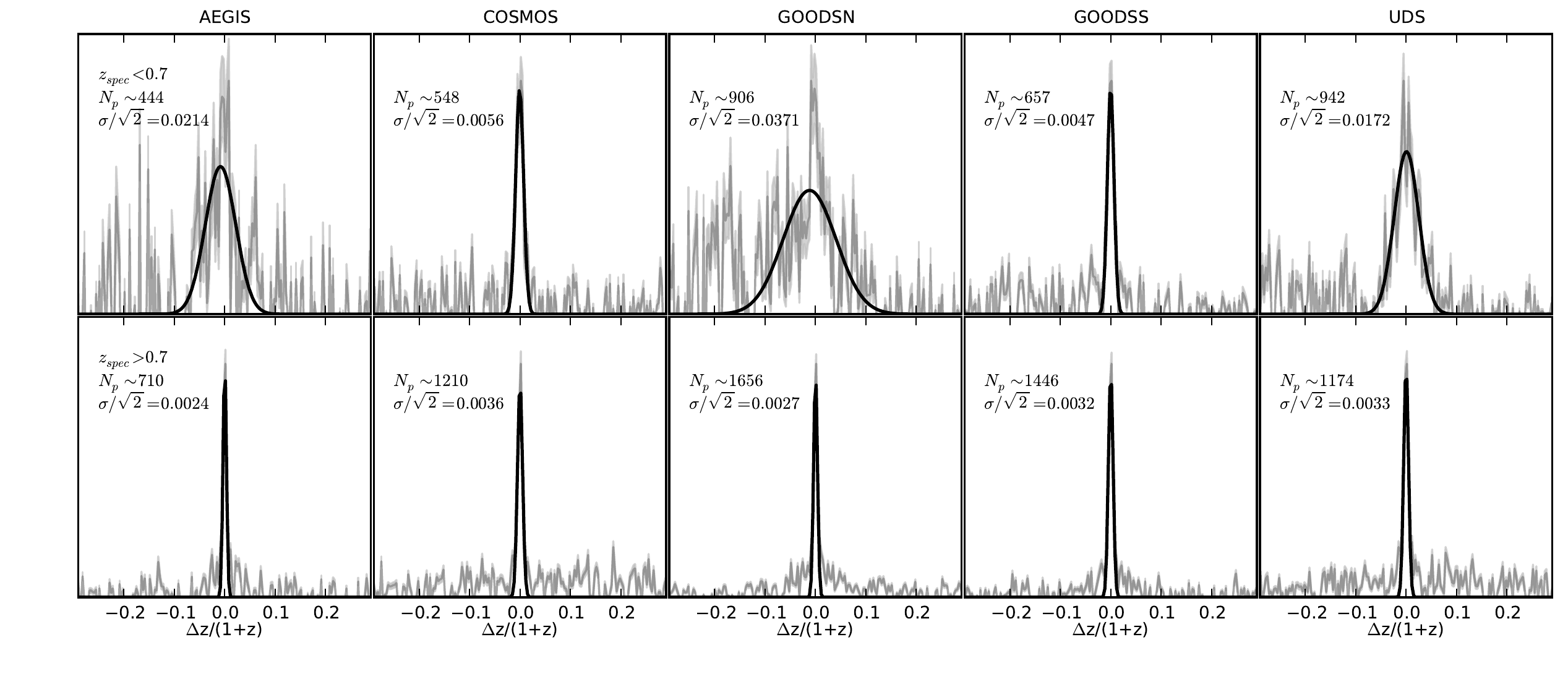}
\caption{\footnotesize Distribution of $\Delta z_{grism}/(1+z_{mean})$ after correcting for projected pairs for each of the five fields in our sample, divided in two bins of redshift: $z<0.7$ (top) and $z>0.7$ (bottom). Tracer galaxies are limited to those within the redshift bin, with magnitudes $\JH<24$, and with grism use flags set to one. The companion galaxies have the same selection, except that there is no constraint on their redshifts. The number of pairs which contribute to each measurement is calculated by integrating under the best fit Gaussian (solid black line). As a result of the different geometries of the fields, they produce different numbers of pairs. We remove the y-axis labels and scale the plots by the maximum histogram value for clarity. The high redshift objects have redshift accuracy $\sigma\sim0.0025$ to 0.0035, consistent with the comparison to spectroscopic redshifts. The low redshift objects have significantly larger scatter.}
\label{fig:pairs_all}
\end{figure*}

In Figure \ref{fig:pairs_all} we show the distribution of $\Delta z_{grism}/(1+z_{mean})$ after correcting for projected pairs for each of the five fields in our sample, divided in two bins of redshift: $z<0.7$ (top) and $z>0.7$ (bottom). At $z>0.7$, we find that the grism redshifts produce narrow peaks with $\sigma_{grism}\sim0.0025$ to 0.0035, consistent with the results from \S~\ref{sec:ground_z}. This is a factor of three (COSMOS) to 12 (AEGIS) improvement in redshift accuracy over the photometric redshifts in these fields. 

At $z<0.7$, before H$\alpha$ enters the wavelength coverage of the G141 grism, there are no strong features in the NIR spectra of galaxies.\footnote{There are relatively strong absorption features (the TiO bands), but the weak S{\sc iii} [9068,9530] doublet lines are typically the only emission features that can be detected.} As a result, we would expect that the grism spectra will add very little to the overall fits in constraining the redshift. In fact, as seen in Figure \ref{fig:pairs_all}, in AEGIS, GOODS-N and UDS, the pairwise grism accuracy $\sigma_{z<0.7}$ is the same as for the photometric redshifts in those fields. In COSMOS and GOODS-S, the fields with the best photometric redshifts, the addition of the grism improves the redshifts by $\sim30\%$.  In the following, we limit our analysis of the grism accuracy to objects with $z>0.7$.

\begin{figure*}
\centering
\includegraphics[width = \textwidth]{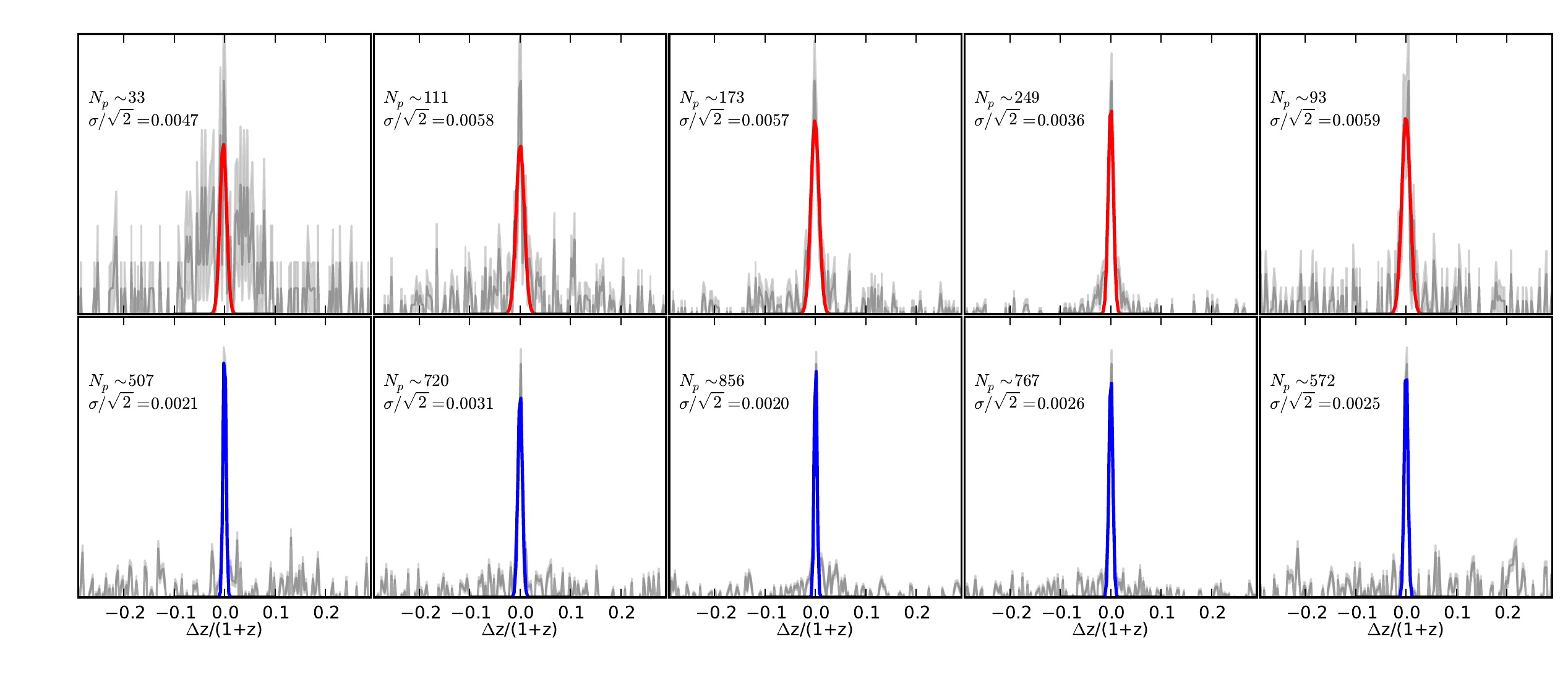}
\caption{\footnotesize Distribution of $\Delta z_{grism}/(1+z_{mean})$ after correcting for projected pairs for each of the five fields in our sample separated into quiescent (top row) and star-forming (bottom row) by $UVJ$ rest-frame color \citep{Whitaker12}. The tracer and companion galaxies are selected in the same way: we require $z_{grism}>0.7$, $\JH<24$, $use\_grism = 1$ and rest-frame color such that it falls within appropriate color selection. The number of pairs which contribute to each measurement is calculated by integrating under the best fit Gaussian (solid blue/red line). We remove the y-axis labels and scale the plots by the maximum histogram value for clarity.}
\label{fig:pairs_color}
\end{figure*}

The accuracy of photometric redshifts depends on the color of the galaxy. As shown in \citet{pairs} and \citet{Bezanson15}, red galaxies have more accurate photometric redshifts as a result of the strong breaks present in their SEDs, while blue SEDs are relatively featureless. However, the addition of low resolution spectroscopic data can counteract this trend with the addition of emission lines for blue galaxies. In Figure \ref{fig:pairs_color} we show the distributions of $\Delta z_{grism}/(1+z_{mean})$ for galaxies at $z>0.7$ in the five fields, divided into star
forming and quiescent samples using the \citet{Whitaker12} UVJ selection. The accuracy of the redshift measurements of quiescent galaxies is $\sigma_{grism}\sim0.006$, which is a factor of several better than the photometric redshift accuracy in most fields.
For star forming galaxies, the characteristic $1\sigma$ errors are $\sim0.002$ to 0.0035, a factor of two lower than those for quiescent galaxies across all fields, independent of the accuracy of the photometric redshifts. As discussed above, based on morphological considerations alone, this is the likely limit of the accuracy of grism redshifts for extended sources.

\subsubsection{Duplicate Grism Redshifts}
\label{sec:z_duplicates}

The third and final method we use to assess the accuracy of the redshifts is repeat measurements. As discussed above, duplicate spectra are not co-added and each 2D spectrum is fit separately in combination with the corresponding photometry. 
There are a total of 4520 objects which have at least one repeat observation and 476 objects with more than two observations down to $\JH=24$. The maximum number of repeat observations is four for most fields and as many as six in GOODS-S. We limit this analysis to $\JH<24$ and we use the visual inspections to remove objects with {\tt use\_grism=0}. These selections leave 4185 redshift measurement pairs. A full list of repeat observations is provided as part of the data release (see below). 

The top panels of Fig.\ \ref{fig:duplicates_z_err} compare the two duplicate redshift measurements. The agreement is excellent, with very few outliers. Next, we use the duplicate measurements to assess the quality of the redshift uncertainties, that is, whether the differences between two repeat observations can be explained by the formal uncertainties in these measurements. This can be quantified by the ratio between the redshift difference of repeat observations $\Delta z$ and the total error in $\Delta z$, $\sigma_{\Delta z} = \sqrt{\sigma_{z1}^2 + \sigma_{z2}^2}$. The individual redshift errors are asymmetric because the $p(z)$ distributions are typically asymmetric. We define a symmetric error by taking the mean of the lower and upper $1\sigma$ errors. We show the distribution of $\Delta z/\sigma_{\Delta z}$ for each of the five fields in the bottom row of Figure \ref{fig:duplicates_z_err}. If the redshift errors are correct, we would expect that these distributions would have a width of unity. We specifically focus on the $z>0.7$ regime where the grism spectra likely dominate the redshift estimates. In all fields we find that the width of the distribution is close to unity. In AEGIS and UDS, the widths are slightly larger, suggesting that errors are underestimated by 5 - 20\%.

We caution that the two redshift measurements are not entirely independent because the underlying photometric catalogs are the same; it is  therefore unsurprising that we see very few ``catastrophic'' outliers in the top panels of Fig.\ \ref{fig:duplicates_z_err}. However, the grism spectra themselves are fully independent, and generally come from opposite sides of the detector; they will therefore have a different effect on the fit depending on factors such as the background level, contamination, coverage fraction, etc.

In light of these considerations we further explore the redshift errors in Fig.\ \ref{fig:duplicates_summary}. Here we consider how the accuracy of the redshift errors depends on the properties of the galaxy, in particular its redshift, magnitude, the error in the redshift ($\sigma/(1+z)$), and the number of pixels that the spectrum covers. In each panel we see that in the parameter space where we expect that the grism spectra would contribute most information to the redshift fits (bright magnitude, $z>0.7$, small error), the scatter in $\Delta z/\sigma$ is $\sim1$ to 1.2, indicating again that the grism errors underestimate the true errors by at most $\sim 20$\%. Where the redshifts are dominated by the photometry -- faint magnitude, low redshift, large error -- $\Delta z/\sigma<1.0$, indicating that the photometric data begin to dominate the redshift measurement.

\subsubsection{Summary}

We find that all methods to determine the redshift accuracy are in good agreement: the comparison to spectroscopic redshifts, the pair analysis, and the analysis of duplicates all imply typical redshift uncertainties of $\approx 0.003\times (1+z)$. Quiescent galaxies have larger redshift uncertainties than star forming galaxies, but even the errors for quiescent galaxies (typically $0.006\times (1+z)$) are well below those from the photometry alone. Most importantly, the duplicate analysis shows that the formal redshift uncertainties are generally very good (within 10\%--20\% of the actual error), independent of magnitude.

\begin{figure*}
\centering
\includegraphics[width=\textwidth]{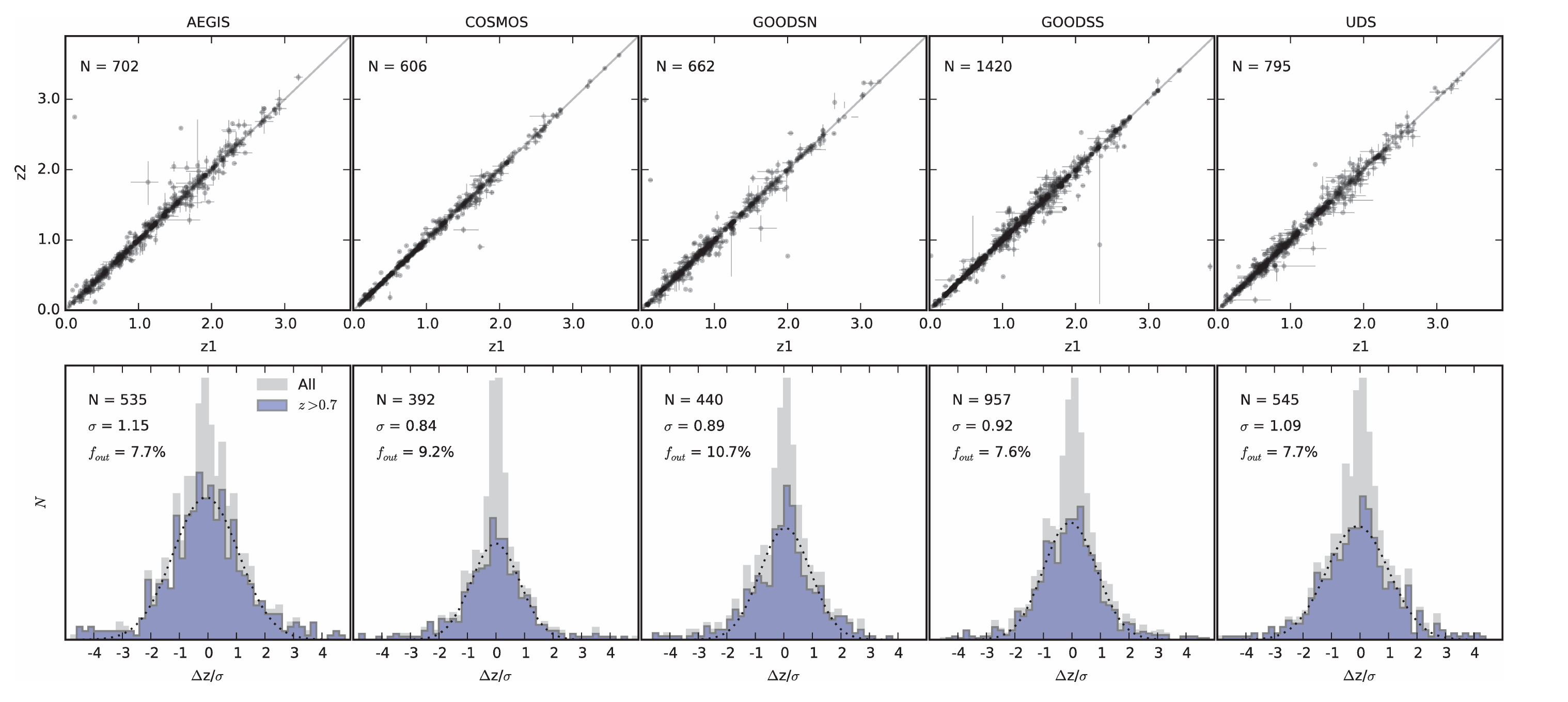}
\caption{\footnotesize {\it Top: } Comparison between duplicate redshift measurements. {\it Bottom: }Distributions of $\Delta z/\sigma$, where $\Delta z$ is the redshift difference between two duplicate measurements and $\sigma$ is the total error in $\Delta z$. We show both the distribution for all objects (gray histogram) and the distribution for $z>0.7$ objects (blue histogram). The number of $z>0.7$ objects for each field is listed in the upper left corner of the panel, along with the best-fit Gaussian $\sigma$ for the distribution. Overall, these distributions indicate that the redshift errors are accurate and possibly underestimated by 5 - 20\%.}
\label{fig:duplicates_z_err}
\end{figure*}

\begin{figure*}
\centering
\includegraphics[width = \textwidth]{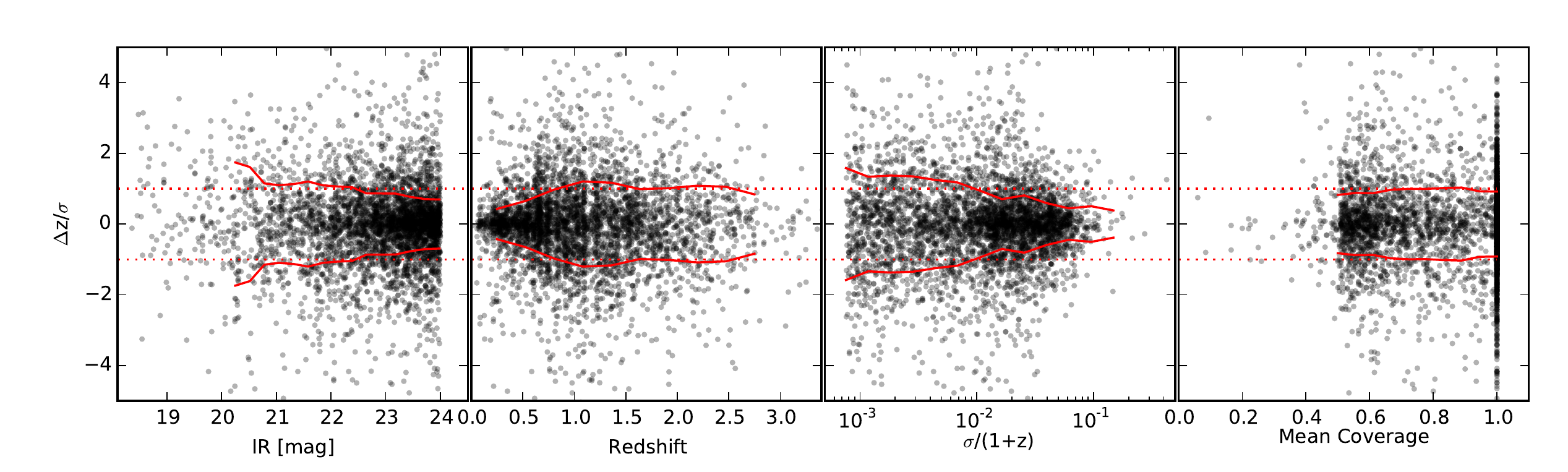}
\caption{\footnotesize $\Delta z/\sigma$ as a function of $\JH$ magnitude, redshift, redshift error $\sigma/(1+z)$ and mean spectral coverage for all duplicate pairs in 3D-HST. If the redshift errors are correct, we would expect that the widths of these distributions would be unity (dotted lines). The solid lines indicate the sliding box NMAD scatter is in each panel. }
\label{fig:duplicates_summary}
\end{figure*}

\section{Emission Line Fits}
\label{sec:line_fits}
\subsection{Methodology}

Following the approach of the redshift fits, we also fit the emission line fluxes directly in the 2D spectrum.  With the redshift fixed to $z_\mathrm{grism}$ we generate a 2D model spectrum for each of the model species listed in Table \ref{table:em_lines} that would fall within the grism passband (with unresolved line widths $\sigma=100$~\kms), and we adopt the 2D continuum template determined from the earlier redshift fit (\S\ref{sec:redshift_fits}).  With parameters for the individual template normalizations, we use \texttt{emcee} sampler \citep{ForemanMackey12} to determine the marginalized posterior distribution functions (PDFs) of parameters for the individual template normalizations, which can be converted directly to line fluxes and observed-frame equivalent widths in physical units (i.e., $\mathrm{erg}~\mathrm{s}^{-1}~\mathrm{cm}^{-2}$ and \AA, respectively).  As with the redshift fit in Eq.~\ref{eq:redshift_chisq}, this method provides the benefit of fitting in the natural units of instrumental count rates and preserved per-pixel uncertainties.  

\begin{figure*}
\centering
\includegraphics[width = 0.85\textwidth]{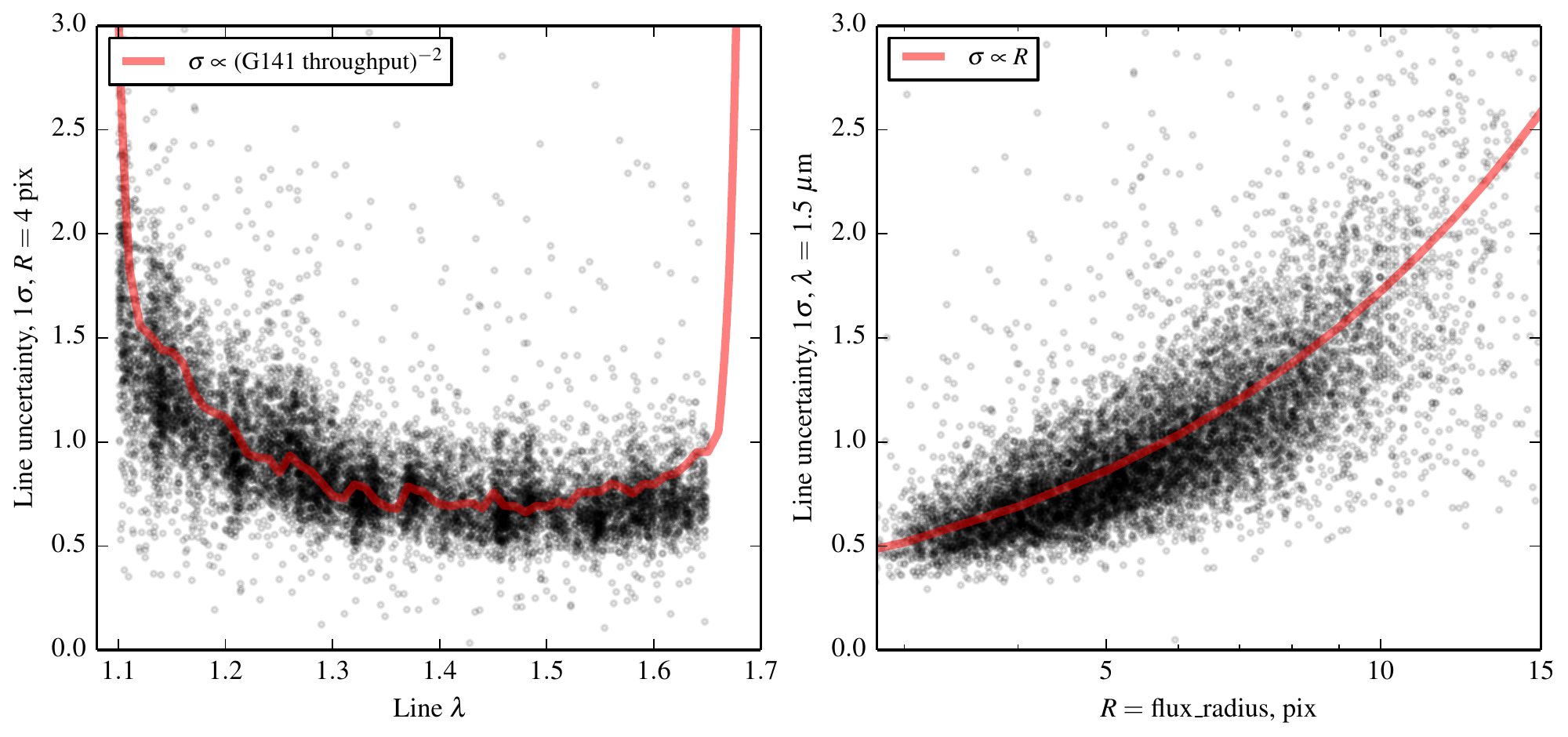}
\caption{\footnotesize Emission line sensitivity determined from the individual MCMC line fits as a function of wavelength (left) and object size $R$ parameterized by the SExtractor ``FLUX\_RADIUS'' value from the catalog (right).  The values on the vertical axes are given in units of $10^{-17}~\mathrm{erg}~\mathrm{s}^{-1}~\mathrm{cm}^{-2}$.  In both panels, the gross trends caused by the effect in the opposite panel have been divided out.  For a typical resolved galaxy ($R=5$ pix) a line at $1.5~\mu\mathrm{m}$ will have a 1-$\sigma$ line uncertainty of $8\times10^{-18}~\mathrm{erg}~\mathrm{s}^{-1}~\mathrm{cm}^{-2}$ in the 2-orbit 3D-HST G141 grism spectra.}

\label{fig:line_sensitivity}
\end{figure*}

The MCMC chains provide a robust estimate of the uncertainties in the fit, which are primarily determined by the wavelength dependence of the grism throughput and by the object size (i.e., the area of the effective aperture of the 2D spectrum fit).  The dependence of the derived uncertainties provided in the emission line catalog on these two characteristics is shown in Fig.\ \ref{fig:line_sensitivity}: the G141 grism is somewhat more sensitive at red wavelengths and line sensitivity rapidly decreases for large extended galaxies.  Overall the line sensitivity can be parameterized by 
\[
1\sigma = 8 \times10^{-18} \left(\frac{G(\lambda)}{G(1.5~\mu\mathrm{m})}\right)^{-2}\left(\frac{R}{5~\mathrm{pix}}\right)~\mathrm{erg}~\mathrm{s}^{-1}~\mathrm{cm}^{-2},
\]
where $G(\lambda)$ is the wavelength dependent throughput of the G141 grism\footnote{The G141 throughput curve can be obtained with PySynphot:
\url{http://ssb.stsci.edu/pysynphot/docs/}} and $R$ is the {\tt SExtractor} FLUX\_RADIUS in pixels.

The line fluxes are implicitly normalized to the broad band photometry of \citet{Skelton14}, as the spectra are scaled to match the photometric data. The fluxes are therefore ``total", and do not refer to a particular aperture, although an implicit assumption is that the equivalent widths of the lines do not increase or decrease strongly outside of the segmentation map.
No absorption corrections are necessary, and in that sense our methodology is different from most measurements in the literature \citep[e.g.,][]{Kriek15,SteidelKBSS}. The standard method is to measure the flux of a bright line with respect to an idealized continuum, parameterized by a constant or a linear function defined in a narrow wavelength region to the blue and red of the line. For H$\alpha$, H$\beta$, and other Balmer lines a correction then needs to be made after the measurement, to account for absorption in the stellar continuum. In our methodology, the stellar continuum model is not a low order polynomial but the best-fitting stellar population synthesis model that came out of the redshift fit. It therefore uses all the information in the broad band photometry and the grism spectrum. No post-measurement absorption corrections are necessary, as the Balmer absorption lines are present in the model, at the appropriate resolution.

\begin{table}[ht]
\centering
\caption{Emission Lines}
\label{table:em_lines}
\begin{tabular}{llcc}
\hline \hline
\noalign{\smallskip}
~Line & Catalog ID & Rest wavelength [\AA] & Ratio    \\
\noalign{\smallskip}
\hline
\noalign{\smallskip}
~Ly$\alpha$ & Lya & 1215.400 &\nodata \\
~\ion{C}{4} & CIV & 1549.480 & \nodata \\
~\ion{Mg}{2} & MgII & 2799.117 & \nodata \\
~\ion{Ne}{5}  & NeV & 3346.800 & \nodata \\
~\ion{Ne}{6}  & NeVI & 3426.850 & \nodata \\
~[\ion{O}{2}]  & OII & 3729.875 & \nodata\\
~[\ion{Ne}{3}]  & NeIII & 3869.000 & \nodata \\
~\ion{He}{1} & HeIb & 3889.500 & \nodata \\
~H$\delta$ & Hd & 4102.892 & \nodata\\
~H$\gamma$ & Hg & 4341.680 & \nodata\\
~[\ion{O}{3}] & OIIIx & 4364.436 & \nodata\\
~\ion{He}{2} & HeII & 4687.500 & \nodata\\
~H$\beta$ & Hb & 4862.680 & \nodata\\
~[\ion{O}{3}] & OIII & 5008.240, 4960.295 & 2.98:1 \\
~\ion{He}{1} & HeI & 5877.200 & \nodata\\
~[\ion{O}{1}] & OI & 6302.046 & \nodata \\
~H$\alpha$ & Ha & 6564.610 & \nodata \\
~[\ion{S}{2}] & SII & 6718.290, 6732.670 & 1:1\\
~\ion{S}{3} & SIII & 9068.600, 9530.600 & 1:2.44\\
\noalign{\smallskip}
\hline
\noalign{\smallskip}
\end{tabular}
\end{table}

\subsection{Results}

\begin{figure*}
\centering
\includegraphics[width = 0.9\textwidth]{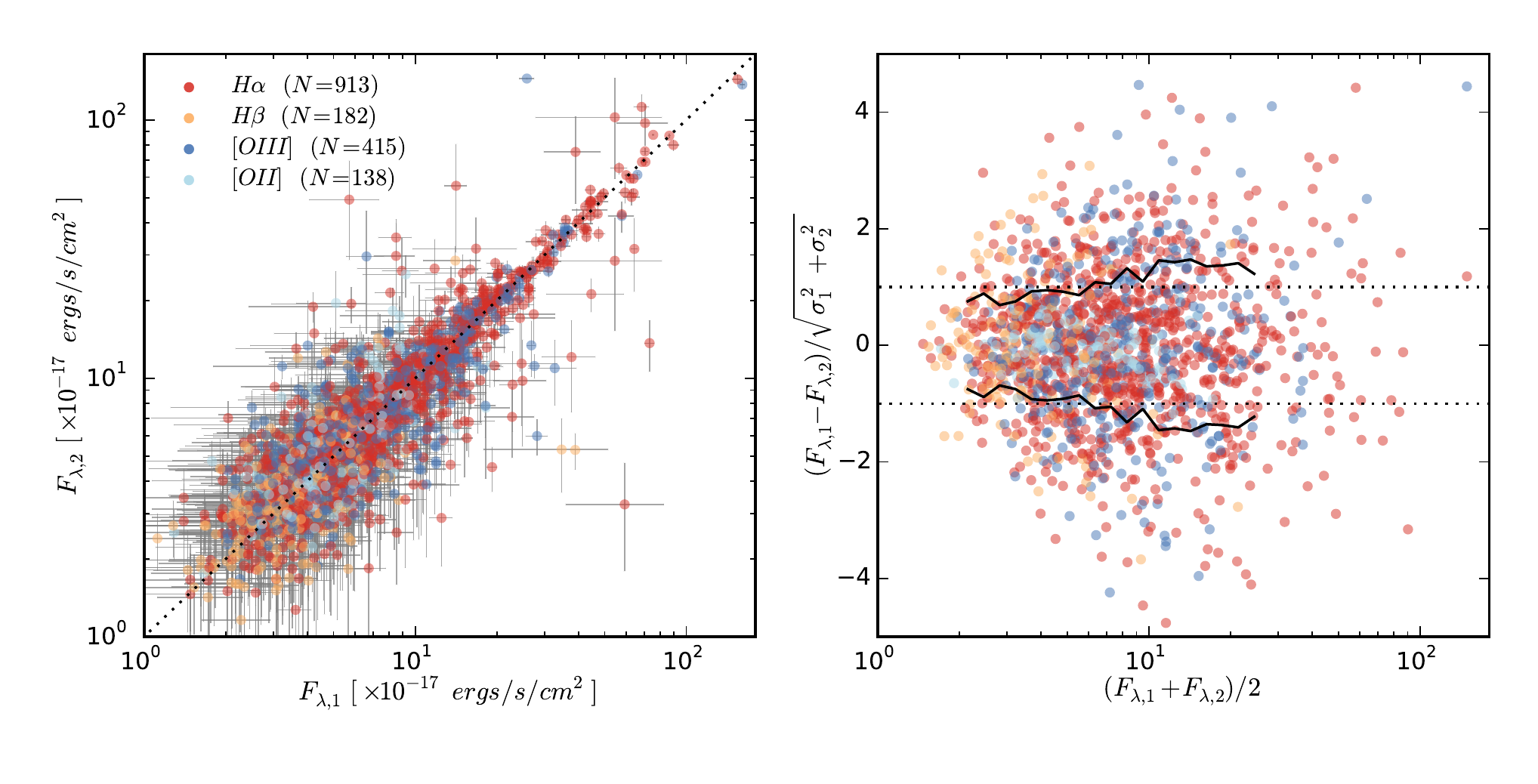}
\caption{\footnotesize {\it Left:} Comparison of line flux measurements for objects with multiple grism spectra. Points are color-coded by the emission line with the number of measurements for each species listed in the legend. We require $\JH<24$, $S/N>2$ in the line in both spectra and apply the redshift use flags. Error bars are $1\sigma$. {\it Right: }$\Delta F_{\lambda}/\sigma$ for the emission line flux measurements as a function of mean line flux. The sliding box NMAD scatter as a function of flux (solid black line) is $\sim1$ at all fluxes, indicating that the line flux errors are correct.}
\label{fig:lines_dup}
\end{figure*}

\begin{figure*}
\centering
\includegraphics[width = 0.9\textwidth]{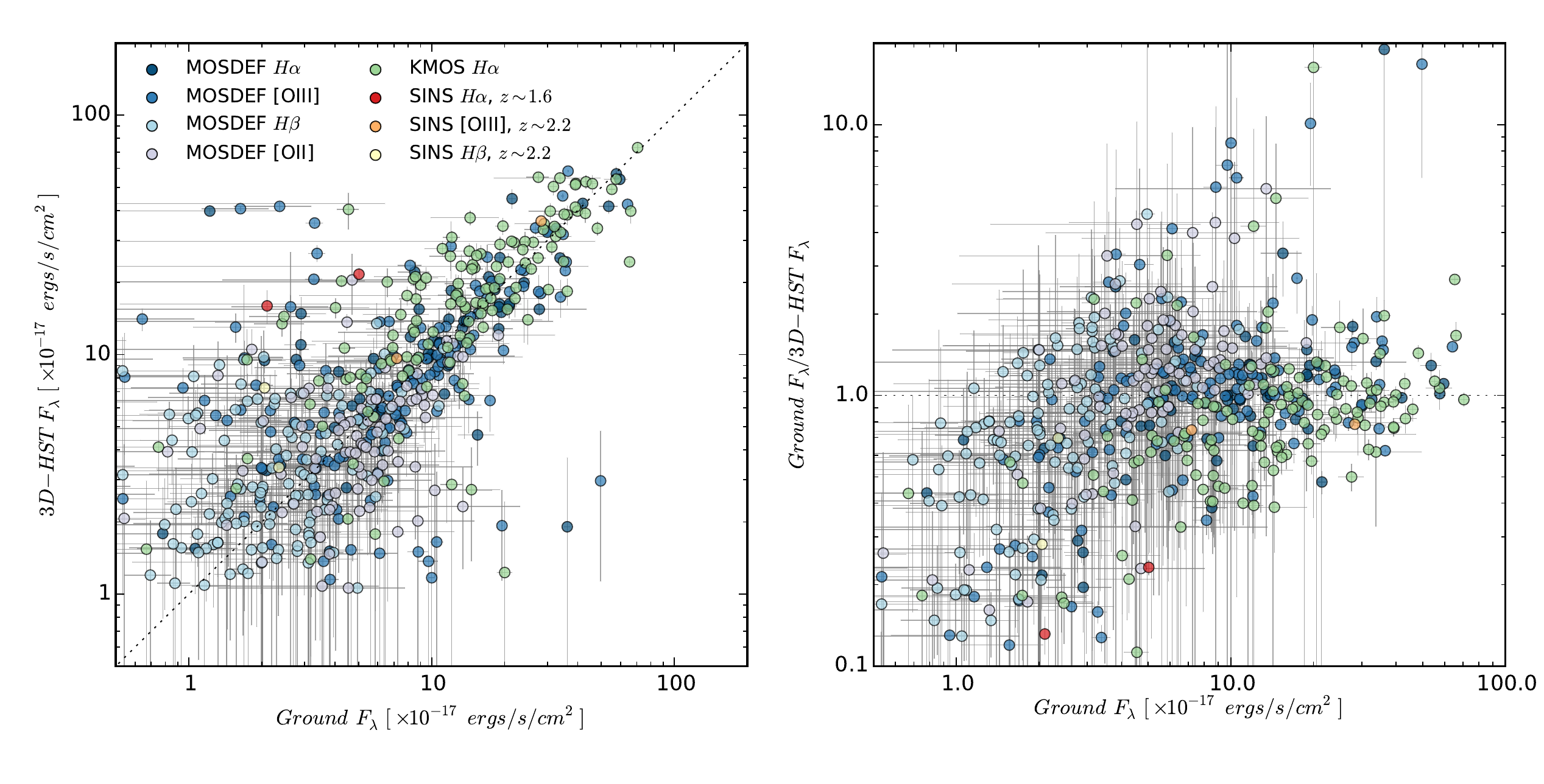}
\caption{\footnotesize Emission line flux measurements from 3D-HST compared to those from KMOS$^{3D}$, MOSDEF and SINS/zC-SINF. For the comparison, $H\alpha$ and NII from the ground-based observations are coadded as are the [O{\sc ii}] and [O{\sc iii}] doublets. {\it Left:} Ground-based flux vs. 3D-HST flux for emission lines in common between the surveys. {\it Right:} Ratio between the ground-based and 3D-HST grism fluxes. The scatter for bright lines ($>10^{-16}$) is a factor of 1.8, and the mean ratio is 0.9.}
\label{fig:lines_ground}
\end{figure*}

Emission lines are fit for all objects down to $\JH=26$, however the analysis in this section is limited to $\JH=24$ where we take advantage of the grism redshift use flags. If none of the emission lines in Table \ref{table:em_lines} fall within the grism wavelength range (for the best-fit redshift), an emission line fit is not produced. We test the accuracy and precision of the emission line flux measurements by comparing the fits for duplicate objects within the survey and by comparing our measurements to those from ground-based surveys.

We begin by comparing the emission line fluxes measured from repeated observations of the same object. Unlike the redshift fits, the emission line measurements are based on the 2D grism spectra alone and they are therefore truly independent measurements. Repeat spectra are typically taken at different angles and the spectra fall on different parts of the detector. Mismatch of repeat line flux measurements can indicate problems with the background subtraction, the flat-fielding, and a myriad of other effects. In Figure \ref{fig:lines_dup} we show the flux measurements from objects with multiple grism spectra.  The measurement follow the $1:1$ line tightly, with the scatter and errors increasing with decreasing flux (left panel). In the right panel of Figure \ref{fig:lines_dup} we analyze the errors of the line fluxes in a manner similar to the analysis of the redshift errors in \S\ref{sec:redshift_fits} and Fig.\ \ref{fig:duplicates_summary}. The sliding box NMAD scatter, $\sigma_{NMAD}$ (solid black line), is approximately unity across all fluxes, demonstrating that the formal errors are an excellent approximation of the actual uncertainties. The emission line flux errors are calculated on the basis of the interlaced G141 image background errors, which include terms for the contamination subtraction. The fact that $\sigma_{NMAD}\sim1$ shows that these errors properly account for the line flux uncertainties and that there are no systematic errors introduced in our data reduction. The outlier fraction in Figure \ref{fig:lines_dup} is strikingly small. Even though we apply no quality flags to select spectra with clean emission lines other than the line $S/N>2$ cut and the redshift use flag, the fraction of objects with $|\Delta F_{\lambda}/\sigma|>3$ is only 3.9\%. 

An external check on the emission line fluxes, shown in Figure \ref{fig:lines_ground}, is provided by a comparison between our measurements and those from MOSDEF and KMOS$^{3D}$, as well as from the SINS/zC-SINF survey with VLT/SINFONI \citep{Schreiber09,Mancini11,Newman14}. Compared to the 3D-HST spectra, ground-based spectra are affected by (rapidly varying) atmospheric emission and absorption.  While the IFU data from KMOS$^{3D}$ and SINS/zC-SINF allows one to recover well the full 2D spatial emission, slit losses may affect the multi-slit spectra from MOSDEF. Significant effort has been made by the survey teams to correct the ground-based line fluxes used here for such losses. Close pairs of lines in the ground based spectra are coadded to compare to the lower-resolution 3D-HST measurements. The grism and ground-based line fluxes match well and follow the 1:1 line overall (left panel) with a scatter of a factor of a factor of $1.8$ for lines brighter than $10^{-16}$\,ergs\,s$^{-1}$\,cm$^{-2}$ in both datasets (right panel). This agreement can be considered good when considering the large differences in observing and analysis methods; attributing an equal uncertainty to both datasets, the per-measurement error is a factor of 1.5.

At low line fluxes there may be a systematic effect, such that the 3D-HST line measurements are slightly higher than the ground-based ones. This could be due to the effects of the large errors in this regime: the fact that the objects were selected in 3D-HST and subsequently observed from the ground may introduce an asymmetry. It could also be due to  uncertainties in aperture corrections, the fact that our line measurements are corrected for the underlying stellar absorption, or other effects.

\section{Catalogs}
\label{sec:catalogs}

The results from the redshifts and emission line fits are assembled into several different catalogs. For the majority of users these catalogs will probably be the main, or only, gateway to the 3D-HST dataset. In this section we describe the catalogs produced from the survey and the applications for which each of them may be appropriate. 

\subsection{Redshift and Emission Line Catalog}

The first type of catalogs we produce are simply concatenations of the outputs of all redshift and emission line fits. These catalogs contain repeat fits for the same photometric object. The fits are done for each extracted 2D spectrum of each object separately (in conjunction with the photometric information) for a total of 98,668 individual spectra down to $\JH = 26$ (except for the UDS field where the fits reach approximately 0.5\,mag fainter). In these catalogs, each row corresponds to the outputs from a single spectrum. Each spectrum has a unique identifier of the format {\tt aegis-01-G141\_00001}, listing the field, the pointing number (zero-padded two-digit integer), the grism name (G141 for 3D-HST) and the photometric identification number of the object (padded five-digit integer).The objects are ordered by pointing number and, within that pointing, by photometric identifier. A list of all duplicate spectra is also provided (see \S~\ref{sec:linematched}).

Two concatenated catalogs are produced, one containing the redshift fits and one containing the emission line fits. Both catalogs have the same length. The column names and the corresponding descriptions are listed in Tables~\ref{table:z_cat_cols} and \ref{table:line_cat_cols}. We produce catalogs for each field separately as well as a master catalog which contains all objects in the survey.

The concatenated catalogs provide information for {\it all} objects fitted as part of the current release. The $\JH$ magnitude is included as a column in the catalog, however we do not preselect objects in any way for this catalog and we specifically do not exclude duplicate observations. These catalogs can be used to identify all the information available for a given object in the photometric catalogs.

\begin{table*}[!th]
\centering
\caption{Redshift Catalog Columns}\label{table:z_cat_cols}
\begin{tabular}{p{2cm}p{2cm}l}
\hline \hline
\noalign{\smallskip}
Column name & Default & Description \\
\hline
\noalign{\smallskip}
\tt{phot\_id} & \nodata & Unique identifier from \citet{Skelton14}\\
\tt{spec\_id} & 00000 & Unique identifier for the spectrum which was used in this measurement\\
\tt{jh\_mag} & \nodata & SExtractor MAG\_AUTO $\JH$ magnitude of the objects, described in \S~\ref{maskred.sec} \\
\tt{z\_spec} & -1 & Spectroscopic redshift, when available, see \citet{Skelton14} for sources and quality\\
\tt{z\_peak\_phot} & -1 & Photometric redshift. Same as {\tt z\_peak} from the EAZY catalogs of \citet{Skelton14}\\
\tt{z\_phot\_l95} & -1 & Photometric redshift at the lower 95\% confidence limit\\
\tt{z\_phot\_l68} & -1 & Photometric redshift at the lower 68\% confidence limit\\
\tt{z\_phot\_u68} & -1 & Photometric redshift at the upper 68\% confidence limit\\
\tt{z\_phot\_u95} & -1 & Photometric redshift at the upper 95\% confidence limit\\
\tt{z\_max\_grism} & -1 & The redshift where the p(z | grism, phot) is maximized, should be used as default grism redshift\\
\tt{z\_peak\_grism} & -1 & Integral of p(z | grism, phot)*z*dz, integrated over the whole redshift range \\
\tt{z\_grism\_l95} & -1 & Grism redshift at the lower 95\% confidence limit \\
\tt{z\_grism\_l68} & -1 & Grism redshift at the lower 68\% confidence limit \\
\tt{z\_grism\_u68} & -1 & Grism redshift at the upper 68\% confidence limit \\
\tt{z\_grism\_l95} & -1 & Grism redshift at the upper 95\% confidence limit \\
\tt{f\_cover} & -1 & Fraction of spectrum within the image (0=bad, 1=good)\\
\tt{f\_flagged} & -1 & Fraction of flagged pixels (1=bad, 0=good)\\
\tt{max\_contam} & -1 & Maximum contamination\\
\tt{int\_contam} & -1 & Contamination integrated over the spectrum (= flux\_contam/flux\_object)\\
\tt{f\_negative} & -1 & Fraction of pixels with negative flux after contamination correction \\
	&	&		(if big could indicate a problem with the contamination correction)\\
\tt{flag1} & -1 & User assigned flag for the redshift quality\\
\tt{flag2} & -1 &  User assigned flag for the redshift quality\\
\tt{use\_grism}$^{\dagger}$ & -1 & Flag defining objects with the most reliable grism-derived redshifts.  See \S~\ref{sec:inspections}\\
\tt{use\_phot}$^{\dagger}$  &  \nodata	 & Photometric use flag from \citet{Skelton14}: 1 = use; 0 = do not use\\ 
\tt{z\_best\_s}$^{\dagger}$    &   \nodata & Source of the best redshift: 1 = ground-based spectroscopy; 2 = grism; 3 = photometry; 0 = star \\
\tt{z\_best\_best}$^{\dagger}$  &  -1 & Best available redshift measurement (-1 for stars) \\
\tt{z\_best\_l95}$^{\dagger}$  &  -1 & Lower 95\% confidence limit derived form the z\_best $p(z)$\\
\tt{z\_best\_l68}$^{\dagger}$  &  -1 & Lower 68\% confidence limit derived form the z\_best $p(z)$\\
\tt{z\_best\_u68}$^{\dagger}$  & -1 & Upper 68\% confidence limit derived form the z\_best $p(z)$\\
\tt{z\_best\_u95}$^{\dagger}$  & -1  & Upper 95\% confidence limit derived form the z\_best $p(z)$\\
\noalign{\smallskip}
\hline
\noalign{\smallskip}
\multicolumn{3}{l}{$^{\dagger}$ This column is only present in the line-matched catalogs.}
\end{tabular}
\end{table*}

\begin{table*}[!th]
\centering
\caption{Emission Line Catalog Columns}\label{table:line_cat_cols}
\begin{tabular}{p{2cm}p{2cm}l}
\hline \hline
\noalign{\smallskip}
Column name & Default & Description \\
\hline
\noalign{\smallskip}
\tt{number} & \nodata & Unique identifier from \citet{Skelton14} \\
\tt{gris\_id} & 00000  & Unique identified for the spectrum which was used in this measurement\\
\tt{jh\_mag} & \nodata  & SExtractor MAG\_AUTO $\JH$ magnitude of the objects, described in the text\\
\tt{z} &  -1 & Grism redshift used in the emission line fit, identical to {\tt z\_max\_grism} in the redshift catalog \\
\tt{s0} &  -99 & Normalization coefficient {\tt s0}, see description in text\\
\tt{s0\_err} & -99 & Error for normalization coefficient {\tt s0}\\ 
\tt{s1} &  -99 & Normalization coefficient {\tt s1}, see description in text\\
\tt{s1\_err} & -99 & Error for normalization coefficient {\tt s1}\\ 
\tt{X\_FLUX} & -99 & Emission line flux in units of $10^{-17}$ $ergs~s^{-1}~cm^{-2}$\\
\tt{X\_ERR} & -99  & Error in the emission line flux  in units of $10^{-17}$ $ergs~s^{-1}~cm^{-2}$\\
\tt{X\_SCALE} & -99 & Multiplicative scaling factor to correct the flux of the emission line to the photometry\\ 
\tt{X\_EQW} &  -99 & Emission line equivalent width in \AA\\
\noalign{\smallskip}
\hline
\noalign{\smallskip}
\multicolumn{3}{l}{NOTE: X = emission line name, as given in Table~\ref{table:em_lines}.}
\end{tabular}
\vspace{+0.5cm}
\end{table*}

\subsection{Line-matched Catalogs}
\label{sec:linematched}

We also produce redshift and emission line catalogs that are matched to the photometric catalogs of \citet{Skelton14}. These catalogs, one for each field, as well as a master catalog containing all fields, have the same length as those in the v4.1 photometric release with one entry per object from \citet{Skelton14}. The column names and the corresponding descriptions are listed in Tables~\ref{table:z_cat_cols} and \ref{table:line_cat_cols}.  The rows corresponding to objects in the photometric catalog that do not have grism spectra are set to default values, also listed in the tables. Duplicate objects appear only once in these catalogs; the selection of the primary object among duplicates is described in \S~\ref{sec:inspections}.

The line-matched catalog contains a total of 79,609 unique objects with fits or $38.2\%$ of the photometric catalog. Of these, 22,548 objects have magnitudes brighter than $\JH=24$ and it is only these $\JH<24$ objects that have been visually inspected, and have a {\tt use\_grism} flag assigned as described in \S~\ref{sec:inspections}. The bright $\JH<24$ objects constitute  $10.8\%$ of the objects in the photometric catalog. We caution against blindly using our redshifts and emission line fits for faint objects with $24<\JH<26$. Even though checks of faint sub-samples have allowed us to verify that our methods do not break down in this parameter space, the vast majority of these spectra have not been inspected. 

In addition to the redshift and emission line catalogs, we create a row-matched listing of all duplicate spectra of a given object. We also make available the SEextractor catalog with $\JH$ fluxes measured from the $J_{125}$ + $JH_{140}$ + $H_{160}$ images.  

Using the grism redshift fits in the line-matched catalogs, we refit the stellar population parameters, rest-frame colors and star-formation rates as described in \citet{Skelton14} and \citet{Whitaker14}. The outputs from these fits are made available as part of the release.

\subsection{``Best'' Catalog}

Finally, we create a ``best'' redshift catalog, by merging the grism redshift fits with the photometric redshifts from \citet{Skelton14}. The best redshift is:
\begin{enumerate}
\item {\tt z\_spec} if it exists from the \citet{Skelton14} compilation of spectroscopic redshifts.
\item {\tt z\_max\_grism} if there is no spectroscopic redshift and {\tt use\_grism = 1}.
\item {\tt z\_phot} if  there is no spectroscopic redshift and {\tt use\_grism < 1}.
\end{enumerate}
We emphasize that we only use the photometric redshift if there is no grism spectrum that can be used (either because an object was not observed or because the spectrum has problems, as detailed above). Even if a grism spectrum appears to contain only noise we use it in the fit; as discussed earlier the error weighting in the fitting procedure ensures that the resulting redshift is nearly completely determined by the photometry in such cases. Using the best redshifts, we also create merged catalogs of the stellar population parameters, rest-frame colors and star formation rates.

\begin{table}[!th]
\centering
\caption{Best Redshift Catalog}\label{table:z_best}
\begin{tabular}{lll}
\hline \hline
\noalign{\smallskip}
Column name & Description \\
\hline
\noalign{\smallskip}
\tt{field}	& Field identifier (aegis/cosmos/goodsn/goodss/uds)\\
\tt{phot\_id}	& Unique identifier from \citet{Skelton14}\\
\tt{z\_best\_s}    & Source of the best redshift: \\
            & 1 = ground-based spectroscopy; \\
            & 2 = grism; \\
            & 3 = photometry; \\
            & 0 = star \\
\tt{use\_phot}	 & Photometric use flag from \citet{Skelton14}: \\
            & 1 = use; 0 = do not use\\ 
\tt{use\_grism} & Grism use flag as defined in \S~\ref{sec:inspections}\\
\tt{z\_best} & Best available redshift measurement (-1 for stars) \\
\tt{z\_l95} & Lower 95\% confidence limit derived form the z\_best $p(z)$\\
\tt{z\_l68} & Lower 68\% confidence limit derived form the z\_best $p(z)$\\
\tt{z\_u68} & Upper 68\% confidence limit derived form the z\_best $p(z)$\\
\tt{z\_u95} & Upper 95\% confidence limit derived form the z\_best $p(z)$\\
\noalign{\smallskip}
\hline
\noalign{\smallskip}
\end{tabular}
\vspace{+0.5cm}
\end{table}

\begin{figure*}
\centering
\includegraphics[width = 0.9\textwidth]{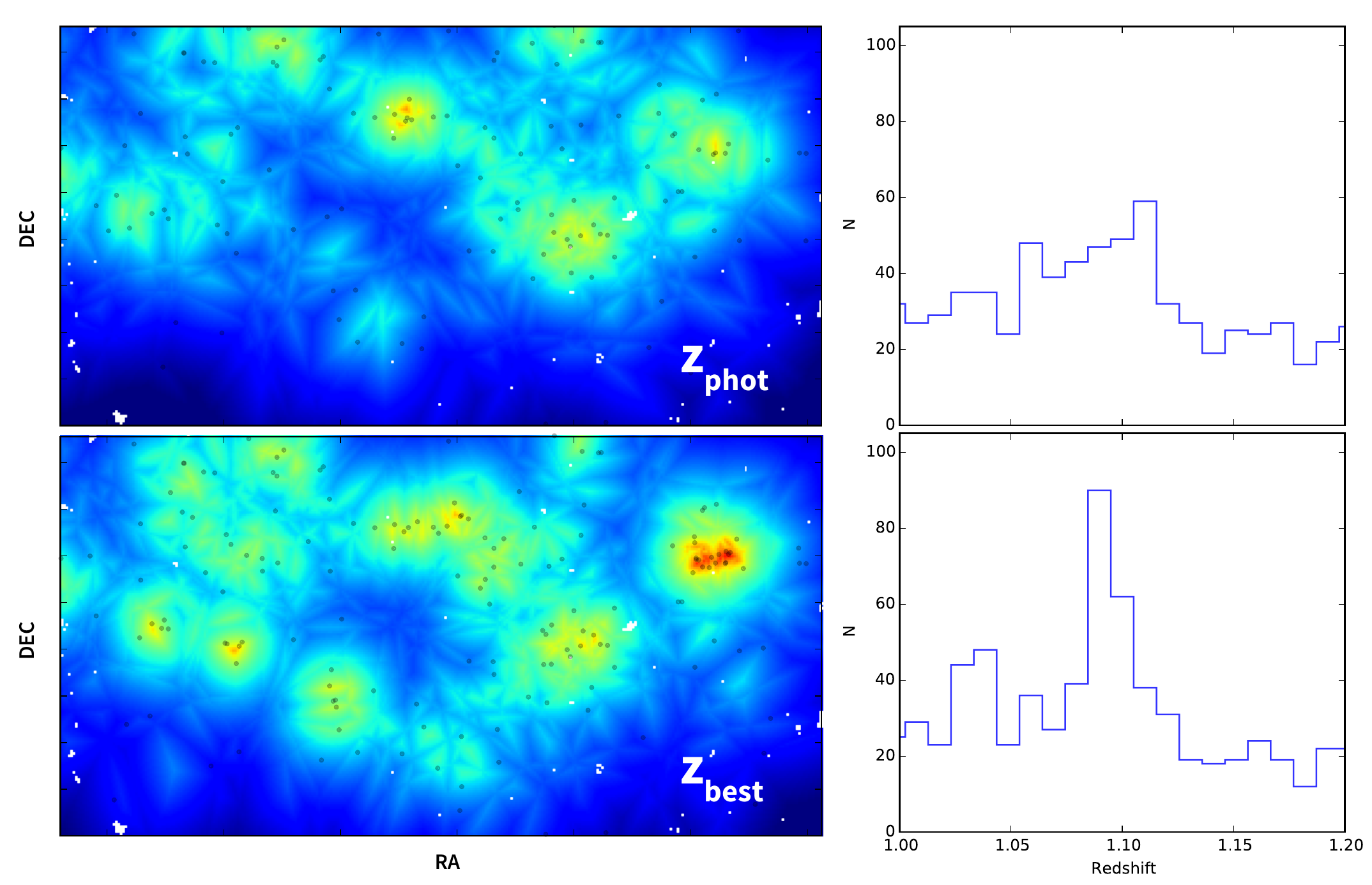}
\caption{\footnotesize Example of the ability of the grism redshifts to identify overdensities and characterize the environment of galaxies. The panels show the distribution of galaxies in the UDS field, in a narrow redshift bin between 1.07 and 1.11. {\it Left: } Smoothed 5th nearest neighbour maps using the {\tt z\_phot} and {\tt z\_best} redshifts for the $\JH<24.$ sample. {\it Right: } Redshift histograms. The overdensity at $z=1.09$ is clearly defined in the grism redshift distribution.}
\label{fig:overdensity}
\end{figure*}

\section{Properties of the 3D-HST Data Products}
\label{sec:properties}

Here we briefly summarize what 3D-HST contributes to existing datasets and catalogs that are based on deep, ``blank'' fields. The immediate contributions of the grism spectroscopy are a uniform, complete redshift catalog with relatively small and well-understood uncertainties; emission line fluxes; and 2D emission line maps. Furthermore, the combination of these data with stellar masses determined from SED fits, UV+IR star formation rates, and WFC3 morphologies constitutes the most complete dataset to date for studies of ``normal'' galaxies out to $z\sim 3$.

\subsection{Redshifts and Redshift Distribution}

\begin{figure}[th]
\includegraphics[width = 0.5\textwidth]{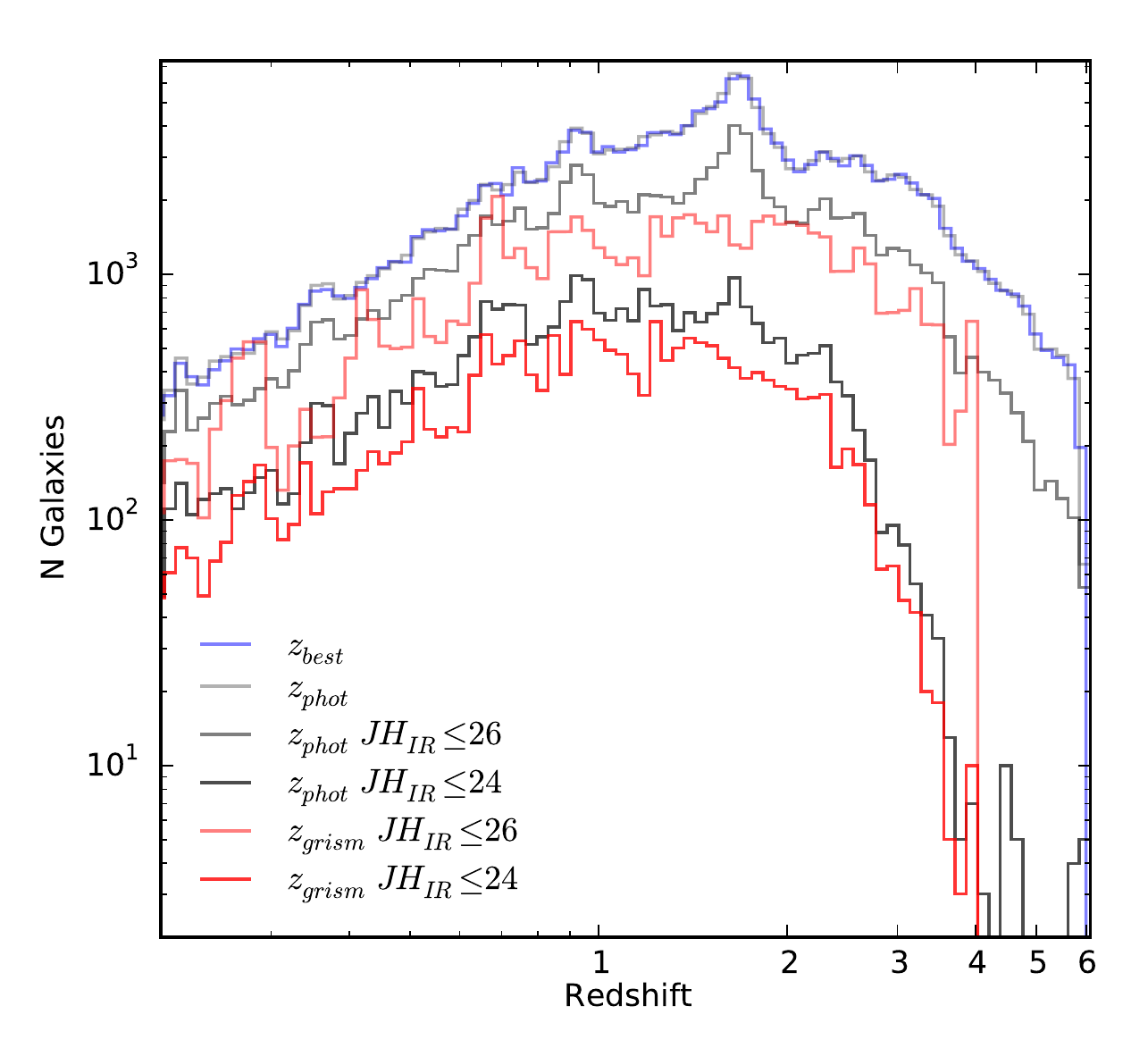}
\caption{\footnotesize Redshift distributions of the catalogs in this paper. Distributions that are derived from the full photometric + grism fits are shown in red/pink. Distributions that are based on the photometry only are shown in black/grey. The grism data produces more pronounced peaks in the redshift distributions, as expected. Note that the (spurious) broad peak at $z\sim 1.6$ in the photometric redshift distributions is not present in the grism redshift distributions.}
\label{fig:redshifts_all}
\end{figure}

The accuracy of the redshifts is discussed in detail in Section 5: we find that it is $\approx 0.003 \times (1+z)$ for most galaxies, with some dependencies on magnitude, rest-frame color, and redshift (mostly reflecting an underlying dependency on whether a bright emission line is in the observed wavelength range). Crucially, the formal uncertainty in the redshift is generally an excellent measure of the actual error (see Fig.\ \ref{fig:duplicates_z_err} and \ref{fig:duplicates_summary}). The error corresponds to a velocity uncertainty of $\sim 1000$\,km/s.

\begin{figure*}
\centering
\includegraphics[width = \textwidth]{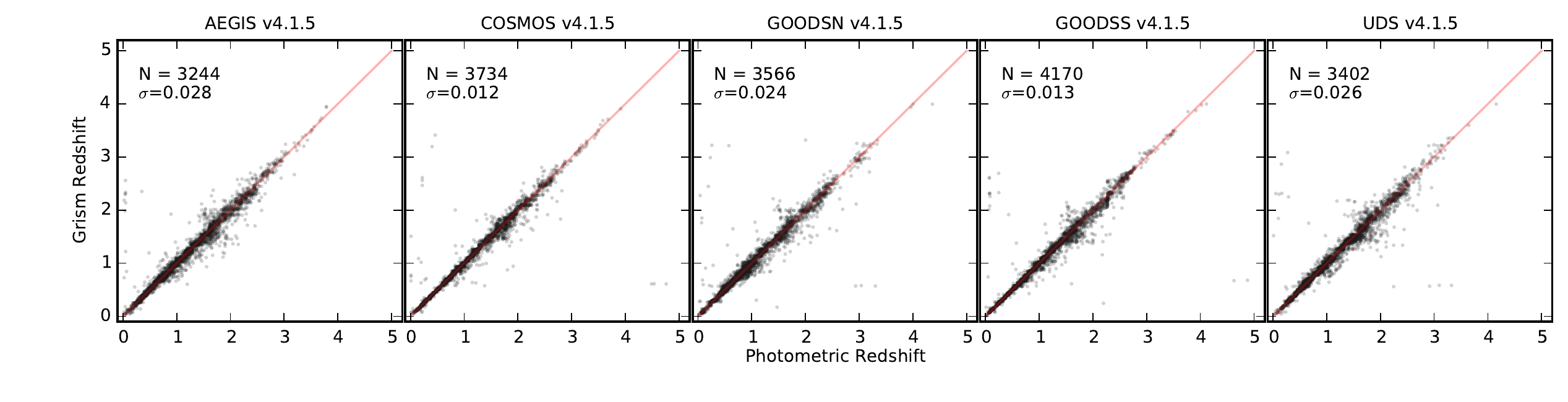}
\caption{\footnotesize Comparison between grism and photometric redshifts. Horizontal features are real structures in redshift space. Vertical features indicate spurious ``attractors'' in photometric redshift determinations.}
\label{fig:redshift_phot_grism}
\end{figure*}

The redshift accuracy that is achieved makes it possible to identify overdensities, and characterize the environment of galaxies, with much better contrast than with photometric redshifts alone. This is illustrated in Fig.\ \ref{fig:overdensity}, which shows the spatial distribution of galaxies in the UDS in small redshift bins between $z=1.07$ and $z=1.11$. The left panels show smoothed $5^{th}$ nearest neighbor density maps based on photometric redshifts (top) and grism redshifts (bottom), and the right panels show the corresponding redshift histograms. The structure at $z=1.09$ is clearly defined as a narrow grism redshift peak, but is spread out in the photometric redshifts.

Figure \ref{fig:redshifts_all} shows the redshift distributions based on all catalogs presented in this paper. The distribution shows a broad peak between $z=1$ and $z=2$, due to a combination of the observed-frame magnitude limits, the luminosity function of galaxies, and volume effects. For a particular magnitude limit the distributions of grism redshifts (red or pink) is always below that of photometric redshifts (black or grey), due to the fact that not all objects have a usable grism spectrum. The grism redshift distribution for $JH_{\rm IR}<24$  shows more pronounced peaks than the photometric redshift distribution; this is because physically-associated galaxies in groups and clusters have more accurate redshifts in the grism catalog. The same behavior is seen in the fainter sample with $JH_{\rm IR}<26$, but we emphasize that the grism redshifts for these faint objects were not inspected.
Strikingly, the pronounced photometric redshift peak at $z\sim 1.6$ in both the bright and the faint sample is not visible in the grism redshift distribution. This should be regarded as a success of our methodology: this peak is a well-known (but not well understood) artifact in photometric redshift measurements \citep[see, e.g.,][]{Skelton14,Brammer08}.

The differences between photometric redshifts and grism redshifts are shown explicitly in Figure~\ref{fig:redshift_phot_grism}. Horizontal features in this Figure are overdensities that are more clearly identified in the grism redshift distribution. Vertical features indicate ``attractors'' in photometric redshift; the most prominent of these is the broad peak at $z\sim 1.6$.  Note that galaxies with $z_{\rm phot} = z_{\rm grism}$ do not necessarily have highly accurate photometric redshifts; these can also be cases where the grism spectrum does not add significant information to the fit and both redshifts are essentially determined by the photometry alone. The accuracy of photometric redshifts is discussed in \citet{Bezanson15}.

\subsection{Spectral Features}

As discussed in Section \ref{sec:line_fits} the catalogs contain flux and equivalent width measurements, with well-calibrated uncertainties, for every emission line of Table 4 that falls in the observed wavelength range for a particular object.  The emission lines are measured for every extracted spectrum down to $JH_{IR}=26$, but we only supply use flags for galaxies with $JH_{IR}<24$.
Several papers and projects have used early versions of these catalogs; as an example, both the MOSDEF \citep{Kriek15} and KMOS$^{3D}$ \citep{Wisnioski15} surveys have used 3D-HST line measurements to select objects for follow-up spectroscopy with ground-based spectrographs. 

A comprehensive study of the line fluxes is beyond the scope of the present paper; the H$\alpha$ emission line luminosities and stellar absorption features are analyzed in \citet{Fumagalli12,Fumagalli15a} and \citet{Fumagalli15b,Whitaker13} respectively. Here we illustrate the relation between the strengths of various emission lines and other galaxy properties in a series of 2D stacks. These stacks are created by ordering the G141 spectra by a particular parameter, such as redshift or stellar mass. Then a 2D surface is generated with (observed or rest-frame) wavelength on the $x$-axis and the sorting parameter on the $y$-axis.\footnote{See http://www.sdss.org/science/ for an example of such a 2D stack of 46,420 SDSS quasars, created by X.\ Fan.} Rather than show all spectra they are binned in small intervals of the sorting parameter, such that each line corresponds to the median of many spectra.

\begin{figure*}
\centering
\includegraphics[width = \textwidth]{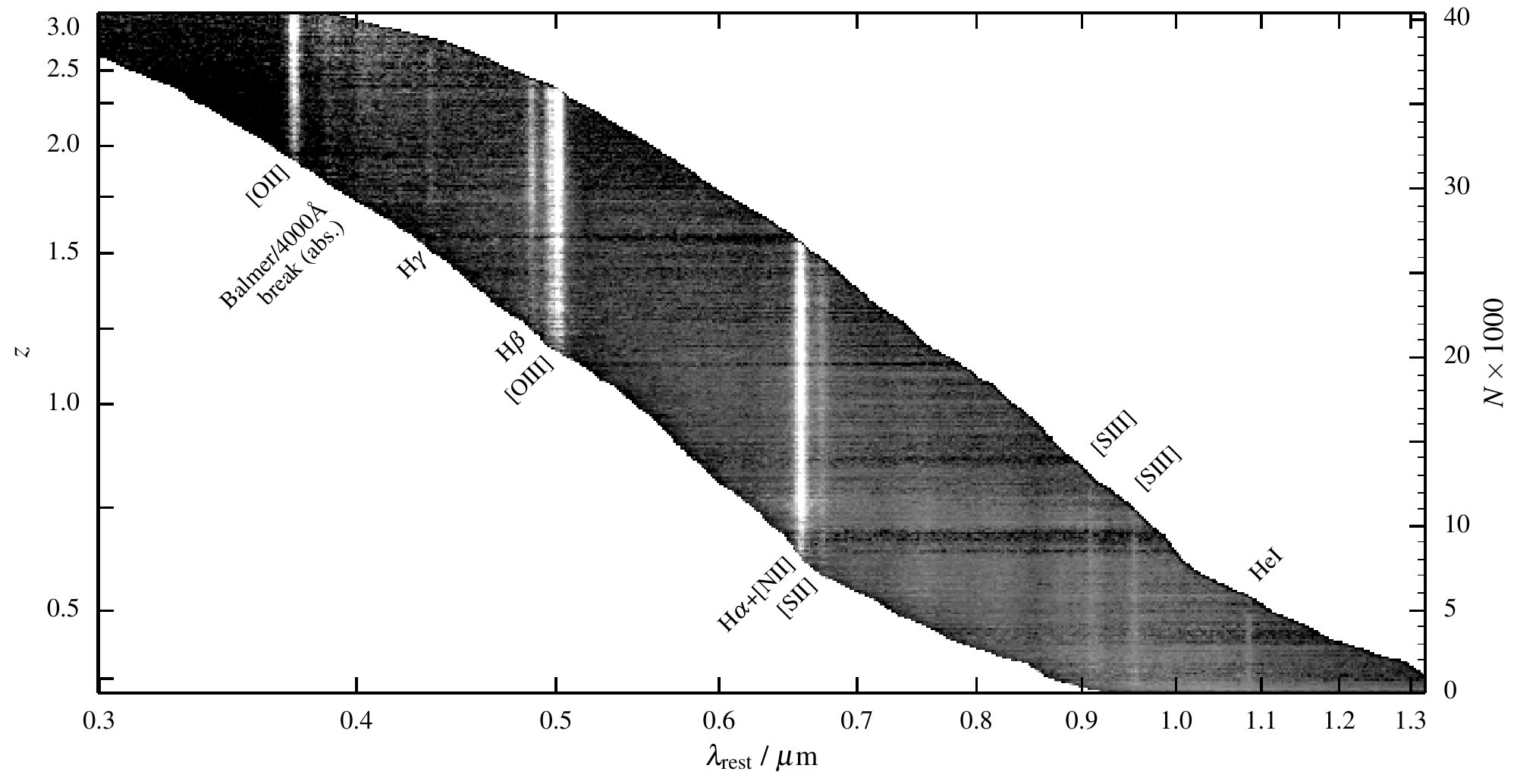}
\caption{\footnotesize Overview of $\sim 40,000$ 3D-HST G141 grism spectra with $H_{160}<25$.  Each pixel row shown is the median of 100 individual 1D spectra sorted by redshift and shifted to the rest frame; ticks on the right axis mark every 1000 galaxies, and tick labels on the left axis indicate the corresponding redshift.  Each spectrum is normalized by the object's $JH_{140}$ flux.  Absorption and emission lines that move through the G141 passband at different redshifts are indicated.}
\label{fig:moap_all}
\end{figure*}

\begin{figure*}
\centering
\includegraphics[width = \textwidth]{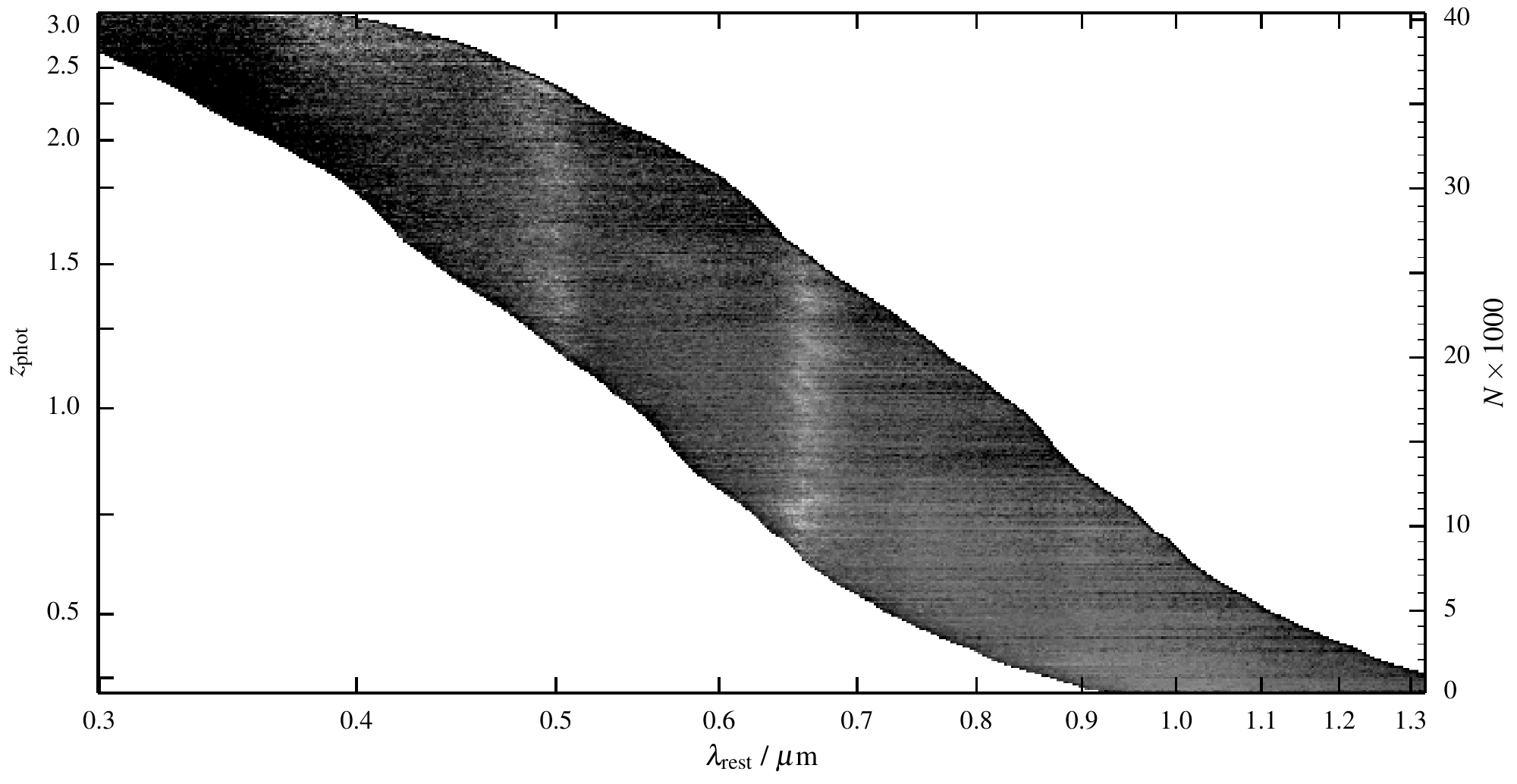}
\caption{\footnotesize Same as Fig.\ \ref{fig:moap_all}, but using photometric redshifts rather than grism redshifts to order the spectra. The differences between this Figure and Fig.\ \ref{fig:moap_all} graphically illustrate the improvement in the redshift accuracy when going from photometric redshifts to grism redshifts.}
\label{fig:moap_zphot}
\end{figure*}

\begin{figure*}
\centering
\includegraphics[width = \textwidth]{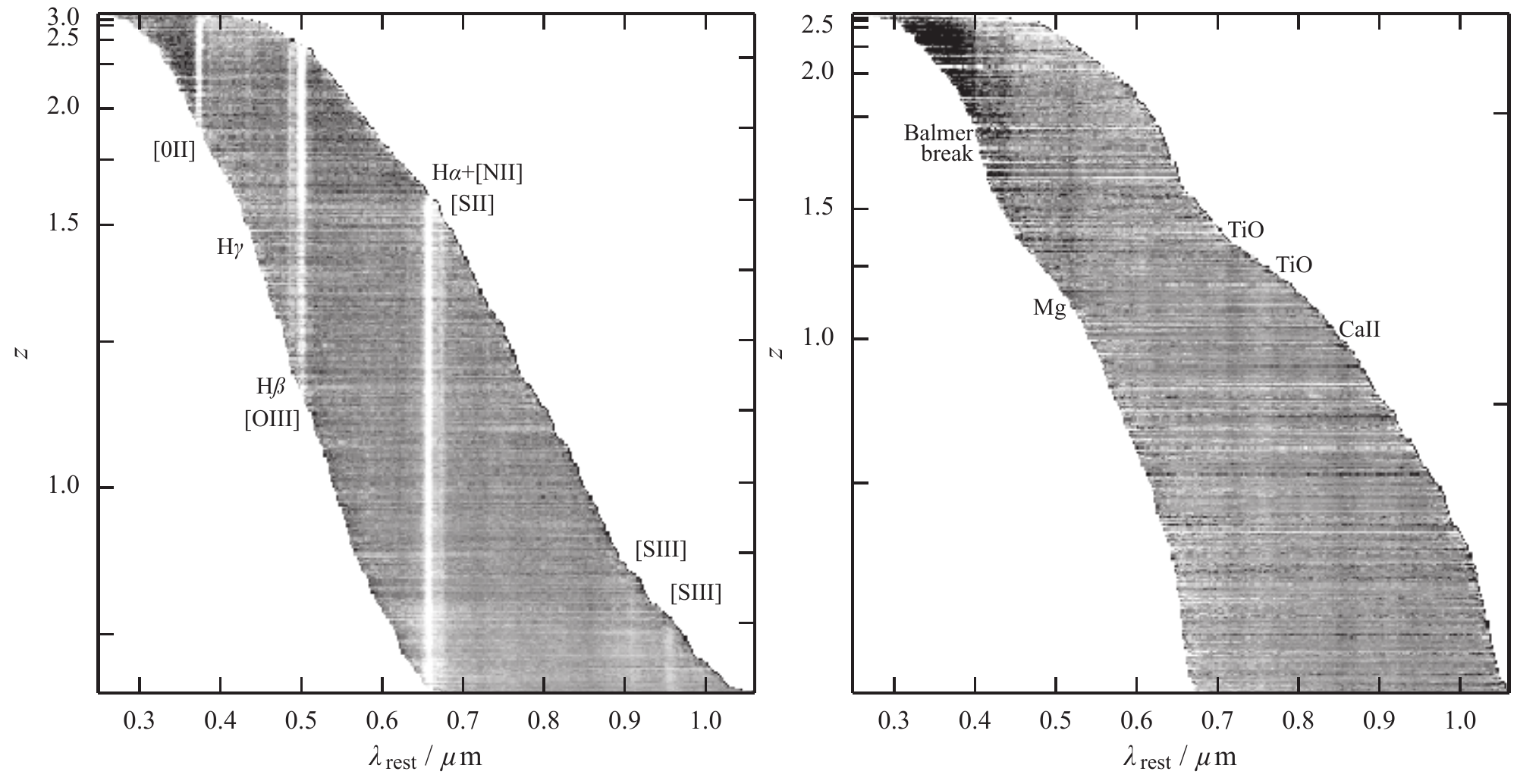}
\caption{\footnotesize Same as Fig.\ \ref{fig:moap_all} but split by emission line properties and only showing galaxies with $z>0.605$. Galaxies in the left panel have at least one emission line with a S/N ratio greater than 3. Galaxies in the right panel have a relatively bright magnitude limit ($H_{160}<23$) and no detected emission lines with a S/N ratio greater than 2. As in Fig.\ \ref{fig:moap_all} each tickmark on the right vertical axis corresponds to 1000 spectra. The survey contains $>2000$ spectra of relatively bright quiescent galaxies.}
\label{fig:moap_split}
\end{figure*}

\begin{figure*}
\centering
\includegraphics[width = \textwidth]{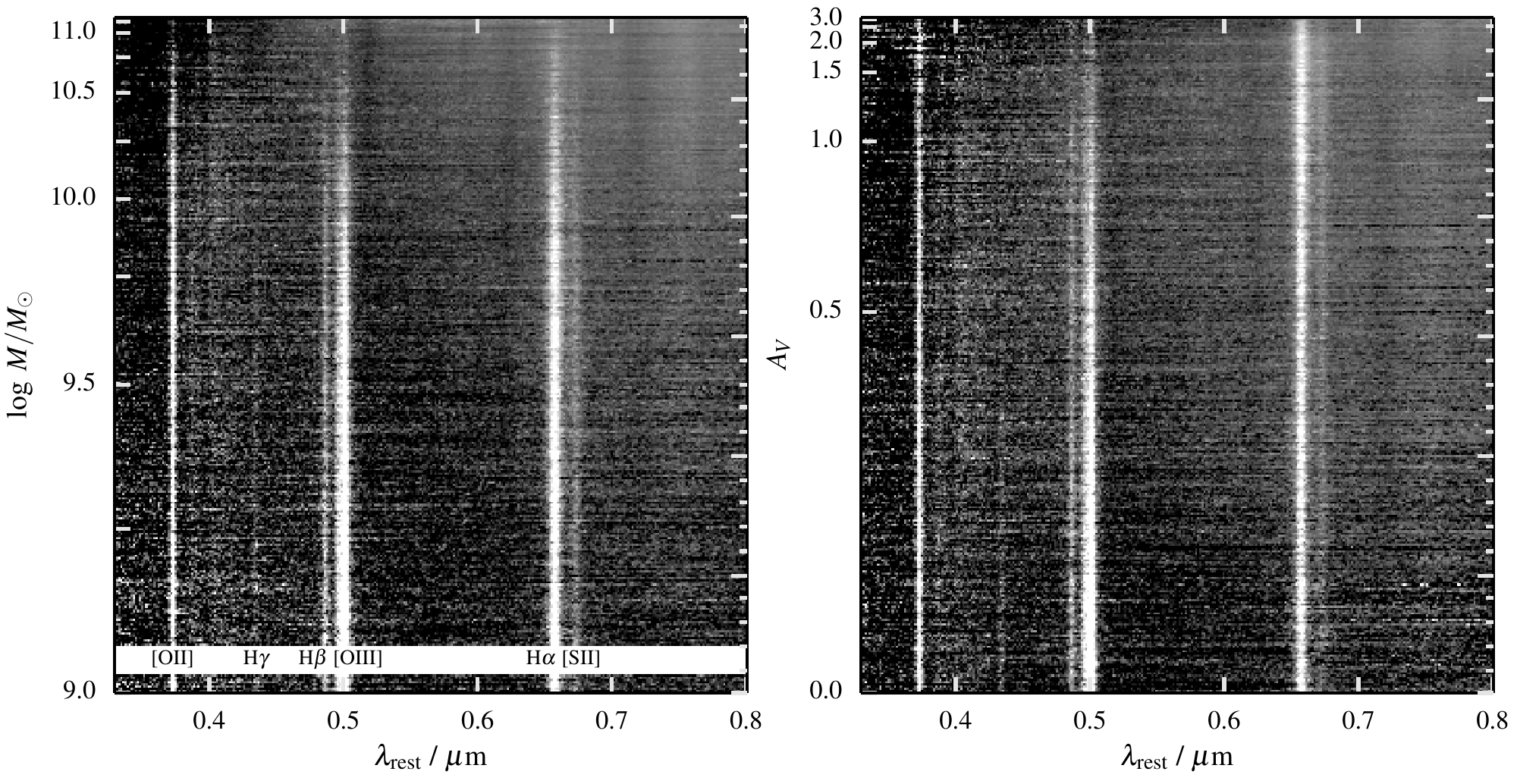}
\caption{As Fig.\ \ref{fig:moap_all} but with objects sorted by $M_{\ast}$ (left panel) and continuum dust extinction ($A_V$, right panel), both determined from stellar population synthesis fits to the broad band photometry.  Here galaxies with a range of redshifts contribute to each row, providing rest-frame spectra from 3300--8000~\AA.  There are clear trends: higher mass galaxies have weaker emission lines and stronger absorption lines, and galaxies with higher continuum extinction have stronger Balmer decrements (see text).}
\label{fig:moap_mass_av}
\end{figure*}

Figure \ref{fig:moap_all} shows the ``basic" 2D stack where redshift is the sorting parameter and the horizontal axis is rest-frame wavelength. The selection is $0.15<z<3.3$ and $H_{160}<25$, with some additional constraints on the quality of the spectra. Each line is the median of 100 individual 1D spectra. Each of the 100 spectra was normalized by the object's $JH_{140}$ flux prior to taking the median. Therefore, the intensity of emission and absorption lines in Fig.\ \ref{fig:moap_all} (and subsequent figures) corresponds approximately to their equivalent widths, and not to their fluxes or luminosities. Redshift is shown on the left vertical axis; the cumulative number of spectra is shown on the right axis. Figure \ref{fig:moap_zphot} shows the same spectra as Fig.\ \ref{fig:moap_all}, but ordered by their photometric redshift rather than grism redshift. The differences between this Figure and Fig.\ \ref{fig:moap_all} is a qualitative demonstration of the improved redshift accuracy that is enabled by the grism spectroscopy. The accuracy of photometric redshifts is discussed in detail in \citet{Bezanson15}.

In Fig.\ \ref{fig:moap_split} galaxies are split by their emission line properties. Galaxies in the left panel have $z>0.605$, $H_{160}<24$, and the S/N ratio of [O{\sc ii}], [O{\sc iii}], and/or H$\alpha$ greater than 3. In the right panel galaxies have $z>0.605$, $H_{160}<23$, and a S/N ratio of all lines $<2$.
Several well-known emission lines can readily be identified in the left panel of Fig.\ \ref{fig:moap_split} as they shift into and out of the observed wavelength range of the G141 grism. At $z\gtrsim 0.7$, the prominent H$\alpha$ line and the S{\sc ii} [6718,6733] doublet are visible. Over the redshift range $0.7<z<1.5$ the H$\alpha$ equivalent width gradually increases, as discussed by \citet{Fumagalli12}. At $z\gtrsim 1.3$, H$\beta$ and the [O{\sc iii}] [4959,5007] doublet fall in the observed wavelength range.  Again, we see that the equivalent widths of emission lines (in this case [O{\sc iii}]) increase with redshift in broad band flux-limited samples. The small redshift range where H$\alpha$ and H$\beta$ are both covered by the G141 grism was utilized by, e.g., \citet{Price14}, who study the reddening of H{\sc ii} regions as measured by the Balmer decrement in 3D-HST galaxies. Finally, at $z\gtrsim 1.9$ the [O{\sc ii}] doublet enters the grism wavelength range.

Absorption features (right panel of Fig.\ \ref{fig:moap_split}) are not measured automatically in our analysis. Such measurements are highly dependent on the spectral resolution and the precise definitions for the line and continuum wavelength regions. We note that our spectral resolution is too poor to use common definitions such as the Lick system \citep{Worthey94}. We {\em can} use our data for full spectrum fitting techniques \citep[see][]{Conroy14}, as demonstrated in \citet{vanDokkum10} and \citet{Whitaker13}. Prominent absorption features include multiple TiO molecular bands at low redshift, the Mg$_2$ $\lambda 5170$ feature at $1.2\lesssim z
\lesssim 2.2$, and the Balmer break at the highest redshifts.

We show 2D stacks with a physical galaxy property (rather than redshift) as the sorting parameter in Fig.\ \ref{fig:moap_mass_av}. Each line in these stacks was created from spectra at a wide range of redshifts; this is the reason why the rest-frame wavelength coverage is much larger than in Fig.\ \ref{fig:moap_all}. In the left panel the galaxies are sorted by their stellar mass, as determined from stellar population synthesis models \citep[see][for a detailed discussion]{Skelton14}. The dependence on mass is striking: at low masses the galaxies have strong emission lines, and at high masses the emission lines are weak or absent. The spectra also become gradually redder with increasing mass, and the Balmer break becomes more prominent. In the right panel the sorting parameter is the continuum extinction $A_V$, again determined from fits to the broad band SEDs. The spectra again become redder with increasing $A_V$, as expected. Rather strikingly, the Balmer decrement H$\alpha$/H$\beta$ increases strongly with $A_V$. \citet{Price14} demonstrate this effect using earlier 3D-HST catalogs and a narrow redshift range, and it is very clear in this 2D stack which uses a larger number of spectra and combines data from a wide range of redshifts.

\subsection{Spatially-resolved Emission Lines}

Arguably the most unique contribution of 3D-HST is the fact that all emission lines are {\em imaged} at HST's superb resolution. For each object, the grism effectively places images at different wavelengths next to each other on the detector, with each subsequent image 23\,\AA\ (in interlaced space) separated from its neighbours. As a result, if an object is particularly bright in a single emission line, the grism will produce a complete image of the object in the light of that line \citep[see][for a more in-depth explanation]{Nelson15}.  The only datasets that can achieve something comparable are obtained with laser guide star assisted adaptive optics (AO) observations with integral field units (IFUs) on large telescopes \citep[e.g.,][]{Genzel06}. These AO observations yield only one object at a time, and even though the diffraction-limited performance of VLT and Keck is superior to that of HST, the AO-delivered PSF generally has a much poorer Strehl ratio than the HST PSF.

As the lines are broad for large galaxies it is generally not trivial to obtain these maps from extracted spectra \citep[see][]{Nelson12}. We therefore provide continuum-subtracted maps in the data release, which can be directly used. Examples of these maps are shown in the bottom panels of the 2D grism spectra shown in Fig.\ \ref{fig:examples}. The power of these spatially-resolved line emission maps is demonstrated in several 3D-HST papers. \citet{Nelson12,Nelson13,Nelson15} study the spatial distribution of H$\alpha$ emission in galaxies at $z\sim 1$, in different bins of mass and star formation rate. \citet{Brammer12arc} show the G141 spectrum of a spectacular lensed galaxy with very strong emission lines. The 2D spectrum (Fig.\ 2 in that paper) demonstrates that the grism provides images of the arc in the light of a range of different emission lines. \citet{Wuyts13} compare the spatial distribution of H$\alpha$ emission in a sample of relatively bright galaxies to that of the rest-frame UV emission.

\section{Summary}
\label{sec:summary}

In this paper we have described the observations and data products of
the 3D-HST Treasury program. This is a companion study to the photometric analysis
presented in \citet{Skelton14}, and together these two papers present a comprehensive
photometric and spectroscopic wide-field dataset for studies of the distant Universe.
All data products are available through the 3D-HST website.\footnote{\url{http://3dhst.research.yale.edu/}}.

These datasets, together with structural parameters and star formation rates presented elsewhere \citep{vdWel14,Whitaker14}, accomplish an important goal of observational extragalactic astronomy: a census of stars and star formation in reasonably bright galaxies out to $z\sim 2.5$. The main source of uncertainty is shifting from errors in counting to errors in interpreting: systematic uncertainties in stellar masses, star formation rates, and gas-phase metallicities are beginning to dominate in the regime discussed in this paper. There are excellent prospects to extend the work described in this paper to fainter objects and larger areas: the {\it James Webb Space Telescope}, {\it WFIRST} and {\it Euclid} will use multi-slit and slitless near-IR spectroscopy to characterize the galaxy population in regions of parameter space that are beyond the capabilities of the WFC3 camera on {\it HST}.

\acknowledgements
This work is based on observations taken by the 3D-HST Treasury Program (GO 12177 and 12328) as well as GO 11600 and GO 13420 with the NASA/ESA Hubble Space Telescope, which is operated by the Association of Universities for Research in Astronomy, Inc., under NASA contract NAS5-26555. RB and KEW gratefully acknowledge support by NASA through Hubble Fellowship grants \#HST-HF-51318.001  and \#HST-HF2-51368 awarded by the Space Telescope Science Institute, which is operated by the Association of Universities for Research in Astronomy, Inc., for NASA, under contract NAS5-26555.

\bibliography{grism_release}

\appendix{}

\section{Removing Individual Reads from FLT Frames}

Our process for removing bad reads is detailed in Sect.\ \ref{sec:read_removal} in the main text. Briefly, we use the fact that the WFC3 IR camera does multiple non-destructive reads per exposure (12 in the case of our G141 exposures). Particular problems such as satellites passing in the field of view, or Earthshine, only affect one or a few of these 12 independent samples, and we can reconstruct a clean exposure by removing the offending reads. 
Here we provide a list of all the removed reads (Table \ref{table:popped_reads}), as well as examples of the removal of the effects of Earthshine and satellite trails.

\begin{figure*}
\centering
\includegraphics[width = \textwidth]{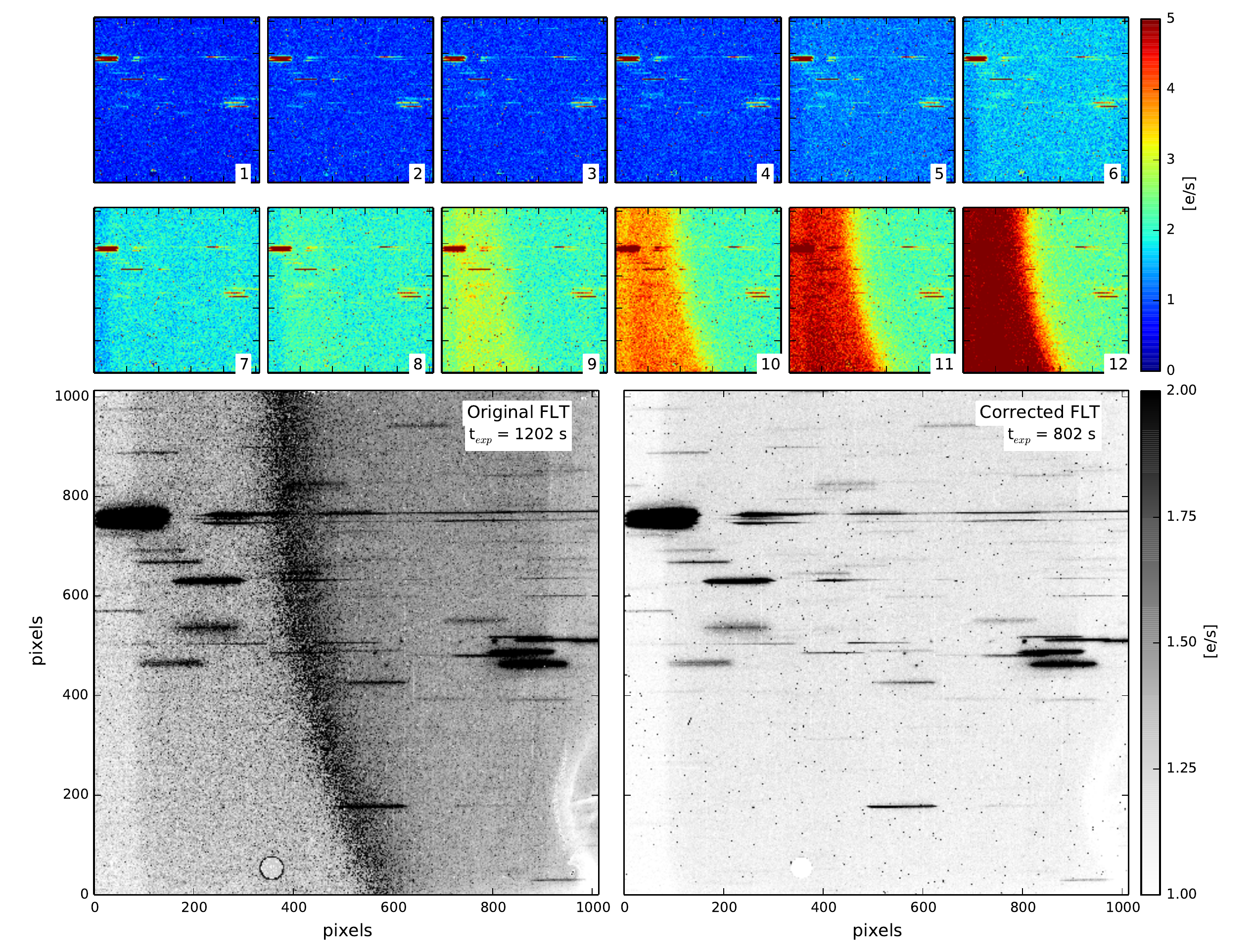}
\caption{\footnotesize Main left panel: a pipeline-processed \textsc{flt} file which is affected by Earthshine. Top rows: individual reads, which show that only the last four reads in the exposure are affected. Main right panel: corrected \textsc{flt} file after removing the problematic reads.}
\label{fig:earthshine}
\end{figure*}

Figure \ref{fig:earthshine} shows the effect of Earthshine on an exposure, and demonstrates how we remove it.
The example in this figure is one of the \textsc{flt} images of pointing AEGIS-01: ibhj39uuq\_flt.fits. The original \calwf\ pipeline-processed \textsc{flt} exposure is in the left main panel: the Earthshine produces a highly structured background, and an apparently unusable exposure. The top two rows show the differences between each sequential pair of non-destructive reads of the WFC3 detector, converted to units of e$^{-}$\,s$^{-1}$. It is clear that only the last four reads are affected. The  In this example, we remove the last four reads in the sequence. The right main panel shows the corrected \textsc{flt} image after removing the last four reads. The corrected \textsc{flt} has 30 percent lower exposure time, however, without the correction, the full exposure would have been unusable. 

\begin{figure*}
\centering
\includegraphics[width = \textwidth]{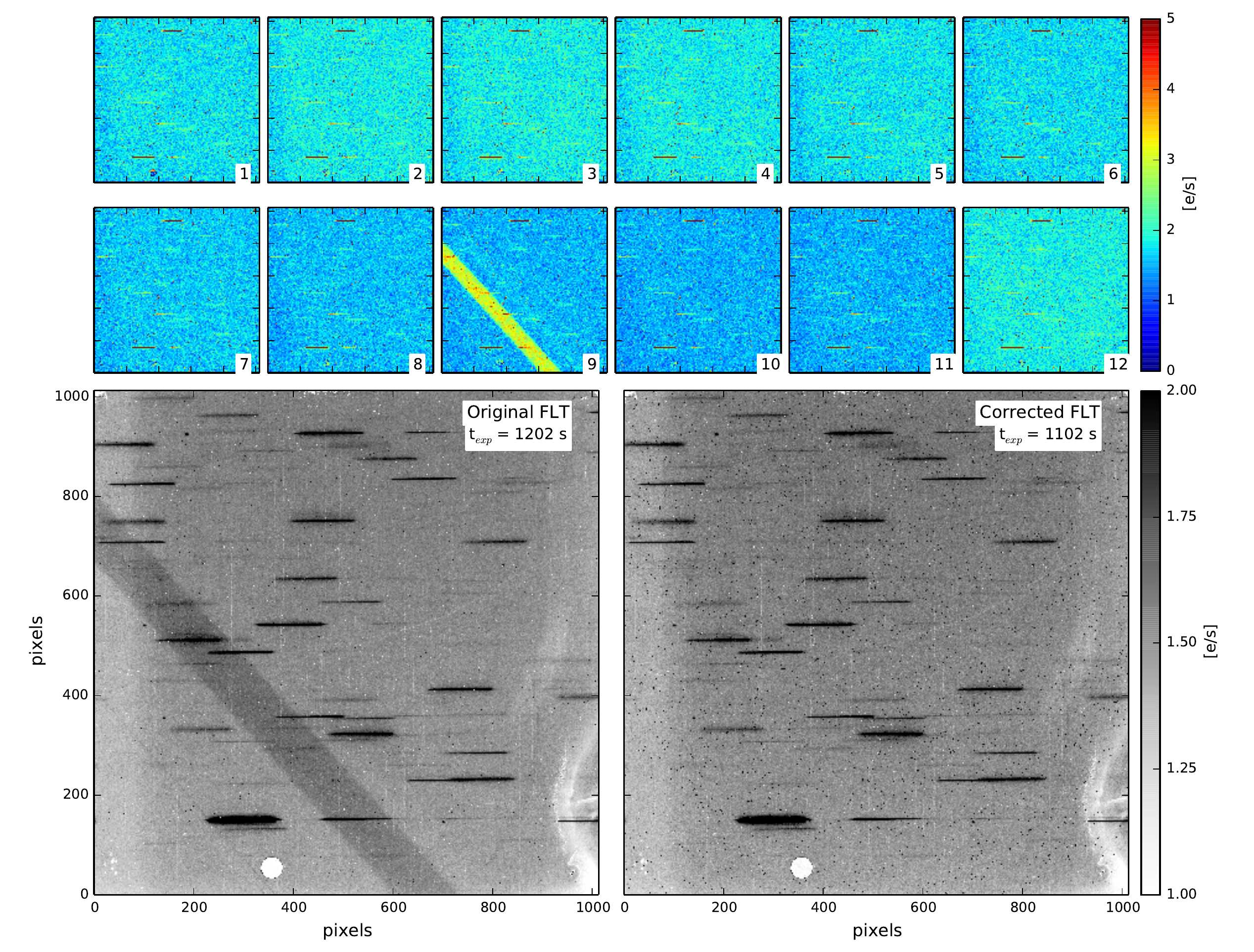}
\caption{\footnotesize Example of satellite removal. The structure of the figure is the same as Fig.~\ref{fig:earthshine}. The sample obtained in the 9$^{\rm th}$ read was removed in the corrected frame.}
\label{fig:satellite}
\end{figure*}

The process of removing satellite trails is illustrated in Fig.\ \ref{fig:satellite}, which has the same structure as Fig.\ \ref{fig:earthshine}. Here the example is one of the \textsc{flt} images of pointing GOODSS-10: ibhj10vmq\_flt.fits. A satellite moved across the observed field between reads 8 and 9. After removing the sample obtained in read 9 the corrected \textsc{flt} file shows no trace of the satellite trail. As only a single read was removed the exposure time of the corrected frame is only 100 seconds shorter than the uncorrected one.

\begin{table}[ht]
\centering
\caption{Removed Reads}\label{table:popped_reads}
\begin{tabular}{lcllcclcllc}
\cline{1-5} \cline{7-11}
\cline{1-5} \cline{7-11}
\noalign{\smallskip}
Field & Pointing & FLT & Removed Reads & Defect  & & Field & Pointing & FLT & Removed Reads & Defect  \\
\noalign{\smallskip}
\cline{1-5} \cline{7-11}
\noalign{\smallskip}
AEGIS    &    01        &    ibhj39uuq    &    [9,10,11,12]  &  E & &  GOODS-N &    11   &    ib3701s4q    &    [1,2,3]    		&  E  \\
              &         &    ibhj39viq    &    [11,12,13]    &  E & &        &         &    ib3701skq    &    [1,2,3]    		&  E  \\
              &    04   &    ibhj42oqq    &    [10]    		 &  S & &         &    12   &    ib3702u8q    &    [1,2,3,4,5,6,7]    & E   \\
              &    21   &    ibhj59cvq    &    [3]    		 &  S & &         &         &    ib3702uoq    &    [1,2,3,4,5,6,7,8]  &  E  \\
              &    28   &    ibhj66dfq    &    [3]    		 &  S & &         &    13   &    ib3703uzq    &    [1,2,3,4,5,6,7]    & E   \\
              &    30   &    ibhj69hgq    &    [10,11,12]    &  E & &         &         &    ib3703vfq    &    [11,12,13]    		&  E  \\
\noalign{\vspace{1.5pt}}
\cline{1-5}
\noalign{\vspace{1.5pt}}
COSMOS  &    03        &    ibhm31rcq    &    [3]    &   S & &         &         &    ib3703vmq    &    [1,2,3,4,8]    	&  E  \\
        &    08        &    ibhm36ksq    &    [11]   &   S & &         &    14   &    ib3704wrq    &    [1,2]    			&  E  \\
        &    10        &    ibhm38dfq    &    [5]    &   S & &         &         &    ib3704x8q    &    [6,7,8,9,10,11,12,13]    &  E  \\  
        &    17        &    ibhm45y1q    &    [6]    &   S & &          &    17   &    ib3719v7q    &    [5]    			&  S   \\
        &    23        &    ibhm51x2q    &    [5]    &   S & &          &    21   &    ib3705y1q    &    [1,2,3,4,5,6]    	&   E \\
        &    25        &    ibhm53o3q    &    [1]    &   S & &         &         &    ib3705y5q    &    [4]    			&  S   \\
\noalign{\vspace{1.5pt}}
\cline{1-5}
\noalign{\vspace{1.5pt}}
GOODS-S     &    10        &    ibhj10vmq    &    [9]    &   S & &         &         &    ib3705ylq    &    [4,5,6,7,8,9,10]   &  E  \\
                  &    30        &    ibhj30bzq    &    [6]    &  S  & &     &    22   &    ib3706b2q    &    [2,3,4,5,6,7]    	&  E  \\       
                  &    36        &    ibhj36j3q    &    [10]    &  S  & &        &         &    ib3706biq    &    [12,13]    		&  E  \\ 
                  &    37        &    ibhj37uvq    &    [11]    &   S & &   &         &    ib3706bpq    &    [1,2,3]    		&  E  \\
\noalign{\vspace{1.5pt}}
\cline{1-5}
\noalign{\vspace{1.5pt}}
UDS    &    05        &    ibhm05fjq    &    [7]    &   S & &         &    23   &    ib3707caq    &    [1,2]    			&  E  \\
           &    13        &    ibhm13kfq    &    [6,7]    &  S  & &         &         &    ib3707cqq    &    [7,8,9,10,11,12]   &  E  \\
           &    14        &    ibhm14vhq    &    [9]    & S  & &          &         &    ib3747zyq    &    [11]    			&  S   \\
           &    18        &    ibhm18oeq    &    [2]    &  S  & &         &         &    ib3747a5q    &    [5]    			&  S   \\
           &    19       &    ibhm19pjq    &    [9]    &  S  & &         &    24   &    ib3708i5q    &    [7,8,9]    		& E   \\
           &    20        &    ibhm20l1q    &    [10]    &  S & &          &         &    ib3708ipq    &    [2,3,4,12]    		&  E  \\
           &    24        &    ibhm24f8q    &    [3]    &  S  & &         &    28   &    ib3724riq    &    [13]    			&  S  \\
           &    26        &    ibhm26n0q    &    [9]    &  S  & &         &    31   &    ib3709j3q    &    [7,8,9,10]    		&  E  \\
           &    27        &    ibhm27qiq    &    [11]    & S  & &          &         &    ib3709joq    &    [8,11,12,13]    	&  E  \\
         & & & & & &         &         &    ib3749o5q    &    [3]    			& S    \\
        & & & & & &         &         &    ib3749oqq    &    [11]    			&  S   \\
        & & & & & &         &    32   &    ib3710nmq    &    [12,13]    		&  E  \\
        & & & & & &         &    33   &    ib3711bkq    &    [4]    			&  E  \\
        & & & & & &         &    36   &    ib3726bpq    &    [13]    			& S    \\
        & & & & & &         &         &    ib3726c5q    &    [10]    			& S    \\
        & & & & & &         &    44   &    ib3716psq    &    [12,13]    		&  E  \\

\noalign{\smallskip}
\cline{1-5} \cline{7-11}
\noalign{\smallskip}
\end{tabular}
\end{table}

\end{document}